\documentclass[11pt,tightenlines,eqsecnum,floats,aps,prd,showpacs,amsmath,superscriptaddress,amssymb,nofootinbib,longbibliography]{revtex4-1}
\usepackage{graphicx, wrapfig}
\usepackage{amssymb}
\usepackage[usenames, dvipsnames]{color}
\usepackage{mathrsfs}
\usepackage{subfigure}
\usepackage{graphicx}
\usepackage{verbatim}
\usepackage{cancel}
\usepackage[normalem]{ulem}
\usepackage[hidelinks]{hyperref}
\hypersetup{
  colorlinks   = true, 
  urlcolor     = blue, 
  linkcolor    = blue, 
  citecolor   = blue 
}
\def\t{\tilde}
\def\f{\frac}
\def\h{\hat}
\def\v{\vec}

\def\dd{\textrm{d}}
\def\d{\textrm{d}}

\newcommand*{\backin}{\rotatebox[origin=c]{-180}{$\in$}}%

\def\pp{p_{\phi}}
\def\dpp{\delta p_{\phi}}
\def\dph{\delta\phi}
\def\qz{\mathring{q}}
\def\pz{\mathring{\pi}}
\def\u{\mathfrak{A}}
\def\dd{\textrm{d}}
\def\etat{\tilde\eta}

\def\Nbstar{N_{{\rm B}\,\star}}

\def\dphi{\partial_\phi}

\def\l{\left}
\def\r{\right}

\usepackage{enumerate}

\newcommand{\ig}{\includegraphics}
\newcommand{\be}{\nopagebreak[3]\begin{equation}}
\newcommand{\ee}{\end{equation}}
\newcommand{\bfig}{\nopagebreak[3]\begin{figure}}
\newcommand{\efig}{\end{figure}}
\newcommand{\bea}{\nopagebreak[3]\begin{eqnarray}}
\newcommand{\eea}{\end{eqnarray}}

\newcommand{\bmult}{\nopagebreak[3]\begin{multline}}
\newcommand{\emult}{\end{multline}}

\begin{document}
\title{Non-Gaussianity in loop quantum cosmology}
\author{Ivan Agullo}\email{agullo@lsu.edu}
\affiliation{Department of Physics and Astronomy, Louisiana State University, Baton Rouge, LA 70803, U.S.A.
}
\author{Boris Bolliet}\email{boris.bolliet@manchester.ac.uk}
\affiliation{Department of Physics and Astronomy, Louisiana State University, Baton Rouge, LA 70803, U.S.A.
}
\affiliation{Jodrell Bank Centre for Astrophysics, The University of Manchester, Alan Turing Building, Oxford Road, Manchester, M13 9PL.}

\author{V. Sreenath}\email{vsreenath@iucaa.in}
\affiliation{Department of Physics and Astronomy, Louisiana State University, Baton Rouge, LA 70803, U.S.A.
}
\affiliation{Inter-University Centre for Astronomy and Astrophysics, Post Bag 4, Ganeshkhind, Pune 411007, India.}

\begin{abstract}
 We extend the  phenomenology of loop quantum cosmology (LQC) to second order in perturbations.  Our motivation is twofold. On the one hand, since  LQC predicts a cosmic bounce that  takes place at the Planck scale, the second-order contributions could be large enough to jeopardize the validity of the perturbative expansion on which previous results rest. On the other hand,  the  upper bounds on primordial non-Gaussianity obtained by the Planck Collaboration are expected to play a significant role on  explorations of the  LQC phenomenology.  We  find that the bounce in LQC produces an enhancement of non-Gaussianity of several orders of magnitude, on length scales that were larger than the  curvature radius at the bounce.  Nonetheless, we find that one can still rely on the perturbative expansion to make predictions about primordial perturbations. We discuss the consequences of our results for LQC and its predictions for the cosmic microwave background.

\end{abstract}


\maketitle

\newpage
\tableofcontents
\newpage

\section{Introduction} \label{sec:introduction}

The origin of the large scale cosmic structure can be traced back to quantum  vacuum fluctuations in the early universe, which were amplified by a  dynamical  gravitational field. The inflationary paradigm provides a theoretical framework to materialize this idea, and to make concrete predictions that  can be confronted with observations 
(see \cite{Steinhardetal,ResponsetoSteinhardetal} for a recent debate about the pros and cons of inflation). But despite the many interesting aspects of the inflationary scenario, the picture of the early universe that it provides remains  incomplete (for a list of open questions, see, e.g., \cite{Brandenberger:2012aj}). Among the  most important open issues is the fact that inflationary models suffer from the initial big bang singularity \cite{Borde:2001nh}, that makes us uncertain about the way inflation begins and  about  the initial state of the universe at the onset of inflation. This  point is particularly relevant, since the predictions for the cosmic microwave background (CMB) and large scale structure {\em depend} on what the initial state was. It would be more satisfactory to have a scenario in which inflation arises in a well-defined manner, free of singularities, and in which the dynamics of the pre-inflationary universe could be incorporated.

The idea that the universe did not begin with a big bang but rather it  {\em bounced}, transitioning from a contracting phase to an expanding one,  is an attractive possibility. Bouncing models have been considered since the early days of relativistic cosmology, e.g.\ by  de Sitter in 1931 \cite{1931Natur.128..706D}, and  more recently this idea has emerged in  more precise terms within different scenarios, including loop quantum cosmology (LQC) \cite{Ashtekar:2006rx,Ashtekar:2006uz,Ashtekar:2006wn,Ashtekar:2007em},  string theory-related models \cite{Brandenberger:2016vhg}, higher-derivative scalar-tensor theories \cite{Chamseddine:2016uef,Liu:2017puc}, etc.  In this paper, we focus on cosmological  bounces as predicted by loop quantum cosmology, although some of our results shall apply to other  models as well. 

In LQC (see \cite{Bojowald:2008zzb, Ashtekar:2011ni,MenaMarugan:2011va,Banerjee:2011qu,Agullo:2013dla,Calcagni:2012vw,Barrau:2013ula,Ashtekar:2015dja,Agullo:2016tjh,Grain:2016jlq,ElizagaNavascues:2016vqw,Wilson-Ewing:2016yan,Barrau:2016nwy} for review articles), the cosmic bounce is caused by quantum gravitational effects. This scenario has been used to provide a detailed  quantum gravity extension of the inflationary scenario \cite{Agullo:2012sh,Agullo:2013ai} in which  trans-Planckian issues of the inflationary paradigm are addressed from first principles. After the bounce,  as the value of matter energy density and curvature invariants  become smaller than the Planck scale, quantum gravitational effects quickly become irrelevant. In the presence of a scalar field $\phi$ and an appropriate potential $V(\phi)$, the matter content of the universe  becomes dominated by this potential soon after the bounce, and the universe generically enters an inflationary phase \cite{Ashtekar:2009mm,Ashtekar:2011rm,Bolliet:2017czc}. In this scenario,  scalar and tensor cosmological perturbations begin their evolution in the quantum vacuum at early times, and  then evolve across the bounce, until the onset of inflation, and beyond. One then can use this evolution to {\em compute} the state of perturbations at the onset of inflation, and to obtain predictions  for the CMB. The propagation across the bounce   leaves an imprint in scalar and tensor perturbations. 
If the state of perturbations at the onset of inflation  
happens to be completely different from the Bunch-Davies initial conditions normally postulated in standard inflation, existing observational constraints would  jeopardize the viability of the LQC proposal for the pre-inflationary universe \cite{Bolliet:2015raa}. On the other hand, if the resulting state is close enough to the Bunch-Davies vacuum at the onset of slow-roll, but still contains some differences, new effects would be predicted for the CMB temperature distribution.

In the last few years, a research program has been dedicated to quantitatively  analyze these possibilities (see  \cite{Bojowald:2011iq,Ashtekar:2011ni,MenaMarugan:2011va,Banerjee:2011qu,Agullo:2013dla,Calcagni:2012vw,Barrau:2013ula,Ashtekar:2015dja,Agullo:2016tjh,Grain:2016jlq,ElizagaNavascues:2016vqw,Wilson-Ewing:2016yan,Barrau:2016nwy,Barrau:2014maa,Mielczarek:2008pf}, and references therein).
More concretely, the primordial power spectra of perturbations have been analyzed in detail by different groups,  following different strategies. The main conclusions are that the bounce can leave an imprint on the largest scales probed by CMB, while still being compatible with current observational constraints. Concrete predictions have been obtained for the amplitude of the scalar and tensor power spectrum, spectral indices, and tensor-to-scalar ratio.\\

In this paper we argue that the analyses done so far for the primordial power spectrum  provides only a first step towards a complete comparison of the predictions of LQC with observations. In order to declare  the viability of the theoretical framework and  the compatibility of its predictions  with observations, one  has to  go to the next order in the perturbative expansion and show, first, that the next-to-leading order contribution introduces only small corrections, in such a way that the perturbative expansion on which  the computation rests is meaningful. But this is not enough, since  these corrections, although small enough to maintain the validity of perturbation theory,  could still give rise to large non-Gaussianity and violate observational upper bounds \cite{Ade:2015ava}. Such analysis was  done for the standard theory of inflation in \cite{Maldacena:2002vr}, and it was shown that higher order corrections and non-Gaussianity generated {\em during the slow-roll era} are indeed  small, consistent with CMB data. But the situation could be  different in presence of a cosmic bounce that takes place at a higher curvature. Non-Gaussianity arises from self-interactions between perturbations, and these are mediated by gravity. One  expects, from general arguments, that  these interactions would become `stronger' at higher curvatures. Since the bounce in LQC takes place at the Planck scale, there exists the possibility that the resulting non-Gaussianity is too large. Here we extend the analysis 
of scalar perturbations  in LQC to second order and investigate the non-Gaussianity generated by the LQC bounce. This goes in three main steps. Firstly, since LQC is based on a canonical approach to quantization, we re-write perturbation theory of cosmological perturbations at second order in a purely phase space, or Hamiltonian language. Secondly, we extend the existing theoretical framework to quantize cosmological perturbations  in LQC, the so-called dressed metric approach, to second order in perturbations. Finally, as the approximations that are available during inflation and that make the computation of non-Gaussianity tractable\footnote{Namely, the slow-roll approximation and the availability of analytical approximation for the evolution of perturbations based on the quasi-de Sitter symmetry of the inflationary spacetime.} are simply not  applicable in the pre-inflationary era, we have developed a numerical code to compute non-Gaussianity in an arbitrary spatially flat Friedmann-Lemaitre-Robertson-Walker (FLRW) spacetime. Our code is dubbed \verb|class_lqc| and is available in  an online repository\footnote{website: \href{https://github.com/borisbolliet/class_lqc_public}{https://github.com/borisbolliet/class\_lqc\_public}}. It uses the numerical infrastructure of  \verb|class|  \cite{2011arXiv1104.2932L,2011JCAP...07..034B}.

We show that the non-Gaussianity generated by the bounce in LQC are several orders of magnitude larger than those generate by inflation alone, for length scales that were larger than the (spacetime) curvature radius at the bounce. However, we show that  these higher order correlations do not invalidate the perturbative expansion. We compare our results with observations and re-evaluate the range of values of the parameter of the theory that make both, the power spectrum {\em and} the non-Gaussianity compatible with observations. These results opens new possibilities for observational signatures in the CMB and large scale structure arising from the bounce. 

The rest of the paper is organized as follows. In section II, we develop the classical Hamiltonian theory of cosmological perturbations at next-to-leading order in perturbations, and devote section III to their quantization within the dressed metric approach in LQC. In section IV we show the numerical evaluation of the three-point correlation function, and  describe ``shape'' of the resulting scalar non-Gaussianity. 
In section IV, we also explore the dependence of our results on different freedoms in the theory, namely the `initial' value of the scalar field, the value of the energy density (or equivalently, the Ricci curvature) at the bounce, the scalar field potential $V(\phi)$, and the initial state for perturbations, respectively. 
We complement this numerical analysis with an analytical justification of the main features of the non-Gaussianty in section V. In section VI, we calculate the  leading order corrections to power spectrum and discuss the validity of perturbation theory. 
Finally, in section VII, we conclude with a summary of the results and their implications in the light of observational data.

Although the effects of non-Gaussianity in the CMB arising from LQC have been discussed in previous analyses \cite{Agullo:2015aba,Zhu:2017onp}, these works do not incorporate the non-Gaussianity generated during the bounce. Rather, they focus on contributions to non-Gaussianity  originated  {\em during  inflation}, as a consequence of  the fact that  perturbations reach the onset of inflation in an excited state. Since these excitations were generated by the LQC-bounce, the non-Gaussianity they induce during inflation is a by-product of LQC. Here we provide the framework, the numerical tools, and the computation of the \textit{full} non-Gaussianity in LQC.

Throughout this paper we use {\em reduced} Planck units, in which energy and time are measured in units of the reduced Planck mass $M_{P\ell}= \sqrt{\hbar/(8\pi G)}$, and reduced Planck time $T_{P\ell}={\sqrt{8\pi G\, \hbar}}$.  However, we will keep explicitly $\hbar$ and $G$ in our analytical expressions, in order to make the physical origin of our results more transparent.

\section[Hamiltonian formulation of second-order perturbation theory]{Hamiltonian formulation of second-order perturbation theory around spatially-flat  FLRW backgrounds}\label{sec:classperts}
Let us consider general relativity minimally coupled to a scalar field $\Phi$ on a spacetime manifold $M=\mathbb{R} \times \Sigma$. In this paper we are interested in $\Sigma$ having the $\mathbb{R}^3$ topology, although the extension to other choices is straightforward. In the Arnowitt-Deser-Misner, or Hamiltonian formulation, the phase space $\Gamma$  is made of quadruples of fields defined on  $\Sigma$, i.e.,   $(\Phi(\vec{x}),P_{\Phi}(\vec{x}),q_{ij}(\vec{x}),\pi^{ij}(\vec{x}))$, where  $q_{ij}(\vec{x})$ is a Riemannian metric that describes the intrinsic geometry of $\Sigma$, and $\pi^{ij}(\vec{x})$, its conjugate momentum, describes  the extrinsic geometry of $\Sigma$. (Latin indices $i,j$ run from 1 to 3.) The only non-zero Poisson brackets between these canonical variables are
\be \label{PB} \{ \Phi(\vec{x}),P_{\Phi}(\vec{x}')\}=\delta^{(3)}(\vec{x}-\vec{x}')\, ,\hspace{1cm}
\{ q_{ij}(\vec{x}),\pi^{k l}(\vec{x}')\}=\delta_{(i}^k\delta_{j)}^{l}\delta^{(3)}(\vec{x}-\vec{x}')\, .\ee
where $\delta_{(i}^k\delta_{j)}^{l}\equiv \frac{1}{2} (\delta_{i}^k\delta_{j}^{l}+\delta_{j}^k\delta_{i}^{l})$ is the symmetrized Kronecker delta.  Additionally, this phase space $\Gamma$  carries the four constraints of general relativity, the so-called scalar and vector (or diffeomorphism) 
constraints
\bea \label{scons} \mathbb S (\vec{x})&=&\f{2\kappa}{\sqrt{q}}\l( \pi^{ij}\pi_{ij} -\f{1}{2}\pi^2 \r) - \f{\sqrt{q}}{2\kappa} ~^{(3)}R + \f{1}{2\sqrt{q}} P_{\Phi}^2 + \sqrt{q} \, V(\Phi) + \f{\sqrt{q}}{2} D_i \Phi D^i \Phi \approx 0 \, ,  \\ \label{vcons}
\mathbb V_i(\vec{x})&=&  -2 \sqrt{q}\,  q_{ij} \, D_k(q^{-1/2} \pi^{kj}) +P_{\Phi} \, D_i \Phi \approx 0\, , \eea
where $\kappa=8\pi G$ and $V(\Phi)$ is a potential for the field $\Phi$. In these expressions, $q$, $^{(3)}R$, and $D_i$ are the determinant, the Ricci scalar, and the covariant derivative associated with $q_{ij}$, respectively.\footnote{In terms of the ordinary derivative associated with a reference frame, the components of vector constraint  read $\mathbb V_i(\vec{x})=-2\partial_k (q_{ij}\pi^{jk}) + \pi^{jk}\partial_i q_{jk} + P_{\Phi} \partial_i\Phi\approx 0$.} 

The Hamiltonian that generates time evolution in $\Gamma$ is a combination of constraints 
\be \label{ham} \mathcal H=\int \d^3x \, \Big[N(\vec{x}) \,  \mathbb S (\vec{x})+N^i(\vec{x}) \, \mathbb V_i (\vec{x})\Big]\, ,\ee
where the Lagrange multipliers $N(\vec{x})$ and $N^i(\vec{x})$ are the so-called lapse and shift. They can be chosen to depend on the  phase space variables. We now apply this formalism to  the early universe.

One of the main assumptions in  cosmology is that the primordial universe is described by a solution to Einstein's equations that is very close to a FLRW geometry. In the Hamiltonian language, this means that we want to focus on a {\em sector} of the phase space $\Gamma$ of general relativity, consisting of a small neighborhood around the homogeneous and isotropic subspace, $\Gamma_{\rm FLRW}\in \Gamma$. In this neighborhood, the canonical variables can be written as
\bea \label{pert} \Phi(\vec{x})&=&\phi+\dph(\vec{x}) \, , \nonumber \\
P_{\Phi}(\vec{x})&=&\pp+\dpp(\vec{x}) \, , \nonumber \\ 
q_{ij}(\vec{x})&=&\qz_{ij}+\delta q_{ij}(\vec{x}) \, , \nonumber \\
\pi^{ij}(\vec{x})&=&\pz^{ij}+\delta \pi^{ij}(\vec{x})\, ,
\eea
where $\dph(\vec{x}), \dpp(\vec{x}),\delta q_{ij}(\vec{x}),\delta \pi^{ij}(\vec{x})$ describe small perturbations around  the homogenous and isotropic background variables $\phi, \pp, \qz_{ij}, \pz^{ij}$.

\subsection{Background}

 The  variables $\phi, \pp, \qz_{ij}, \pz^{ij}$ are {\em chosen} to describe a spatially flat FLRW universe. This implies the following. First of all, because we are dealing here with homogenous fields and $\Sigma$  has the non-compact $\mathbb{R}^3$ topology, the spatial integrals involved in the definition of the Hamiltonian and the symplectic form, diverge. But this is a spurious infrared divergence, which can be eliminated by restricting the integrals to some finite, although arbitrarily large cubical coordinate volume $\mathcal{V}_0$. This infrared regulator will appear only in intermediate expressions, and physical predictions will not depend on it, therefore allowing us to take $\mathcal{V}_0\to \infty$ {\em at the end} of the calculation. Secondly, the basic Poisson brackets of these background variables  are
\be \label{bpb}  \{\phi, \pp\}=\frac{1}{\mathcal{V}_0}\, ,  \hspace{0.5cm} \{\qz_{ij}, \pz^{kl} \}=\frac{1}{\mathcal{V}_0} \, \delta_{(i}^k\delta_{j)}^{l}\, .\ee 
The rest of Poisson brackets between background variables, as well as the `mixed' brackets involving both background and perturbation fields, all vanish. Thirdly, homogeneity and isotropy allow us to choose a gauge in which the metric variables take the manifestly homogeneous and isotropic form 
\be \label{back} \qz_{ij}=a^2\, \delta_{ij} \, ,  \hspace{1cm} \pz^{ij}=\f{\pi_a}{6\, a}\, \delta^{ij}  \, ,\ee
where $ \delta_{ij} $ is the Euclidean metric on $\Sigma$ and $ \delta^{ij}$ its inverse, and numerical factors have been chosen to make $a$ and $\pi_a$ canonically conjugated variables, $\{a,\pi_a\}=\frac{1}{\mathcal{V}_0}$. %
Furthermore, homogeneity makes the vector constraint to vanish identically, since the spatial derivatives of background variables are all zero. Therefore, the background degrees of freedom are subject only to the scalar constraint  (\ref{scons}), which takes the form 
\be  \label{Fcons} \mathbb S^{(0)}=  -\f{\kappa\,\pi_a^2}{12\,a} +\f{\pp^2}{2\,a^3}\, +a^3\, V(\phi)\approx0\, .\ee
This is the familiar Friedmann constraint. And finally, dynamics is generated by the Hamiltonian 
\be \label{backH0} \mathcal H_{_{\rm FLRW}}=\int \d^3x \, N \, \mathbb S ^{(0)}=\mathcal{V}_0 \, N\left[-\f{\kappa\,\pi_a^2}{12\,a} +\f{\pp^2}{2\,a^3}\, +a^3\, V(\phi)\right] \, .
\ee
Only uniform lapses $N$ contribute to the right hand side of (\ref{backH0}). Commonly used choices are (i) $N=1$, which corresponds to using  proper---or cosmic---time $t$, (ii) $N=a$ that corresponds to conformal time $\eta$, (iii) or $N=a^3$ associated with the so-called harmonic time $\tau$. Friedmann equations are easily obtained from Hamilton's equations of motion which, in cosmic time, read
\be \label{eoma} \dot a=\{a, \mathcal H_{_{\rm FLRW}}\}=- \kappa \f{\pi_a}{6\, a} \, ,\hspace{0.5cm} \dot \pi_a=\{\pi_a, \mathcal H_{_{\rm FLRW}}\}=- \left[\f{\kappa}{12 a^2}\, \pi_a^2-\f{3}{2} \f{1}{a^4}\pp^2+3a^2\, V(\phi)\right]\, ,\ee
\be \label{eomphi} \dot \phi=\{\phi, \mathcal H_{_{\rm FLRW}}\}= \f{\pp}{a^3}  \, , \hspace{1cm} \dot{\pp}=\{\pp, \mathcal H_{_{\rm FLRW}}\}=- a^3 \, \f{\d V(\phi)}{\d\phi} \, .\ee
These equations can be combined into the more familiar set of second-order differential equations 
\be \label{eoms} \ddot \phi+3\f{\dot a}{a}\, \dot\phi+\f{\d V(\phi)}{\d\phi}=0 \, , \hspace{0.5cm} \f{\ddot a}{a}=-\f{\kappa}{2}\, (\f{1}{3}\rho+P)\, ,\ee
where $\rho\equiv \f{1}{2}\dot \phi^2+V(\phi)$ and $P\equiv \f{1}{2}\dot \phi^2-V(\phi)$ are the energy and pressure density of $\phi$, respectively.

By solving (\ref{eoms}) one directly obtains the {\em spacetime} background metric  $ds^2=- dt^2+\qz_{ij}(t) \,dx^i dx^j=-dt^2+a(t)^2 \,d\v{x}^2$, and the scalar field $\phi(t)$. These are the background fields upon which perturbations propagate. \\ 

{\bf Remark:} From now on, we choose to raise and lower {\em all} indices with the FLRW background metric $\qz_{ij}$ and its inverse $\qz^ {ij}$.

\subsection{Perturbations}\label{sec:IIB}
Perturbation fields are defined by equations (\ref{pert}). The Poisson brackets of the physical fields (\ref{PB}) together with those of the background variables (\ref{bpb}), imply
\be \label{pertPB} \{ \dph (\vec{x}),\dpp (\vec{x}')\}=\delta^{(3)}(\vec{x}-\vec{x}')-\frac{1}{\mathcal{V}_0}\, 
,\hspace{0.5cm} \{ \delta q_{ij}(\vec{x}),\delta \pi^{k l}(\vec{x}')\}=\delta_{(i}^k\delta_{j)}^{l}\, \Big(\delta^{(3)}(\vec{x}-\vec{x}')-\frac{1}{\mathcal{V}_0}\Big)\, .\ee
The distribution appearing in the right hand side, $\delta^{(3)}(\vec{x}-\vec{x}')-\frac{1}{\mathcal{V}_0}$, is simply the Dirac delta  on the space of purely inhomogeneous fields.\footnote{This can be checked by smearing the left hand side of  (\ref{pertPB}) with arbitrary functions $f(\v x)$ and $g(\v{x})$, and noticing that the presence of the term $-1/\mathcal{V}_0$ removes the homogeneous components of 
those functions.
Thus, only the inhomogeneous components of $f(\v x)$ and $g(\v x)$, defined as $f_{\rm inh}(\v x)\equiv  f(\v x)-1/\mathcal{V}_0\int \d x^3 f(\v x)$ and similarly for  $g(\vec{x}')$, contribute to the right hand side of (\ref{pertPB}). 
Note also that at second order, the equations of motion for perturbations are {\em non-linear}. This implies that perturbation will pick a homogenous contribution throughout the evolution, even if the initial data is purely inhomogeneous. Therefore, strictly speaking, perturbations cannot be assumed to be purely inhomogeneous at this order in perturbations. However, the Poisson brackets (\ref{pertPB}) imply that the homogenous part of the perturbations will Poisson-commute with its conjugate momentum, and hence will have no dynamics in our formulation. This is equivalent to saying that, in perturbation theory, this homogenous mode is neglected, since it is assumed to always be much smaller that the background fields. This is the reason why, in practice, one can treat perturbations as purely inhomogeneous even at second order.}

We have a total of 7  degrees of freedom (per point of space) in configuration variables---6 in $\delta q_{ij}(\vec{x})$ (gravity) and  one in $ \dph (\vec{x})$ (matter)---and 7 more in the conjugate momenta. But perturbations are subject to the 4 constraints (\ref{scons}), hence leaving a total of 3 physical degrees of freedom in configuration variables, and a total of 6 in the phase space of perturbations---recall that each first class constraint actually removes two degrees of freedom in phase space. In order to isolate these physical fields, 
it is  convenient to first decompose $\delta q_{ij}(\vec{x})$ and  $\delta \pi^{ij}(\vec{x})$ in a  way that is adapted to the symmetries of the background metric $\qz_{ij}$. This  leads to the well-know scalar-vector-tensor decomposition of metric perturbations. 
This decomposition can be achieved either in position or Fourier space. We choose to do it in Fourier space (see, e.g, \cite{0264-9381-11-2-011,Agullo:2012fc} for earlier references), with the aim of complementing the more extended  analysis in position space (see, e.g.,\ \cite{Weinberg_2008}, and \cite{Domenech:2017ems} for a recent study of non-Gaussianity in position space, also in the canonical framework). We start by expanding the metric perturbations in Fourier modes
\be \delta q_{ij}(\vec{x})=\frac{1}{\mathcal{V}_0}\sum_{\vec{k}} \delta \t{q}_{ij}(\vec{k})\, e^{i\, \v{k}\cdot\v{x}}\, , \hspace{1cm} \delta \pi^{ij}(\vec{x})=\frac{1}{\mathcal{V}_0}\sum_{\vec{k}} \delta \t{\pi}^{ij}(\vec{k})\, e^{i\, \v{k}\cdot\v{x}}\, .\ee
Since the perturbation fields in position space are real, one has $ \delta \t{q}_{ij}^{ \star}(\vec{k})= \delta \tilde q_{ij}(-\vec{k})$, and similarly for $ \delta \t{\pi}^{ij}(\vec{k})$, where the star indicates complex conjugation. 

The Poisson brackets (\ref{pertPB}) translate to
 \be\label{cpb} \{ \delta \tilde q_{ij}(\vec{k}),\delta \tilde \pi^{k l}(\vec{k'})\}=\mathcal{V}_0\, \delta_{(i}^k\delta_{j)}^{l}\,  \delta_{\vec{k},-\vec{k}'}\, ,\ee
for any non-zero $\v{k}$ and $\v{k}'$.

The matrices $ \delta \tilde q_{ij}(\vec{k})$ belong to the vector space of $3\times 3$ symmetric matrices. The scalar-vector-tensor decomposition is obtained by writing $ \delta \tilde q_{ij}(\vec{k})$
 in a convenient basis in this space, namely
\begin{align}\label{matrixbases}
 {A}^{{(1)}}_{ij}\, &=\, \f{\qz_{ij}}{\sqrt{3}} \hspace{1in} &
 {A}^{(2)}_{ij}\, &=\,\sqrt{\f{3}{2}}\,\l(\hat{\bm k}_i\,\hat{\bm k}_j - \f{\qz_{ij}}{3}\r) \nonumber\\
 {A}^{(3)}_{ij}\, &=\, \f{1}{\sqrt{2}}\, \l(\, \hat{\bm k}_i\, \h x_j\, +\, \hat{\bm k}_j\, \h x_i \,\r)
 \hspace{1in} &{A}^{(4)}_{ij}\, &=\,\f{1}{\sqrt{2}}\, \l(\, \hat{\bm k}_i\, \h y_j\, +\, \hat{\bm k}_j\, \h y_i \,\r)\nonumber\\
 {A}^{(5)}_{ij}\, &=\,\f{1}{\sqrt{2}}\, \l(\, \hat x_i\, \h y_j\, +\, \hat x_j\, \h y_i \,\r) 
 \hspace{1in}& {A}^{(6)}_{ij}\, &=\, \f{1}{\sqrt{2}}\, \l(\, \hat x_i\, \h x_j\, -\, \hat y_i\, \h y_j \,\r),\nonumber
\end{align}
where $\hat{\bm k}$ is the unit vector in the direction of $\v{k}$, and $\hat{\bm k}, \h x, \h y$ form an orthonormal set of unit vectors (with respect to $\qz_{ij}$). These six matrices 
form an orthonormal basis, with respect to the inner product ${A}^{{\star \, (n)}}_{ij}{A}_{{(m)}}^{ij}=\delta_{nm}$. Now, we expand the perturbation fields in this basis:
\be \delta \tilde q_{ij}(\vec{k})=\sum_{n=1}^{6} \t \gamma_n(\vec{k}) \,  {A}^{{(n)}}_{ij}(\v k)\, , \hspace{1cm} \delta \tilde \pi^{ij}(\vec{k})=\sum_{n=1}^{6} \t \pi_n(\vec{k}) \,  {A}_{{(n)}}^{ij} (\v k)\, . \ee
These equations  can be seen as the definition of   $\t \gamma_n(\vec{k})\equiv {A}_{(n)}^{ij}\delta \tilde q_{ij}(\vec{k})$ and $\t \pi_n(\vec{k})\equiv  {A}^{(n)}_{ij}\delta \tilde \pi^{ij}(\vec{k})$. Consider the group of rotations around the direction $\hat{\bm k}$, i.e.\ the $\mathrm{SO}(3)$ subgroup  that leaves $\hat{\bm k}$ invariant---but  rotates $\hat x$ and $\h y$. It is evident  from their definition that ${A}^{(1)}_{ij}$ and ${A}^{(2)}_{ij}$ are unaffected by these rotations, ${A}^{(3)}_{ij}$ and ${A}^{(4)}_{ij}$ transform as vectors, and ${A}^{(5)}_{ij}$ and ${A}^{(6)}_{ij}$ as  two-covariant tensors. For this reason $\t \gamma_n$ and $\t \pi_n$ are called scalar modes for $n=1,2$,  vector modes for $n=3,4$, and tensor  modes for $n=5,6$. The canonical Poisson brackets (\ref{cpb}) are equivalent to
\bea \{\t \gamma_n(\v k),\t \pi_m (\v{k}')\}&=&{A}_{(n)}^{ij} {A}^{(m)}_{rs} \times\, \{ \delta \tilde q_{ij}(\vec{k}),\delta \tilde \pi^{rs}(\vec{k'})\}=\mathcal{V}_0 \, \delta_{nm} \, \delta_{\vec{k},-\vec{k}'}\, , \nonumber \\
\{\t \gamma_n(\v k),\t \gamma_m (\v{k}')\}&=&0
\, , \nonumber \\
\{\t \pi_n(\v k),\t \pi_m (\v{k}')\}&=&0\, .
\eea
Note that  the conjugate variable of $\t \gamma_n(\v k)$ is $\t \pi_m (-\v{k})=\t \pi^{\star}_m (\v{k})$.

\subsection{Physical degrees of freedom}

There are two common strategies to isolate physical degrees of freedom in perturbations from pure gauge ones,  namely gauge fixing or working with the so-called gauge invariant variables. Gauge invariant variables are combinations of $ {\delta \t \phi}$ and $\t \gamma_n$'s that are invariant under the Hamiltonian flow generated by some of the constraints. More precisely, when working at linear order in perturbations, gauge invariant variables are  defined to be invariant under the flow generated by the terms in the constraints (\ref{scons}) that are {\em linear} in perturbations, and these variables  are commonly used in the literature (see, e.g.,  \cite{0264-9381-11-2-011}, and section III.C of \cite{Agullo:2012fc}). However, finding gauge invariant perturbations at second order is more tedious \cite{Domenech:2017ems},  since one must involve second-order constraints in their definition. The gauge fixing strategy is more efficient, and  more common in the literature (see, e.g.,\ \cite{Maldacena:2002vr}), and we shall follow it in this paper. 

Recall also that in making predictions for primordial perturbations, the important point is to write the answer in terms of the comoving curvature perturbations $\mathcal{R}$ (see, e.g.,\ \cite{Maldacena:2002vr} for its definition at higher order in perturbations). This is because Fourier modes of $\mathcal{R}$  {\em remains  constant from the time they  exit the Hubble radius during inflation until they re-enter towards the end of the radiation era}. This property of $\mathcal{R}$ is crucial, since it allows us to connect the inflationary predictions  with observables in the late time universe, even if we are uncertain about the evolution of the universe immediately after  inflation. Therefore, irrespective of what strategy one decides to follow---gauge invariant variables or gauge fixed ones---the important point is to write the answer in terms of $\mathcal{R}$ at the end of inflation. 

However, performing all computations using $\mathcal{R}$ presents some difficulties. When the  universe is dominated by a scalar field $\phi$, the variable $\mathcal{R}$ is ill-defined  whenever $\dot \phi$ vanishes. During inflation this situation does {\em not} occur, because  the evolution of the  scalar field during this period is monotonic, rolling down the potential, as long as the slow-roll conditions are satisfied. In the scenario under consideration in this paper, $\dot \phi$ vanishes just before the onset of inflation, thus making the  variable $\mathcal{R}$ unsuitable for our purposes (see \cite{Agullo:2013ai,Schander:2015eja} for further details). Therefore, in our analysis below we  work with the scalar perturbations $\dph$ in the  spatially flat gauge, and rewrite the answers in terms of comoving curvature perturbation $\mathcal{R}$ {\em at the end of the inflation}, when all modes of interest are in super-Hubble scales.

The spatially flat gauge is defined as the gauge in which the scalar and vector modes of metric perturbations vanish, i.e.,\  $\t \gamma_i=0$ for $i=1,2,3,4$. The physical degrees of freedom are therefore encoded in the scalar  perturbations $\tilde \dph$ and the tensor modes $\t \gamma_5$ and $\t \gamma_6$. This strategy completely fixes the gauge freedom.

We are now ready  to write the Hamiltonian that generates dynamics,  including terms up to third order in perturbations. This will produce equations of motion that incorporate terms up to second order.%

\subsection{Third-order Hamiltonian}\label{sec:3H}

This paper focuses on non-Gaussianity of {\em scalar} perturbations. Therefore, {\em we will not write terms containing tensor modes in this section}. Including them, however, does not add any conceptual difficulty (for a treatment of tensor modes in the context of inflation, see for instance, \cite{Maldacena:2002vr,Sreenath:2013xra,1475-7516-2014-10-021}), although the expressions below become significantly longer. The third order Hamiltonian for scalar perturbations in the spatially flat gauge is obtained as follows:
\begin{enumerate}[(i)]
\item Expand the constraints (\ref{scons}) in perturbations
 \bea \mathbb S(\v x)&=&\mathbb S^{(0)}+\mathbb S^{(1)}(\v x)+\mathbb S^{(2)}(\v x)+\mathbb S^{(3)}(\v x)+\cdots\,  ,\nonumber \\
  \mathbb V(\v x)&=&\mathbb V^{(0)}+\mathbb V^{(1)}(\v x)+\mathbb V^{(2)}(\v x)+\mathbb V^{(3)}(\v x)+\cdots\, ,\eea
where the superscript $(0)$ denotes the terms that are independent of perturbations, $(1)$ the linear terms, $(2)$ and $(3)$ the second- and third-order terms, respectively. Expressions for each of these terms can be obtained directly from (\ref{scons}) and (\ref{vcons}), and are reported in Appendix A. 

Expand also the lapse and shift as $N+\delta N$ and $N^i+\delta N^i$, where $N$ and $N^i$ are the homogenous lapse and shift. For consistency with  the FLRW gauge fixing [Eqn.\ (\ref{back})], we take $N^i=0$. On the other hand, $\delta N(\v x)$ and $\delta N^i(\v x)$ are the inhomogeneous part of the lapse and shift, which may depend on perturbations. 

\item Impose the gauge conditions $\t \gamma_1=0$, $\t \gamma_2=0$ in the constraints  (\ref{scons}).\footnote{From the phase space viewpoint, this is equivalent to introducing two new (second class) constraints.} (Since we are interested in terms involving only scalar perturbations, the  gauge conditions $\t \gamma_3=0$, $\t \gamma_4=0$ are not needed.)

\item Find the lapse $\delta \t N$  and shift $\delta \t N^i$ associated with this gauge fixing by demanding that the gauge conditions are preserved upon evolution; i.e.,\ use the equations
\be \dot{\t{\gamma}}_1=\{  \t\gamma_1, \mathcal H \}=0\, , \hspace{1cm} \dot{ \t \gamma}_2=\{ \t \gamma_2, \mathcal H \}=0\, ,\ee
to obtain $\delta \t N$ and and $\delta \t N^i$ in terms of $ \t \pi_1$, $\t  \pi_2$, $\delta \t \phi$, and ${\delta}\t p_{\phi}$. To write the third order Hamiltonian it is sufficient to keep terms in $\delta \t N$ and and $\delta \t N^i$ up to first order in perturbations.
\item Impose the first order constraints, $\mathbb S^{(1)}(\v x)=0$, $\mathbb V_i^{(1)}(\v x)=0$ to eliminate the conjugated variables $\t \pi_1$, $\t \pi_2$ in favor of $\delta {\t \phi}$ and $\delta \t{p}_{\phi}$, i.e.,\ to find the relations $\t \pi_1=\t \pi_1(\delta {\t \phi}, \delta \t{p}_{\phi})$, $ \t \pi_2=\t \pi_2(\delta {\t \phi}, \delta \t{p}_{\phi})$.
\item  Plug these results in the Hamiltonian (\ref{ham}) and keep terms up to third order in perturbations.  
\end{enumerate}
We performed these calculations using the \verb|Mathematica|  package \verb|xAct|\footnote{http://www.xact.es}\cite{Brizuela2009}. The result is 
\bea \delta \tilde N\, &=&\, -\f{2\,N}{a\,\pi_a}\,(\,\sqrt{3}\,\t\pi_1\, +\,  \sqrt{6}\,\t\pi_2)\, ,\nonumber\\
\delta \t N^i&=&ik^i \t \chi \, , \hspace{0.5cm} {\rm where} \, \,  \,  \t \chi\, =\,N\,  \f{ \sqrt{6}\,\kappa}{k^2 a}\,\t \pi_2\, , \nonumber \\
 \t\pi_1\, &=&  \f{\sqrt{3}\, a^5\, V_{\phi}}{\kappa\,\pi_a}\,\delta {\t \phi}+\, \f{\sqrt{3}\,\pp}{\kappa\, a\, \pi_a}\,\delta \t{p}_{\phi}\, , \nonumber \\
\t \pi_2\, &=&\,\sqrt{\f{3}{2}}\,\biggl[ \biggl(\,\f{\pp}{2} -\, \f{a^5\, V_{\phi}}{\kappa\,\pi_a} \biggr)\delta {\t \phi}\, -\, \f{\pp}{\kappa\, a\, \pi_a}\delta \t{p}_{\phi}\, \biggr]\, ,  \eea
where $k^2\equiv  k_i k_j\, \delta^{ij}=a^2 k_i k^i$ is the so-called comoving wave-number.

Moving back to position space, we obtain the expression for the  Hamiltonian up to third order for scalar perturbations $\mathcal{H}_{\rm pert}=\mathcal{H}^{(2)}+\mathcal{H}^{(3)}$. The second-order Hamiltonian is\footnote{We have, in addition, performed the  canonical transformation $(\dph,\dpp)\rightarrow (\dph,\delta \bar{p}_{\phi}=\dpp-\f{3\,p_{\phi}^2}{\,a\,\pi_a}\dph)$ to eliminate a term proportional to $\dpp \dph$ in the second-order Hamiltonian. From now on we will work with $\delta \bar{p}_{\phi}$, but we will drop the bar to simplify the notation.} 

\begin{eqnarray} \label{hams}
 \mathcal{H}^{(2)}\, =\,\int \d^3 x \, N \, \, \mathbb{S}^{(2)}(\v x)=\, N\f{1}{2}\,\int \d^3 x \,  \biggl[\, \f{1}{\,a^3}\, \dpp^2\, +\, a^3\, (\v \partial \delta\phi)^2\, 
 +\,a^{3}\,  \u\, \dph^2\biggr]\, ,
\end{eqnarray}
with  the potential  ${\u}$ given by
\be \label{potu} {\u}=-9 \f{\pp^4}{a^8\pi_a^2}+\f{3}{2}\kappa\f{\pp^2}{a^6}-\f{6 \, \pp }{a\, \pi_a} V_{\phi}+ V_{\phi \phi} {\color{black} +6\f{\pp \dot{p}_{\phi}}{a^4\, \pi_a}-3\f{\pp^2 \, \dot{\pi}_a}{a^4\, \pi_a^2}-3\, \f{\dot a\, \pp^2}{a^5\, \pi_a}} \, .\ee
The `dot' on background variables  must be understood as $\dot x\equiv  \{x,\mathcal{H}_{\rm FLRW}\}$, and each  subscript $\phi$ for the potential $V$ means a derivative with respect to $\phi$. \\

The third order Hamiltonian is
\begin{eqnarray}\label{eq:H3}
 \mathcal{H}^{(3)}\, &=&\,\int \d^3 x  \left(\,\delta N\, \mathbb{S}^{(2)}(\v x)+\delta N^i \, \mathbb{V}^{(2)}(\v x)+\, N \, \, \mathbb{S}^{(3)}(\v x)\right) 
\nonumber \\
 &=&
 \, N\,\int\, \d^3  x\, \biggl[ 
\left(  \frac{9\,\kappa\,p_{\phi}^3}{4\,a^4\,\pi_a}-\f{27 \, \pp^5}{2\, a^6\pi_a^3} -\, \frac{3\,a^2\,p_{\phi}\,V_{\phi\phi}}{2\,\pi_a}\, 
 +\frac{a^3\,V_{\phi\phi\phi}}{6} \right) \,\delta\phi^3\, \nonumber \\
 &&-\, \frac{3\,p_{\phi}}{2\,a^4\,\pi_a}\,\delta p_{\phi}^2\, \delta\phi\,-\f{9 \, \pp^3}{ a^5 \pi_a^2} \, \dpp\dph^2
-\, \frac{3\,a^2\, p_{\phi}}{2\, \pi_a}\delta\phi\, (\vec{\partial}\delta\phi)^2 +\, \frac{3\,p_{\phi}^2}{N\, a\,\pi_a}\,\delta\phi^2 \partial^2\chi\, +\, \frac{3}{2}\f{a^2\,p_{\phi}}{N^2\,\kappa\,\pi_a}\,\delta\phi\,\partial^2\chi\,\partial^2\chi \nonumber \\
 &&+\,3\, \f{\pp^2}{N\, a\,\pi_a}\, \delta\phi\, \partial^i\chi\partial_i\delta \phi +\f{1}{N}\, \delta p_{\phi}\,\partial_i \delta\phi\, \partial^i\chi\,
 -\, \frac{3}{2}\f{a^2\,p_{\phi}}{N^2\,\kappa\,\pi_a}\, \delta\phi\, \partial_i\partial_j\chi\, \partial^i\partial^j\chi\,
 \biggr].
\end{eqnarray}
By performing a Legendre transformation, it can be checked that these expressions agree with the third-order Lagrangian derived in \cite{Maldacena:2002vr} (recall that, unlike \cite{Maldacena:2002vr}, we use the physical background metric $\qz_{ij}=a^2\, \delta_{ij}$ and its inverse, to lower and raise indices). Note that we have not used the Friedmann constraint (\ref{Fcons}) to derive, or simplify, the second- and third-order Hamiltonians.\\

The second-order Hamiltonian $\mathcal{H}^{(2)}$ provides the {\em free} evolution of perturbations, i.e., it leads to the  {\em linear} equations of motion
\be \label{eqpert}  \dot \dph=\{\dph, \mathcal{H}^{(2)}\} \, \, , \, \,  \dot {\delta p_{\phi}}=\{\delta p_{\phi}, \mathcal{H}^{(2)}\}  \, \,  \longrightarrow \, \, \,  (\Box - { {\u(t)}})\,{{\dph}}(\vec{x},t) =0  
\, ,\ee
where $\Box$ is the d'Alembertian of the FLRW background metric.

The third order piece of the Hamiltonian, $\mathcal{H}^{(3)}$, is the so-called interaction Hamiltonian, which provides self-interactions between perturbations (quadratic terms in the equations of motion).  Some of these interactions are generated by the scalar field's potential $V(\phi)$, but note that most terms in $\mathcal{H}^{(3)}$ are independent of $V(\phi)$, and therefore would be present even if $V(\phi)=0$. These are self-interaction mediated by gravity. %

Finally, the relation between  $\delta \phi$ to the comoving curvature perturbations $\mathcal{R}$, needed to write our results in terms of $\mathcal{R}$ at the end of inflation, is given by \cite{Maldacena:2002vr} 
\bea \label{zetadph} \mathcal{R}(\v x,t)&=&-\f{a}{z} \, \dph+\left[-\f{3}{2}+3\f{V_{\phi}\, a^5}{\kappa\, \pp\, \pi_a}+\f{\kappa}{4}\f{z^2}{a^2}\right]  \left(\f{a}{z} \, \dph\right)^2\nonumber\\ &&-\,\f{3\, a^2}{\kappa\, \pi_a} \f{d}{dt} \left[\f{a}{z}  \dph\right]^2 -9\f{a^4}{\kappa^2 \, \pi_a^2} \f{a^2}{z^2}\left(  \v{\partial}  \dph\right)^2+9\f{a^4}{\kappa^2 \, \pi_a^2} \f{a^2}{z^2} \partial^{-2}\partial_i\partial_j \left(\partial^i \dph\partial^j  \dph\right)\nonumber\\ &&+\, 3\f{a^4 }{\kappa\,  \pi_a} \f{a}{z} \partial_i \chi\partial^i \dph- 3\f{a^4 }{\kappa\,  \pi_a} \f{a}{z}\partial^{-2}\partial_i\partial_j \left[\partial^i\chi\partial^j\dph\right] \, .\eea
where $z\equiv   -\f{6}{\kappa}\f{\pp}{\pi_a}$. Although this relation looks complicated, we will only need to use it  at the end of the inflation, and at that time the terms in the second and third lines become negligible compared to those in the first line. The reason for this is that 
perturbations that can affect  our CMB have wave-lengths much larger than the Hubble radius at the end of inflation. As previously mentioned,  these super-Hubble modes  of  $\mathcal{R}$  become time independent. These two facts---super-Hubble wavelength and time independence---make both the spatial and time derivatives appearing in the second and third line  negligibly small.%

\section{Extension of the dressed metric approach to second order}
\label{sec:EXT}
In this section we obtain the equations that describe the propagation of scalar perturbations in the Planck era of the universe, using LQC.  We use the so-called dressed metric approach, introduced in \cite{Ashtekar:2009mb}, and further developed  in \cite{Agullo:2012fc,Agullo:2013ai} (see also the review articles \cite{Ashtekar:2011ni, Agullo:2013dla, Agullo:2016tjh}).  Here we extend the existing formalism to second order in perturbations. 

In semiclassical cosmology, to account for the CMB temperature fluctuations it has sufficed to consider just the first-order perturbations around a FLRW solution, ignoring their back-reaction. In the Planck era of the  universe, to begin with, one has a quantum gravitational field instead of a smooth  metric. The  question is whether we can find  solutions in loop quantum cosmology that deviate from a quantum FLRW configuration {\em only} by small perturbations, and whose effect on the background quantum geometry can be neglected. Such solutions exist \cite{Ashtekar:2009mb,Agullo:2012fc,Agullo:2013ai} and can be calculated, and they can be used to build  a self-consistent quantum gravity extension of the inflationary scenario  \cite{Agullo:2012sh,Agullo:2013ai}. We first summarize how these solutions are obtained, and then extend previous analyses by including  terms up to second order in perturbations.

Our goal is to find the quantum theory of the classical midi-superspace  made of spatially flat FLRW geometries sourced by a scalar field $\phi$, together with scalar perturbations $\dph(\vec{x})$ propagating thereon. In LQC, dynamics is extracted from the constraint equation (the analog of the Wheeler-deWitt equation) $\hat{\mathcal{H}} \Psi=0$,  where $\hat{\mathcal{H}}  =\hat{\mathcal{H}}_{\rm FLRW}+\hat{\mathcal{H}}_{\rm pert}$ is the operator associated with the Hamiltonian obtained in the previous section, and $\Psi$ is the total wave-function describing both the background  degrees of freedom, $a$ and $\phi$, as well as scalar perturbations $\dph$. In LQC it is convenient to trade the scale factor $a$ for the `volume' $v$, defined as $v \equiv   a^3 \, \mathcal{V}_0 \, 4/\kappa$ and use the lapse  $N_{\tau}\equiv a^3$ (see \cite{Agullo:2016hap}, and references therein, for additional details). The constraint equation $\hat{\mathcal{H}} \Psi(v,\phi, \dph)=0$ takes the form
\be \label{WdW}
 - \hbar^{2}\dphi^2\Psi(v,\phi,\dph) = \Big(\h H^2_{0} - \h H^2_{1}- 2 \mathcal{V}_0\, \hat{\mathcal{H}}_{\rm pert}[N_{\tau}]\Big)\, \Psi(v,\phi,\dph).
\ee
where $\h H^2_{1}\equiv \frac{1}{8}\kappa^{2} \h v^{2} \h V(\phi)$, and $\h H^2_{0}$ is a difference operator, whose explicit form is not important for our discussion (it  can be found, e.g.,\ in equation (2.2) of \cite{Agullo:2016hap}; see also the original references \cite{Ashtekar:2003hd,Ashtekar:2006rx,Ashtekar:2006uz,Ashtekar:2006wn,Ashtekar:2007em}). Both $\h H_{0}$ and $\h H_{1}$ act only on background degrees of freedom, while $\hat{\mathcal{H}}_{\rm pert}$ acts on both, background and perturbations. We are interested in solutions to this equation of the form $\Psi(v,\phi,\dph)= \Psi_0(v,\phi)\otimes \delta \Psi(v,\phi,\dph)$, with  $\Psi_0(v,\phi)$ representing a quantum FLRW gravitational field, and $\delta \Psi(v,\phi,\dph)$ describing inhomogeneous scalar perturbations. 

\subsection{Background} \label{IIIA}

The states $ \Psi_0(v,\phi)$ are chosen to be a normalized solution, with respect to a suitably defined inner product \cite{Ashtekar:2011ni}, of (\ref{WdW}) with $\hat{\mathcal{H}}_{\rm pert}=0$. They describe a quantum FLRW geometry. The  Hilbert space $\mathscr{H}_{\rm FLRW}$ to which the states $ \Psi_0(v,\phi)$ belong to, was studied in detail  in \cite{Ashtekar:2006rx, Ashtekar:2007em, Ashtekar:2011ni} in absence of a potential $V(\phi)$, i.e.,\ with $ \h H_{1}=0$. 

Adding a potential introduces additional subtleties related to the definition of the inner product on the Hilbert space. This issue has been discussed in \cite{Agullo:2016hap}, and the reader is referred there for details. In this paper, we will focus only on bounces that are ``kinetic dominated'', since this is the regime of phenomenological interest for us (see  sections \ref{sec:sum}). For such bounces,  one can check that $\langle  \h H^2_0  \rangle \gg \langle  \h H^2_{1}  \rangle$  during the Planck era \cite{Agullo:2016hap}.\footnote{This epoch is defined as the period for which the quantum gravity corrections to the dynamics are larger than a $0.1\%$.} This makes the  term proportional to $\h H_{1}$ in our quantum equations to produce negligible effects on physical observables (e.g., the primordial power spectrum),  several orders of magnitude smaller than  observational error bars. Hence, although the mathematical subtleties that appear in the  inclusion of $\h H_{1}$ are important from the conceptual and mathematical viewpoint, they are not of direct relevance for phenomenological considerations. Therefore, in this paper we will work with  states $\Psi_0(v,\phi)$ obtained by neglecting $\h H_{1}$ in the Planck era. 

The Hilbert space of the states for the background geometries that we are interested in, $\mathscr{H}_{\rm FLRW}\, \backin\, \Psi_0(v,\phi)$, 
is then made of solutions to the `Schr\"odinger-like' equation
\be \label{Sleq} - i \hbar \, \dphi \Psi_0(v,\phi) = \h H_{0} \, \Psi_0(v,\phi)\, , \ee
with finite norm $||\Psi_{0}||^{2} \, \equiv \, \sum_{v} |\Psi_{0}(v,\phi)|^{2} \, <\, \infty$. This equation is simply  the  positive `square root'  of (\ref{WdW}) with $\h H_{1}=0$ and $\h{\mathcal{H}}_{\rm pert}=0$.  $\mathscr{H}_{\rm FLRW}$  is the  analog of the  space of states of the more familiar example of a scalar field in Minkowski spacetime, that is made of positive frequency solutions to the Klein-Gordon equation. It is useful---although not essential---to think of $\phi$ in  $\Psi_0(v,\phi)$ as a {\em relational time} variable with respect to which the wave-function `evolves'. %

As shown in \cite{Ashtekar:2007em}, states in (a dense subspace of)  $\mathscr{H}_{\rm FLRW}$ are free of curvature singularities, in the sense that curvature invariants are all bounded. The eigenvalues  of the matter energy density and pressure have  also an absolute supremum on $\mathscr{H}_{\rm FLRW}$, given by a fraction of the Planck scale. Furthermore, every state $\Psi_0(v,\phi)$ experiences precisely one `instant' $\phi_{\rm B}$  at which the  expectation value of the  volume of the fiducial box, or of any other finite region of space, attains its minimum, while energy density and curvature reach their maximum. In other words, in this theory a {\em cosmic bounce} replaces the big bang singularity of classical general relativity.

\subsubsection{Effective theory}\label{sec:EE}

To gain physical intuition, consider states $\Psi_0(v,\phi)$ that are sharply peaked in the volume $v$, i.e.,   states with {\em  small relative dispersion} in $v$ (or equivalently, in the scale factor $a$) during the entire `evolution'. Such solutions to (\ref{Sleq}) exist, and  have been studied in  detail \cite{Ashtekar:2006uz,Ashtekar:2006wn,Diener:2013uka,Diener:2014mia}. For these states, it has been shown \cite{Taveras:2008ke,Ashtekar:2011ni} that the expectation value of the scale factor, $\bar a\equiv \langle \h a \rangle$, and the rest of background quantities, can be obtained very accurately from an {\em effective} theory. This effective theory takes the form of  a classical theory whose equations of motion incorporate the leading quantum corrections. The phase space  is four dimensional,  made of quadruples  $(\bar a, \bar \pi_a, \bar \phi,  \bar{p}_{\phi})$, and dynamics on it is generated by the effective Hamiltonian constraint\footnote{We have included the  potential $V(\bar \phi)$ because, as emphasized before, it plays an important role at late times, out of the Planck era. However, within the Planck era  it is completely subdominant in all solutions of interest for this article. Hence, the way we use this effective Hamiltonian is consistent with the previous discussion, where the potential $V$ was neglected in deriving the wave-function $\Psi_0$ in the Planck era.}
\be \label{effH0}\mathcal H_{_{\rm FRW}}^{(\rm eff)}[N]=\mathcal{V}_0\, N\left[ \f{1}{2\,\bar a^3}\, \bar{p}_{\phi}^2-\f{3\, a^3}{\kappa} \, 
\f{1}
{\ell_0^2}
\sin^2{\left( 
\ell_0\f{\kappa}{6}\, \f{\bar{\pi}_a}{\bar a^2}
\right)}+\bar a^3\, V(\bar \phi)
\right]\, ,\ee
where $\ell_0^2\equiv  \f{\Delta_0^3}{48\pi^2}\ell_{P\ell}^2$, and $\Delta_0$  is area gap  in LQC---the lowest non-zero eigenvalue of the area operator. This  Hamiltonian depends on $\hbar$ through $\ell_0$. 
In the limit   $\ell_0\to 0$, it reduces to the classical FLRW Hamiltonian given in (\ref{backH0}). In terms of the energy density $\rho\equiv  \f{1}{2}\, {\bar p}_{\phi}^2\, \bar a^{-6} +V(\bar \phi)$, the  equation $\mathcal H_{_{\rm FRW}}^{(\rm eff)}=0$ becomes
\be \label{effH2}
 \f{1}
{\ell_0^2}
\sin^2{\left( 
\ell_0\f{\kappa}{6}\, \f{\bar{\pi}_a}{\bar a^2}
\right)}=\f{\kappa}{3}\, \rho\, .\ee
The trigonometric function on the left hand side revelas that the energy density is bounded above by $\rho_{\rm sup}=\f{3}{\kappa\ell_0^2}$. Some analyses of black hole entropy in loop quantum gravity \cite{Agullo:2008eg,Agullo:2010zz,Agullo:2008yv} suggest  the value $\Delta_0=5.17$ for the area gap, that in turn makes $\rho_{\rm sup}=0.4092\rho_{Pl}$ (see, e.g., \cite{Bianchi:2012ui} for an alternative view). In this paper we treat $\Delta_0$ as a free parameter, and derive results for the CMB for different values of $\Delta_0$.

The equations of motion (using cosmic time) for the canonical variables  $\bar a$,  $\bar{\pi}_{a}$, $\bar{\phi}$, and $\bar{p}_{\phi} $ that describe the effective geometry, read
\bea \label{HrateLQC} \dot{\bar a} &=&\{\bar a,\mathcal H_{_{\rm FRW}}^{(\rm eff)}\} \hspace{0.5cm}{\Longrightarrow} \hspace{0.5cm} \bar H\equiv  \f{\dot{\bar a}}{\bar a}=-\f{1}
{2 \ell_0}
\sin{\left( 2 \, 
\ell_0\f{\kappa}{6}\, \f{\bar{\pi}_a}{\bar a^2}
\right)} \, ,\\
\label{HrateLQC2} \dot{\bar \pi}_{a}&=&\{{\bar \pi}_a,\mathcal H_{_{\rm FRW}}^{(\rm eff)}\}=\f{3}{2}\f{\bar{p}_{\phi}^2}{\bar a^4 }+ 9\f{a^2}{\kappa}\f{1}{ \ell_0^2}
\sin^2{\left(
\ell_0\f{\kappa}{6}\, \f{\bar{\pi}_a}{\bar a^2}
\right)}-\f{\bar{\pi}_a}{\ell_0} \, \sin{\left( 2 \, 
\ell_0\f{\kappa}{6}\, \f{\bar{\pi}_a}{\bar a^2}
\right)}-3\, \bar a^2\, V(\bar \phi)\, , \nonumber 
 \\
\label{HrateLQC3} \dot{\bar \phi}& =&\{\bar \phi,\mathcal H_{_{\rm FRW}}^{(\rm eff)}\} =\bar{p}_{\phi}/\bar a^3\, , \nonumber
 \\
\label{HrateLQC4} \dot{\bar p}_{\phi}&=&\{\bar p_{\phi},\mathcal H_{_{\rm FRW}}^{(\rm eff)}\}=- \, \bar{a}^3 \, \f{d V(\bar \phi)}{d\bar \phi}\, .\nonumber
\eea
These equations reproduce the classical FRLW dynamics (\ref{eoma})--(\ref{eomphi}) in the limit $\ell_0\to 0$. Equation (\ref{HrateLQC}) implies,  due  the presence of the trigonometric function, that the Hubble rate of the effective geometry is also bounded from above, by $ |\bar H_{\rm sup}|=\f{1}{2 \ell_0}=\sqrt{\f{\kappa}{12}\rho_{\rm sup}}$. %

Now, a relation between energy density and Hubble rate, that generalizes the classical Friedmann  constraint, can be obtained by combining (\ref{effH2}) and (\ref{HrateLQC}). More precisely, using the identity $ \sin^2{(2x)}=4\, \sin^2 x\, (1 - \sin^2 x)$, together with (\ref{effH2}),  equation (\ref{HrateLQC})  takes the form
 \be \label{lqcF} \bar H^2 =\f{\kappa}{3}\, \rho\, \left(1-\f{\rho}{\rho_{\rm sup}}\right)\, .\ee
The term in parenthesis breaks the  linearity between the Hubble parameter $\bar H^2$ and the energy density $\f{\kappa}{3}\, \rho$ that holds in general relativity. Moreover,   $\bar H$  vanishes when $\rho$ reaches its maximum value $\rho_{\rm sup}$; such instant corresponds to a smooth transition between a contracting and an expanding universe, i.e.,\ a cosmic bounce. When $\rho$ is small compared to $\rho_{\rm sup}$, the classical relation $\bar H^2 =\f{\kappa}{3}\, \rho$ is recovered.  

The set of equations (\ref{HrateLQC})--(\ref{HrateLQC3}) can be recast as a system of two second-order differential equations
\bea
\label{EE} 
\f{\ddot{\bar a}}{\bar a}&=&-\f{\kappa}{6}\, \rho\, \left(1-4\f{\rho}{\rho_{\rm sup}}\right)-\f{\kappa}{2}\, P\, \left(1-2\f{\rho}{\rho_{\rm sup}}\right),\nonumber \\
\ddot{\bar \phi}&+&3 \bar H \dot{\bar \phi}+ V_{\bar \phi}=0 \, ,
\eea
where $P\equiv  \f{1}{2}\dot{\bar \phi}-V(\bar \phi)$ is the  pressure density of the scalar field, and the dot indicates derivative with respect to cosmic time $t$.\footnote{Recall that in LQC evolution has been defined, at the fundamental level, in a relational manner. I.e.\ we have studied how the gravitational degree of freedom $a$ evolves with respect to the matter degree of freedom $\phi$. In this sense, the `time' variable  $t$ in this effective theory arises just as a  parameter that changes monotonically with $\bar \phi$, that allows us to `separate' the relation $a(\phi)$ into $a(t)$ and $\phi(t)$. This is the way the ordinary time we use in general relativity `emerges' in loop quantum cosmology.}  These are the so-called \emph{effective equations} of LQC. The solutions to these equations provide an effective FLRW metric  $\bar g_{ab}$ around which the quantum geometry $\Psi_0(v,\phi)$ is sharply peaked.\\

It is important to notice that solutions of the effective equations  are characterized by {\em two parameters}, which can be chosen to be the value of the scalar field at the time of the bounce $\bar \phi(t_B)\equiv  \phi_{\rm B}$ and its energy density at that same time, $\rho(t_B)\equiv  \rho_{\rm B}=\rho_{\rm sup}$. To  understand why we only need two numbers to characterize a solution, even though the phase space we are working with is four dimensional, consider the following. Note first that in a spatially flat FLRW geometry, the scale factor $a$ can be  re-scaled freely without altering the physics. We choose $\bar a_{\rm B}=1$. On the other hand, at the bounce $\dot{\bar{a}}=0$ {\em in all solutions}. Additionally, because the energy density equals $\rho_{\rm sup}$ at the bounce, $\phi_{\rm B}$ determines $\dot {\bar \phi}(t_B)$. Therefore, from the apparently four initial data required to solve the system (\ref{HrateLQC})--(\ref{HrateLQC3}), the value of $\phi_{\rm B}$ and $\rho_{\rm sup}$ (together with the convention $\bar a_{\rm B}=1$), suffices to uniquely characterize a solution. \\

\subsubsection{Generalized effective equations}\label{sec:GEE}

What about states $\Psi_{0}(v,\phi)$ that are {\em not} sharply peaked? They, of course, are not accurately described by the effective equations. In particular, the geometry they describe cannot be approximated in any reasonable sense by a smooth metric tensor.  For those states, quantum fluctuations play an important role. Nevertheless, it has been proven in \cite{Ashtekar:2015iza} that the expectation value of the scale factor $\bar a=\langle \Psi_0|\hat a|\Psi_0\rangle$ is still  accurately described by equations (\ref{EE}), with the only difference that $\rho_{\rm sup}$ must be replaced by the actual value of the energy density at the bounce,  $\rho_{\rm B}$, which  satisfies $\rho_{\rm B} \leq \rho_{\rm sup}$. That is, $\bar a$ bounces at an energy density $\rho_{\rm B}$ smaller than or equal to $\rho_{\rm sup}$ for states $\Psi_0(v,\phi)$ with large dispersion. It turns out that  $\rho_{\rm B}$ {\em decreases} when the relative quantum dispersion in volume $\Delta v/v$ {\em increases}. (The authors of  \cite{Ashtekar:2015iza}  also derive an analytical relation between $\rho_{\rm B}$ and $\Delta v/v$, valid for Gaussian states.) This behavior is sensible: since $\rho_{\rm sup}$ is a supremum, only infinitely sharply peaked states  reach $\rho_{\rm B}=\rho_{\rm sup}$, while  quantum fluctuation can only {\em decrease}  $\rho_{\rm B}$. However, it is remarkable that, even in presence of large quantum fluctuations, the mean values of $ \Psi_0(v,\phi)$ are still very well approximated by  `generalized effective equations' which are identical to the equations (\ref{EE}) with $\rho_{\rm sup}$ replaced by $\rho_{\rm B}$.

\subsection{Perturbations} 

Recall that we are interested in solutions of (\ref{WdW}) of the form $\Psi(v,\phi,\dph)= \Psi_0(v,\phi)\otimes \delta \Psi(v,\phi,\dph)$, where  $\Psi_0(v,\phi)$ is one of the quantum FLRW states described above, and $\delta \Psi$ is a small perturbation around it. 
Intuition tells us that  states of this type exist, as long as $ \delta \Psi(v,\phi,\dph)$ remains a small perturbation throughout the evolution---i.e.,\ as long as the {\em test field approximation} is valid. As we will see below, this is in fact the case.

The states we are looking for  are the `positive frequency' solutions to the constraint equation (\ref{WdW}), i.e.,\ states satisfying \cite{Ashtekar:2009mb}
\be - i \hbar \, \dphi \Psi(v,\phi,\delta\phi) = \sqrt{\h H^2_{0}[N_{\tau}]-2 \mathcal{V}_0 \, \hat{\mathcal{H}}_{\rm pert}[N_{\tau}]\, } \, \Psi(v,\phi,\delta\phi)\, . \ee
Here $\h H_{0}$ represents the Hamiltonian of the `heavy' degree of freedom (background), and $ \hat{\mathcal{H}}_{\rm pert}[N_{\tau}]=\hat{\mathcal{H}}^{(2)}[N_{\tau}]+\hat{\mathcal{H}}^{(3)}[N_{\tau}]$  the Hamiltonian of `light' ones (perturbations). Recall, $N_{\tau}=a^3$ is the lapse associated with harmonic time. We can now expand out the  square-root, and keep only terms linear in  $ \hat{\mathcal{H}}_{\rm pert}$
\be  \label{Eqn} - i \hbar \, \dphi \Psi(v,\phi,\delta\phi) \approx \left[ \h H_{0} - \mathcal{V}_0\, \Big( ( \h H_{0})^{-1/2} \, \hat{\mathcal{H}}_{\rm pert}[N_{\tau}]\, (\h H_{0})^{-1/2} \Big) \right ]\, \Psi(v,\phi,\delta\phi)\, , \ee
where we have chosen a symmetric order to write the operators in the right hand side. Note that the factors that multiply $\hat{\mathcal{H}}_{\rm pert}$ are physically consistent, since in the classical theory $N_{\phi}= \mathcal{V}_0 \, H_0^{-1} N_{\tau}$ is precisely the lapse  associated with the relational time $\phi$. Hence $\mathcal{V}_0\, ( \h H_{0})^{-1/2} \, \hat{\mathcal{H}}_{\rm pert}[N_{\tau}]\, (\h H_{0})^{-1/2}$ is a specific quantization of ${\mathcal{H}}_{\rm pert}[N_{\phi}]$.

Now, introducing our ansatz  $\Psi(v,\phi,\dph)= \Psi_0(v,\phi)\otimes \delta \Psi(v,\phi,\dph)$, and using that $\Psi_0$ satisfies the background equation (\ref{Sleq}), we obtain from (\ref{Eqn})  the equation of motion for $\delta \Psi$
\be \Psi_0\otimes  [ i \hbar \, \dphi \delta \Psi ]=   \hat{\mathcal{H}}_{\rm pert}[N_{\phi}]\, (\Psi_0\otimes \delta \Psi) \, . \ee
The test field approximation has been crucial to derive this equation, but no other simplification has been used. Also, recall that $\hat{\mathcal{H}}_{\rm pert}[N_{\phi}]$ acts on both $\Psi_0$ and  $\delta \Psi$. However, the presence of $\Psi_0$ in the left hand side indicates that
we can take the inner product with  $\Psi_0$ without loosing any information, and obtain 
\be \label{Sqedp} i \hbar \, \dphi \delta \Psi =  \langle \Psi_0| \hat{\mathcal{H}}_{\rm pert}[N_{\phi}]| \Psi_0 \rangle \,  \delta \Psi \, . \ee
where we have used that $\Psi_0$ is normalized. In other words, the information regarding the background FLRW geometry that influences the evolution of perturbations under the test field approximation is simply  the {\em expectation values} of the background operators that appear in $\hat{\mathcal{H}}_{\rm pert}[N_{\phi}]$; no other `moment' of $\Psi_0$ contributes to the dynamics. %

Equation  (\ref{Sqedp})  is a Schr\"odinger equation for $\delta \Psi$, with evolution Hamiltonian $\langle \hat{{\mathcal{H}}}\rangle_{\rm pert}\equiv  \langle \Psi_0| \hat{\mathcal{H}}_{\rm pert}[N_{\phi}]| \Psi_0 \rangle$, were the hat reminds us that this expectation value is only on the background state, and therefore this quantity  is still an operator when acting on perturbations. To solve this dynamics and compute physical observables, we will follow  techniques that are standard in quantum field theory in curved spacetimes. That is, states of perturbations belong to a Fock space $ \mathscr{H}_{\rm pert}$, on which dynamics is dictated by  $\langle \hat{{\mathcal{H}}}\rangle_{\rm pert} $
 in the standard way. (The total Hilbert space is therefore $\mathscr{H}_{\rm FLRW}\otimes \mathscr{H}_{\rm pert}$; this is the quantum analog of the classical phase space $\Gamma_{\rm FLRW} \times \Gamma_{\rm pert}$ of FLRW metrics plus  perturbations propagating thereon.) 

Now, we shall describe the dynamics of perturbations in more detail. As seen in section \ref{sec:IIB}, at the next-to-leading otder in perturbations the Hamiltonian has a quadratic and a cubic piece $\langle \hat{{\mathcal{H}}}\rangle_{\rm pert}=\langle \hat{{\mathcal{H}}}^{(2)}\rangle +\langle \hat{{\mathcal{H}}}^{(3)}\rangle$, where  $\langle \hat{{\mathcal{H}}}^{(2)}\rangle$ and $\langle \hat{{\mathcal{H}}}^{(3)}\rangle$ are the quantum operators associated with the classical expressions (\ref{hams}) and (\ref{eq:H3}), respectively. The quadratic Hamiltonian $\langle \hat{{\mathcal{H}}}^{(2)}\rangle$ provides the free evolution, and  $\langle \hat{{\mathcal{H}}}^{(3)}\rangle$ describes self-interactions between perturbations, which will be introduced perturbatively. 
\subsubsection{Free evolution of perturbation: the power spectrum}\label{sec:freeev}

The free evolution, which is obtained from (\ref{Sqedp}) by using $\langle \hat{{\mathcal{H}}}^{(2)}\rangle$ as the evolution Hamiltonian, can be now re-written in a more familiar form. Moving to the Heisenberg picture,  dynamics is given by the Heisenberg equations 

\bea \partial_{\phi} \hat \dph&=&i\hbar^{-1}\big[\hat \dph, \langle \hat{{\mathcal{H}}}^{(2)}[N_{\phi}]\rangle\big ]\, , \nonumber \\
 \partial_{\phi}  \hat{\dpp}&=&i\hbar^{-1}\big [\hat \dpp, \langle \hat{{\mathcal{H}}}^{(2)}[N_{\phi}\big]\rangle]\, .
\eea
 Now, by simple algebraic manipulations, these equations can be written as the second-order differential equation \cite{Ashtekar:2009mb}   
\be \label{eqnspert} (\t\Box - { \t{\u}})\,{\h{\dph}}(\vec{x},\t\eta) =0  \, \, , \hspace{1cm} \, .\ee
This equation \emph{has the same form as in semiclassical  cosmology}. The difference is that the differential operator  $\t\Box\equiv \t{g}^{ab}\tilde \nabla_a \tilde \nabla_b$  and the potential $\t\u$ are now constructed using the state $\Psi_{0}(v,\phi)$ chosen to describe the quantum FLRW geometry. More precisely,  $\t\Box$ is the d'Alembertian associated with a smooth FLRW metric tensor %
\be \label{dres}\t g_{ab}\dd x^a \dd x^b = \t {a}^2(\t {\eta})\,(-\dd\etat^2+\dd\vec{x}^2)\, , \ee
where $\t{a}$ is given by
\be
\label{ta}  \t a^4 = \f{\langle \h H_0^{-1/2}\h {a}^4 \h H_0^{-1/2}
\rangle} {\langle \  \h H_0^{-1} \rangle} \, , \ee
and the  conformal time $\tilde \eta$ is defined in terms of  the internal time $\phi$ of LQC via
\be \label{teta} \dd\tilde\eta = \mathcal{V}_0\, (\langle \h H_0^{-1} \rangle)^{1/2} \,\, ( \langle \h H_0^{-1/2}
\hat{a}^4 \h H_0^{-1/2}\rangle)^{1/2} \, \dd\phi \, .
\ee
The tensor $\t g_{ab}$ is known as the {\em effective dressed metric}. Furthermore, the {\em dressed potential}  $\t{\u}(\tilde \eta )$ is defined  by
\be \label{qpot} \t{\u} = \f{\langle \h{H}_0^{-\f{1}{2}}\,
\h{a}^2\, \h{\u}\,  \h{a}^2\, \h{H}_0^{-\f{1}{2}}
\rangle}{\langle \hat{H}_0^{-\f{1}{2}}\, \hat{a}^4\,
\hat{H}_0^{-\f{1}{2}}\rangle}\, , \ee
where $\h{\u}$ is the operator associated with the classical potential ${\u}$ defined in (\ref{potu}). All expectation values are evaluated in the state $\Psi_{0}(v,\phi)$. Recall, $\hat H_{0}$ is the Hamiltonian used in the evolution of $\Psi_{0}(v,\phi)$ and $\h a$ is the operator associated with the scale factor. Hence, under the test field approximation, the evolution of $\delta \phi$ at leading order in perturbations is {\em mathematically equivalent} to a quantum field theory of $\delta \phi$ on a curved FLRW spacetime described by $\t g_{ab}$. (\cite{Agullo:2013ai} has analyzed the  validity of the test field approximation by studying the energy-momentum tensor of perturbations.)\\

Now, if  $\Psi_{0}(v,\phi)$ is taken to be one of the sharply peaked state,  then $\t\Box$ becomes the d'Alembertian associated with the LQC effective metric obtained by integration of (\ref{EE}), and the potential  $\t{\u}$ is obtained from the classical expression (\ref{potu}) by just replacing the background variables $a$, $\pi_a$, $\phi$ and $\pp$ by the solution of (\ref{EE}). Hence, for sharply peaked states $\Psi_{0}$, the  evolution of perturbation proceeds in  the same mathematical manner as in semiclassical cosmology, with the difference that the background FLRW metric is not a solution to Einstein equations, but a solution to the LQC effective dynamics (\ref{EE}). %

For other states $\Psi_{0}(v,\phi)$  containing large dispersion in $v$,  the differential operator $\t\Box$ and the potential  $\t{\u}$ are sensitive not only to the mean values of the scale factor and other simple operators, but also about a few specific `moments' of $\Psi_{0}(v,\phi)$, precisely those appearing in equations (\ref{ta}), (\ref{teta}), and (\ref{qpot}). These moments, although non-trivial in appearance, can be computed numerically, and the result can be used to predict observable effects in   the CMB anisotropies.  Such analysis has been carried out in \cite{Agullo:2016hap} using states  $\Psi_{0}(v,\phi)$ with relative  dispersion $\Delta v/v$ as large as $168\%$ in the Planck regime.  Interestingly,  these computations show that, among all the effects that a large dispersion  produces on  the power spectrum, the only one that becomes significant compared to observational error bars is a direct  consequence of $\rho_{\rm B}$ being smaller than $\rho_{\rm sup}$ [see discussion below equation (\ref{EE})]. This means that,  in order to compute the primordial power spectrum in LQC for states $\Psi_{0}(v,\phi)$ with large dispersion, we can simply use the solution to the effective equations (\ref{EE}) after replacing $\rho_{\rm sup}$ by the actual value of the energy at the bounce (i.e., use the generalized effective equations). This is an accurate and  simple  recipe to extend the phenomenology  in LQC to states $\Psi_{0}(v,\phi)$  that are not sharply peaked \cite{Agullo:2016hap}. \footnote{ In  \cite{Agullo:2016hap} wave functions $\Psi_{0}(v,\phi)$ with different ``shapes'' in the $v$ variable and having large relative dispersion in $v$, although not arbitrarily large, were explored. However, the Hilbert space is infinite dimensional, and one could find states for which the conclusions of \cite{Agullo:2016hap} do not apply.}%

{\bf Remark:} To simplify the notation, from now on we will drop the `tilde' on the conformal time of the dressed metric, and the `bar' on solutions to the effective, and generalized effective equations.\\

Once we have the dressed metric $ g_{ab}$ and the dressed potential $\u$, the computation of observable quantities follow the standard procedure.\footnote{Note that, since we have already solved for the background dynamics, we can take the volume of the fiducial cell to infinity, $\mathcal{V}_0\to \infty$, in this section. Not taking this limit would only introduce a discretization of the wave-numbers $\vec{k}$, and the integrals in $\v k$ below would have to be replaced by sums.} First, expand the field operator in terms of creation and annihilation operators 
\be \label{modeexp}
  \h{\dph}({\vec x},\eta) = \int\f{{\rm d}^3 k}{(2\pi)^3} \,  \h{\dph}_{\vec{k}}( \eta)\, e^{i{\vec k}\cdot{\vec x}}= \int\f{{\rm d}^3 k}{(2\pi)^3} \left(\hat A_{\vec k}~\varphi_k(\eta) + \hat A^\dagger_{-\vec k}
~\varphi_{k}^*(\eta)\right) e^{i{\vec k}\cdot{\vec x}},
\ee
where $[\hat A_{\vec k},\hat A^{\dagger}_{\vec k'}]=\hbar \, (2\pi)^3\, \delta^{{(3)}}(\v k+\v k')$, $[\hat A_{\vec k},\hat A_{\vec k'}]= 0 = [\hat A^{\dagger}_{\vec k},\hat A^{\dagger}_{\vec k'}]$, and the set of  mode functions $\varphi_{k}(\eta)$ form a basis of solutions to the equation
\be \label{modeqn}
 \varphi_k^{\prime\prime} + 2\f{a '}{a} \varphi_k^\prime + (k^2 +
a^2 \,\tilde{\u })\,  \varphi_k =0\, ,
\ee
with normalization
\be  \label{norm} \varphi_k \varphi_k'^*-\varphi^*_k \varphi'_k= \frac{i}{ a^2} \, ,\ee
where  $k^2\equiv  k_i k_j\, \delta^{ij}$ is the comoving wave-number, and prime indicates derivative with respect to conformal time. The scalar power spectrum of $\h{\dph}$ is extracted from the two-point function in momentum space via
\be \label{2pleadord}
  \langle 0|\h{\dph}_{\vec k}( \eta) \h{\dph}_{\vec k^\prime}(\eta)|0\rangle \equiv 
(2\pi)^3\delta^{{(3)}}({\vec k}+{\vec k^\prime}) \f{2\pi^2}{k^3} \mathcal P_{\dph}(k, \eta)\, ,
\ee
where $|0\rangle$ is the vacuum annihilated by the operators $\hat A_{\vec{k}}$ for all $\vec{k}$. In terms of mode functions, we have  $\mathcal P_{{\dph}}(k,\eta) = (\hbar\,{k^3}/{2\pi^2})\,|\varphi_k(\eta)|^2$. The power spectrum of comoving curvature perturbations at the end of inflation, is  obtained from $\mathcal P_{{\dph}}$ by using the relation between $\delta \phi$ and $\mathcal{R}$, written in  (\ref{zetadph}), truncated at linear order
\be \label{powspec}\mathcal  \mathcal{P}_{\mathcal{R}}(k)\equiv 
\bigg(\frac{a(\eta_{\rm end})}{z(\eta_{\rm end})}\bigg)^2\mathcal  P_{{\dph}}(k,
\eta_{\rm end})=\bigg(\frac{a(\eta_{\rm end})}{z(\eta_{\rm end})}\bigg)^2\, \f{\hbar\,{k^3}}{{2\pi^2}}\,|\varphi_k( \eta_{\rm end})|^2\, ,\ee 
where $z= -\f{6}{\kappa}\f{\pp}{\pi_a}$.\\

\noindent {\bf Remark:}

An ambiguity appears in the analysis presented in this section, and it deserves some comments. Note that the  potential $\u$ that appears in the classical Hamiltonian of scalar perturbations [equation (\ref{potu})] contains powers of $\pi_a$, the momentum conjugated to the scale factor $a$. In the quantum theory 
one finds the problem that, in loop quantum cosmology, there is no operator associated with $\pi_a$; only complex exponentials  of $\pi_a$---i.e.,\ holonomies of the connection---are defined as  operators. This fact is intrinsic to the quantization strategy used in loop quantum gravity/cosmology, and it is a consequence of diffeomorphism invariance.  

There are several strategies that one can follow in order to compute the dressed potential  in (\ref{qpot}). We spell here three of them, which have been chosen based on the criteria of simplicity. 

(i) Use the classical Friedmann constraint (\ref{Fcons}) to trade $\pi_a$ for $a$, $\phi$ and $p_{\phi}$. There is no loss of generality in using the classical  constraints; it is an identity in the classical theory, which is the departing point for quantization. 

(ii) At a more practical level, when working with sharply peaked states, we can simply replace the expectation values of $\pi_a$ by the solution $\bar \pi_a(t)$ to the effective equations of LQC.

 (iii) %
Again, at the level of effective equations, replace factors $1/\pi_a$ in the classical Hamiltonian by  $-H/(2a^2 \rho)$, where $\rho$ is the energy density in the background. This equation holds in general relativity. In loop quantum cosmology, such relation is also valid after taking advantage  of the freedom in the quantization strategy (see, e.g.,\ \cite{MenaMarugan:2011me}, and references therein for discussions on  quantization ambiguities in LQC). 

In view of the existing freedom, we have  compared the results for the power spectrum and non-Gaussianity by using all three strategies, in order to understand how sensitive  observables are to these  quantization ambiguities. Our results (see section \ref{sec:tests}) show   that
the results of this paper remain the same regardless of the choice we make for $\pi_a$, out of the three strategies explained above. For the sake of simplicity, we will use strategy (ii) in the main calculations presented in the next section.

\subsubsection{Interaction Hamiltonian: the bispectrum}\label{sec:bisp}

The self-interaction of perturbations are described, at the lowest order, by the interaction Hamiltonian $\h{\mathcal{H}}_{\rm int} \equiv  \langle \Psi_0| \hat{\mathcal{H}}^{(3)}[N_{\phi}]| \Psi_0 \rangle$, where the classical expression for $\mathcal{H}^{(3)}$ was given in (\ref{eq:H3}).  As for the linear evolution, we are not free of factor ordering ambiguities, and we choose a symmetric ordering. At second order, therefore, the evolution of perturbations is sensitive to other moments of the state  $\Psi_{0}(v,\phi)$ chosen to describe the quantum FLRW geometry, in addition to the  three already involved in the free evolution, written in (\ref{ta}), (\ref{teta}), and (\ref{qpot}). The new moments follow straightforwardly  from (\ref{eq:H3})---keeping in mind the expression for $N_{\phi}$ and the symmetric ordering---and we do not explicitly write them here. 

To begin with, in the computation of the three-point function of scalar perturbations, we  restrict ourselves to sharply peaked states $\Psi_0 $ for the background geometry. As discussed above, at the practical level this is equivalent to replacing expectation values of background quantities by  solutions to the effective equations  (\ref{EE}).  Furthermore, as described at the end of section \ref{sec:EE}, the leading effects introduced by using more generic states can be accounted for by varying the value of the mean energy density at the bounce  $\rho_{\rm B}$. We  postpone such analysis to section \ref{sec:rhoB}.\\

The equal-time n-point correlation functions of scalar perturbations $\dph$, can be now computed at second order in perturbations by using the standard time-dependent perturbation theory:
\be \langle 0|\h{\dph}(\vec x_1,\eta) \h{\dph}(\vec x_2,\eta) \cdots \h{\dph}(\vec x_n,\eta)|0\rangle =\langle 0| U^{\dagger}(\eta,\eta_0) \, \h{\dph}^{\rm I}(\vec x_1,\eta) \h{\dph}^{\rm I}(\vec x_2,\eta) \cdots \h{\dph}^{\rm I}(\vec x_n,\eta)\,  U(\eta,\eta_0)|0\rangle\, ,\ee
where the superscript $I$ denotes operators in the interaction picture, and  $$U(\eta,\eta_0)= T \exp{\left(-i/\hbar \int_{\eta_0}^{\eta} \d\eta' \, \h{\mathcal{H}}^{\rm I}_{\rm int}(\eta')\right),}$$
is the time evolution operator relative to  $\h{\mathcal{H}}^{\rm I}_{\rm int}$.

The observable quantity we are interested in is the bispectrum $B_{\mathcal{R}}(k_1,k_2,k_3)$ of comoving curvature perturbations evaluated at the end of inflation. It is defined from the three-point correlation function of ${\mathcal{R}}$ in Fourier space, via
\be \label{fnlz}
\langle 0|\h{\mathcal{R}}_{{\vec k}_1} \h{\mathcal{R}}_{{\vec k}_2} \h{\mathcal{R}}_{{\vec k}_3}|0\rangle\equiv  (2\pi)^3\delta^{(3)}(\v{k}_1+\v{k}_2+\v{k}_3) \, B_{\mathcal{R}}(k_1,k_2,k_3) \, .\ee
The bispectrum $B_{\mathcal{R}}(k_1,k_2,k_3)$ has dimensions of ${\rm (length)}^{6}$. The presence of the Dirac-delta distribution is a consequence of the homogeneity  of the background FLRW metric. This delta distribution implies that only triads $(\vec k_1,\v k_2,\v k_3)$ that form a triangle may have a non-zero bispectrum. Additionally, isotropy makes the orientation of this triangle irrelevant. These two facts combined are the reason why $B_{\mathcal{R}}$ depends on the wave-numbers $(\vec k_1,\v k_2,\v k_3)$ only via three real parameters. Common choices are $(k_1,k_2,k_3)$ with $k_3\lesssim k_1+k_2$, or $(k_1,k_2,\mu\equiv  \hat k_1\cdot \hat k_2)$. 

 It is common, and convenient, to quantify the amplitude of the bispectrum in terms of the dimensionless function $f_{_{\rm NL}}(k_1,k_2,k_3)$, defined as
\be \label{fNLdef} B_{\mathcal{R}}(k_1,k_2,k_3)\equiv  -\f{6}{5} \, f_{_{\rm NL}}(k_1,k_2,k_3)\, \times (\Delta_{k_1}\Delta_{k_2}+\Delta_{k_1}\Delta_{k_3}+\Delta_{k_2}\Delta_{k_3})\, ,\ee
or, equivalently, by
\be \label{fNLdef}  f_{_{\rm NL}}(k_1,k_2,k_3)\equiv  -\f{5}{6}B_{\mathcal{R}}(k_1,k_2,k_3) \,\, \times (\Delta_{k_1}\Delta_{k_2}+\Delta_{k_1}\Delta_{k_3}+\Delta_{k_2}\Delta_{k_3})^{-1}\, ,\ee
where $\Delta_{k}\equiv  \frac{2\pi^2}{k^3} \, \mathcal{P}_{\mathcal{R}}(k)$ is the dimensionful power spectrum. (See \cite{Komatsu:2001rj} for the origin of the convention leading to the numerical factor $-5/6$, and see Appendix A of \cite{LoVerde:2007ri} for a summary of different conventions for the sign). Looking at expression (\ref{fNLdef}), we can intuitively think about  $f_{_{NL}}$ as the amount of correlations in ``units" of $\Delta_{k}^2$.

Now, in order to compute the bispectrum $B_{\mathcal{R}}(k_1,k_2,k_3)$ in terms of $\dph$, we use the relation between both variables given in section \ref{sec:3H}
\be \label{Rtodph} \mathcal{R}(\v x,\eta )=- \f{a}{z} \, \dph(\v x,\eta)+\left[-\f{3}{2}+3\f{V_{\phi}\, a^5}{\kappa\, \pp\, \pi_a}+\f{\kappa}{4}\f{z^2}{z^2}\right]  \left(\f{a}{z} \, \dph(\v x, \eta)\right)^2+\cdots \, ,\ee
where,   the dots represent terms producing subdominant contributions to correlation functions at the end of inflation for the wave-numbers $\vec{k}$ that we can observe today (see  equation (\ref{zetadph}) and the discussion after it). With this, we have
\bea \label{3pz}
& & \langle 0|\h{\mathcal{R}}_{{\vec k}_1} \h{\mathcal{R}}_{{\vec k}_2} \h{\mathcal{R}}_{{\vec k}_3}|0\rangle=\left(-\f{a}{z}\right)^3  \langle 0|\h{\dph}_{{\vec k}_1} \h{\dph}_{{\vec k}_2} \h{\dph}_{{\vec k}_3}|0\rangle\nonumber \\ &+& \left(-\f{3}{2}+3\f{V_{\phi}\, a^5}{\kappa\, \pp\, \pi_a}+\f{{\kappa}}{4}\f{z^2}{a^2}\right)\, \left(-\f{a}{z}\right)^4\, \Big[\int \f{d^3p}{(2\pi)^3} \, \langle 0|\h{\dph}_{{\vec k}_1} \h{\dph}_{{\vec k}_2}  \h{\dph}_{{\vec p}}\,  \h{\dph}_{{\vec k}_3-\v p}|0\rangle + (\v k_1 \leftrightarrow \v k_3)+ (\v k_2 \leftrightarrow \v k_3) \, \nonumber \\
&+&\cdots\Big] \, .  \eea
In this equation, $(\v k_i \leftrightarrow \v k_j)$ indicates terms obtained from the first term in the second line  after interchanging $\v k_i$ and $\v k_j$, and the dots  indicate subdominant contributions. To obtain the scalar bispectrum $B_{\mathcal{R}}$ and $f_{_{\rm NL}}$ at leading order we need to compute the three-  and four-point correlation functions of $\h{\dph}_{{\vec k}}$. 

Let us begin with  the three-point  function, appearing  in the first line in (\ref{3pz}). At leading order in the interaction Hamiltonian, it is given by
\bea \langle 0|\h{\dph}_{{\vec k}_1}(\eta) \h{\dph}_{{\vec k}_2}(\eta) \h{\dph}_{{\vec k}_3}( \eta)|0\rangle  &=&\langle 0| \h{\dph}^{\rm I}_{{\vec k}_1}( \eta) \h{\dph}^{\rm I}_{{\vec k}_2} ( \eta)\h{\dph}^{\rm I}_{{\vec k}_3}( \eta)|0\rangle\nonumber \\
&&-\,i/\hbar  \int d \eta' \langle 0|\left[ \h{\dph}^{\rm I}_{{\vec k}_1}( \eta) \h{\dph}^{\rm I}_{{\vec k}_2}(\eta) \h{\dph}^{\rm I}_{{\vec k}_3}( \eta), \h{\mathcal H}^{\rm I}_{\rm int}( \eta')\right]|0\rangle \nonumber \nonumber \\&&+\,\mathcal{O}(\mathcal H^2_{\rm int}) \, . 
\eea
The first term in the right hand side vanishes, $\langle 0| \h{\dph}^{\rm I}_{{\vec k}_1} \h{\dph}^{\rm I}_{{\vec k}_2} \h{\dph}^{\rm I}_{{\vec k}_3}|0\rangle=0$, since  $\h{\dph}^{\rm I}_{\vec k}$ in the interaction picture is a Gaussian field. Hence, the term in the second line gives the leading order contribution. By using the mode expansion (\ref{modeexp}),  we find

\bea \label{bispdph} & &\langle 0|\h{\dph}_{{\vec k}_1}(\eta) \h{\dph}_{{\vec k}_2}(\eta) \h{\dph}_{{\vec k}_3}(\eta)|0\rangle =(2\pi)^3\delta^{(3)}( \vec{k}_1+\vec{k}_2+\vec{k}_3)\, B_{\delta\phi}(k_1,k_2,k_3) \, ,\eea
where
\bea \label{Bphi} && B_{\delta\phi}(k_1,k_2,k_3)=2\, \hbar^2 \, {\rm Im} \Big[\varphi_{\vec{k}_1}( \eta)\varphi_{\vec{k}_2}( \eta)\varphi_{\vec{k}_3}( \eta)\nonumber \\
&\times& \int_{ \eta_0}^{\eta} \d\eta'\, \Big( f_1(\eta')\, \varphi^{\star}_{{k}_1}( \eta')\varphi^{\star}_{{k}_2}(\eta')\varphi^{\star}_{{k}_3}( \eta') 
+f_2( \eta')\, \varphi^{\star}_{{k}_1}( \eta')\varphi^{\star}_{{k}_2}( \eta'){\varphi'}_{{k}_3}^{\star}(\eta') 
+ f_3( \eta')\, \varphi_{{k}_1}^{\star}(\eta'){\varphi'}_{{k}_2}^{\star}( \eta'){\varphi'}_{{k}_3}^{\star}( \eta') \nonumber \\
&+&(\v k_1 \leftrightarrow \v k_3)+ (\v k_2 \leftrightarrow \v k_3) \Big)
 \Big ]+\mathcal{O}(\mathcal H^2_{\rm int}) \, ,\eea
where the functions $f_1( \eta)$, $f_2(\eta)$ and $f_3(\eta)$ are combinations of background functions, given in Appendix B. \\

The terms in the second line of (\ref{3pz}) involve the four-point correlation function of  $\h{\dph}^{\rm I}_{\vec k}$. Applying again time-dependent perturbation theory, we get 
\be \label{4p} \langle 0|\h{\dph}_{{\vec k}_1}(\eta) \h{\dph}_{{\vec k}_2}(\eta) \h{\dph}_{{\vec p}}( \eta) \h{\dph}_{{\vec k}_3-\vec p}( \eta) |0\rangle 
=\langle 0| \h{\dph}^{\rm I}_{{\vec k}_1}( \eta) \h{\dph}^{\rm I}_{{\vec k}_2} ( \eta)\h{\dph}^{\rm I}_{{\vec p}}(\eta)\h{\dph}^{\rm I}_{{\vec k}_3-\vec p}(\eta)|0\rangle +\mathcal{O}(\mathcal H_{\rm int})  \, .
\ee
In this case,  the first term does not vanish, and provides the leading order contribution. There is no need to compute higher order terms, since they are subdominant. The first term, furthermore, does not involve any time integral of the interaction Hamiltonian, and its expression in terms of the mode functions $\varphi_k$   reads
\be \label{4p2} \int \f{\d^3p}{(2\pi)^3} \, \langle 0|\h{\dph}_{{\vec k}_1} \h{\dph}_{{\vec k}_2}  \h{\dph}_{{\vec p}}\,  \h{\dph}_{{\vec k}_3-\v p}|0\rangle = (2\pi)^3 \delta^{(3)}(\v{k}_1+\v{k}_2+\v{k}_3)  \,2\, \hbar^2 \,  |\varphi_{k_1}|^2 |\varphi_{k_2}|^2\,  \, .\ee
Substituting these results in (\ref{3pz}) we obtain the desired expression for $ B_{\mathcal{R}}$

\bea  \label{BR} B_{\mathcal{R}}(k_1,k_2,k_3)&=&\left(-\f{a}{z}\right)^3\, B_{\delta\phi}(k_1,k_2,k_3) \\ &+&  
\left[-\f{3}{2}+3\f{V_{\phi}\, a^2}{\kappa\, \pp\, \pi_a}+\f{\sqrt{\kappa}}{4}\f{z^2}{a^2}\right]  \left(\f{a}{z}\right)^4\, 
2\, \hbar^2\,  \big(|\varphi_{k_1}|^2 |\varphi_{k_2}|^2+|\varphi_{k_1}|^2 |\varphi_{k_2}|^2+|\varphi_{k_2}|^2 |\varphi_{k_3}|^2\big) \nonumber \, , \eea
where all quantities are evaluated at the end of inflation.

\section{Numerical evaluation of  the three-point correlation function}\label{sec:numres}

The main goal of this section is to evaluate the bispectrum $B_{\mathcal{R}}(k_1,k_2,k_3)$, written in  (\ref{BR}), at the end of inflation, for different values of the three momenta $k_1,k_2$, and $k_3$, and to compute the function $f_{_{\rm NL}}(k_1,k_2,k_3)$ from it. This section shows the results of numerical computations, while  in section \ref{sec:analytical} we present  analytical arguments that will help us to better understand their physical origin. 

Scalar perturbations are evolved starting at an early time, to be specified below, across the bounce, and until the modes of interest become super-Hubble during the inflationary phase. The power spectrum and bispectrum will be computed at that time. In order to perform these calculations we need to:

\begin{enumerate} 

\item Specify a potential $V(\phi)$ for the scalar field. 

\item Specify a solution ($ a(\eta)$, ${\pi}_ a(\eta)$, $\phi(\eta)$,  ${p}_{\phi}(\eta)$) to the effective equations  (\ref{HrateLQC})--(\ref{HrateLQC3}) of LQC.  As discussed in the last two paragraphs of section \ref{sec:EE}, these solutions are uniquely characterized by specifying the value of  $\phi$ and  its energy density at the time of bounce.

\item Specify the quantum state of scalar perturbations at some initial time $\eta_0$. 

\end{enumerate}
 
These are the freedoms that we have in our calculation. In this section we choose:
\begin{enumerate} 

\item  The quadratic potential $V(\phi)=\frac{1}{2}m^2\, \phi^2$, with the value of $m$ that is obtained from the Planck normalization \cite{Ade:2015lrj}, $m=6.4\times10^{-6}M_{P\ell}$.

\item  A background effective geometry with  $\phi_{\rm B}=7.62\,M_{P\ell}$ and $\rho_{B}=1\,M_{P\ell}^4$.

\item A Minkowski-like vacuum for perturbations, specified at an early enough time before the bounce such that all Fourier modes of interest are in an adiabatic regime. 
More precisely, we choose  $\varphi_k(\eta_0)=\f{1}{a(\eta_0)\sqrt{2\, k}}$ and $\varphi'_k(\eta_0)=[-i\, k+\frac{a'(\eta_0)}{a(\eta_0)}]\, \varphi_k(\eta_0)$ as initial data for the modes, for $\eta_0=-2.8\times 10^{3}\,T_{P\ell}$ (the bounce takes place at $\eta_0=0$).
\end{enumerate}

In sections \ref{sec:phiB} - \ref{sec:inistate} we analyze  the  way the results vary for other choices.

To carry out the calculation we use the numerical infrastructure of \verb|class| \cite{2011arXiv1104.2932L}, a standard Einstein-Boltzmann solver for cosmological perturbations, written in C. 
First, we solve the background dynamics, and then  we use the result to solve the dynamics of perturbations. We compute the time integrals in (\ref{Bphi}) by writing it as a first order differential equation for the integrands, and we solve them simultaneously with the equation of motion (\ref{modeqn}) for each Fourier mode. This ensures that the time step of the numerical integrator is adapted to achieve the desired accuracy for the bispectrum. For solving the differential equations, we have used the Runge Kutta evolver provided by \verb|CLASS|.

There are other codes aimed at computing  primordial non-Gaussianity (e.g.\ \verb|BINGO| \cite{Hazra:2012yn}, \verb|PyTransport| and \verb|CppTransport| \cite{Dias:2016rjq}, and a code to compute three-point functions involving tensor perturbations \cite{Sreenath:2013xra}). But they are mostly oriented towards computations during the inflationary epoch, and they cannot be used for our purposes.

Before computing the bispectrum, we first summarize our  results for the power spectrum.

\bfig
 \ig[width=0.8\textwidth]{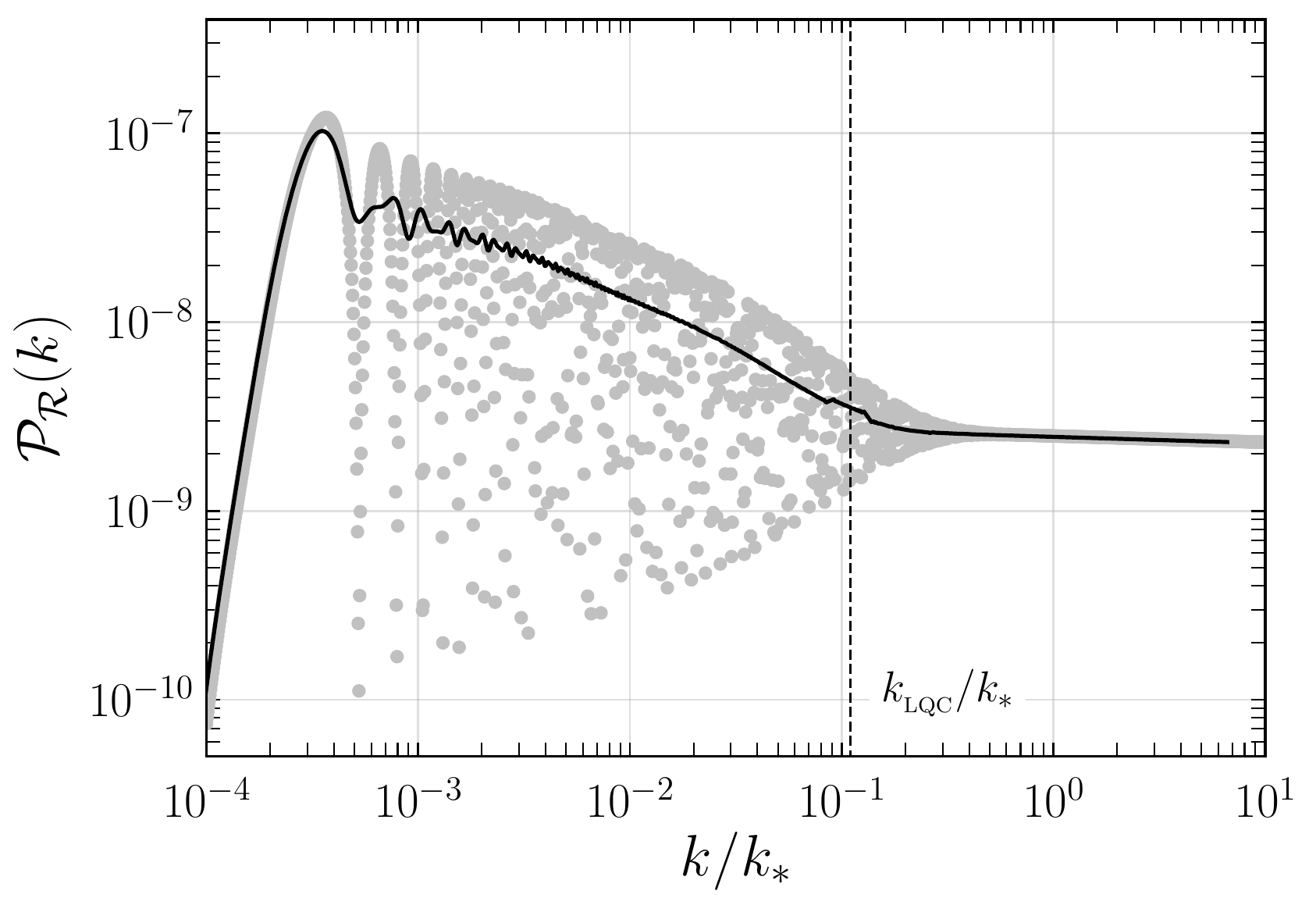}
\caption{Power spectrum for comoving curvature perturbations for $\phi_{\rm B}=7.62 M_{P\ell}$, and $\rho_{\rm B}=1\, M_{P\ell}^4$. Gray dots indicate the numerical value of $\mathcal{P}_{\mathcal{R}}$ for individual values of $k$. The black curve denotes the average of the gray points.  As expected, the spectrum is scale invariant for $k\gg k_{\rm LQC}$. The effects from the bounce appear for $k\lesssim k_{\rm LQC}$.  For the  value of $\phi_{\rm B}$ used in this plot, the number of e-folds between the bounce and  horizon exit for the pivot scale $k_{\star}$, is, $N_{B\star} = 12.3$. This number is large enough to make the effects created by the bounce to  be redshifted to super-Hubble scales at the present time (recall that the observable window is approximately $k \in[k_*/10,1000k_*])$. Section  \ref{sec:phiB} and \ref{sec:rhoB}  contain plots of $\mathcal{P}_{\mathcal{R}}$  for other values of $\phi_{\rm B}$ and $\rho_{\rm B}$ for which the enhancement of the power spectrum occurs for observable scales (see also \cite{Agullo:2015tca}).}
\label{figPS}
\efig
\subsection{The power spectrum}\label{PS}
The mathematical and physical aspects of the primordial power spectrum $\mathcal{P}_{\mathcal{R}}(k)$ in LQC have been  discussed in detail in \cite{Agullo:2012fc,Agullo:2015tca,Schander:2015eja,Bolliet:2015bka}, so we will be brief here. To compute $\mathcal{P}_{\mathcal{R}}(k)$,  we need  to solve the second-order differential equation (\ref{modeqn}) for the set of wave-numbers of interest for observations. The values of $k$ that we can probe in the CMB, range approximately from $k_{\rm min}= k_*/10$, to $k_{\rm max}=1000 k_*$, where $k_*$ is  a pivot, or reference wave-number whose  physical value at  present  is  $k_*/a(t_{\rm today})=0.002\, {\rm Mpc}^{-1}$. We will, however, compute $\mathcal{P}_{\mathcal{R}}(k)$ for values of $k$ smaller than $k_{\rm min}$, because these modes, although not directly observable in the CMB, may indirectly affect the observable power spectrum once non-Gaussianity are taken into account \cite{Agullo:2015aba}.

In order to better understand the form of the power spectrum,  it is convenient to define the re-scaled mode functions $v_k(\eta)\equiv  a(\eta)\, \varphi_k(\eta)$. The wave equation (\ref{modeqn}), when written in terms of $v_k$, takes the form
\be \label{chieq} v_k''(\eta)+(k^2+f(\eta))v_k(\eta)=0\, , \ee
where $f(\eta)\equiv  a(\eta)^2\, \u(\eta)-\f{a''}{a}(\eta)=a^2(\u-\f{R}{6})$, and $R(\eta)$ is the Ricci scalar of the effective spacetime geometry. 
 The potential  $\u$ was defined in (\ref{potu}). It is clear from this equation that whenever  $k^2\gg |f(\eta)|$, the solutions are simple oscillatory functions with time independent frequency equal to $k$. On the contrary,  $v_k(\eta)$ will have a more complicated behavior  when $k^2\lesssim |f(\eta)|$. In particular,  when the function $f(\eta)$ becomes negative, the oscillatory behavior of these modes  changes to an exponentially varying amplitude, that results in a modulation of the amplitude of $v_k(\eta)$, and consequently of the power spectrum. 

During the inflationary era, $f(\eta)$  remains approximately constant, and is proportional to the Ricci scalar $R$, or the square of the Hubble radius. This value sets up the  wave-number scale for which amplification of perturbations takes place. Similarly, the amplification of perturbations around the time of bounce can be characterized  in terms of the {\em  physical scale associated with the  bounce}. This  scale is given by the value of the function $f(\eta)$ at the bounce, which is approximately equal to  $a^2\f{R}{6}$ evaluated at that time (see the definition of $f(\eta)$ above, and take into account that  $\u$ is of the same order as $R/6$ around the bounce). Therefore, we define the bounce scale $k_{\rm LQC}$ as $k_{\rm LQC}\equiv  a({\eta_B}) \sqrt{R_{\rm B}/6}\approx a({\eta_B}) \sqrt{\kappa \, \rho_{\rm B}}$, where the subscript  $B$ indicates quantities evaluated at the bounce.  Qualitatively, we expect  the power spectrum   to  be significantly affected by the bounce for modes with $k\lesssim k_{\rm LQC}$. On the other hand, the bounce is expected to have little effect on $k\gg k_{\rm LQC}$, since these modes are ``too ultraviolet to feel the bounce''.

In figure \ref{figPS} we show the LQC power spectrum $\mathcal{P}_{\mathcal{R}}(k)$  for scalar perturbations computed using the settings specified at the beginning of this section. The scale invariant inflationary prediction is recovered for $k\gg k_{\rm LQC}$.  In contrast, for $k\lesssim k_{\rm LQC}$ there is an extra contribution coming from the propagation of perturbations across the bounce. This contribution breaks  scale invariance, and makes $\mathcal{P}_{\mathcal{R}}(k)$ to grow significantly  for small wave-numbers.  As discussed in section \ref{sec:inistate}, all other choices of initial data for perturbations explored in this paper produce a power spectrum that grows for  $k\lesssim k_{\rm LQC}$. Note, however,  that there exist other choices in the literature for which the spectrum is suppressed, rather than enhanced, on these scales \cite{deBlas:2016puz,Ashtekar:2016wpi,Ashtekar:2016pqn}. We do not consider such states in the analysis presented here.

\bfig
 \ig[width=0.7\textwidth]{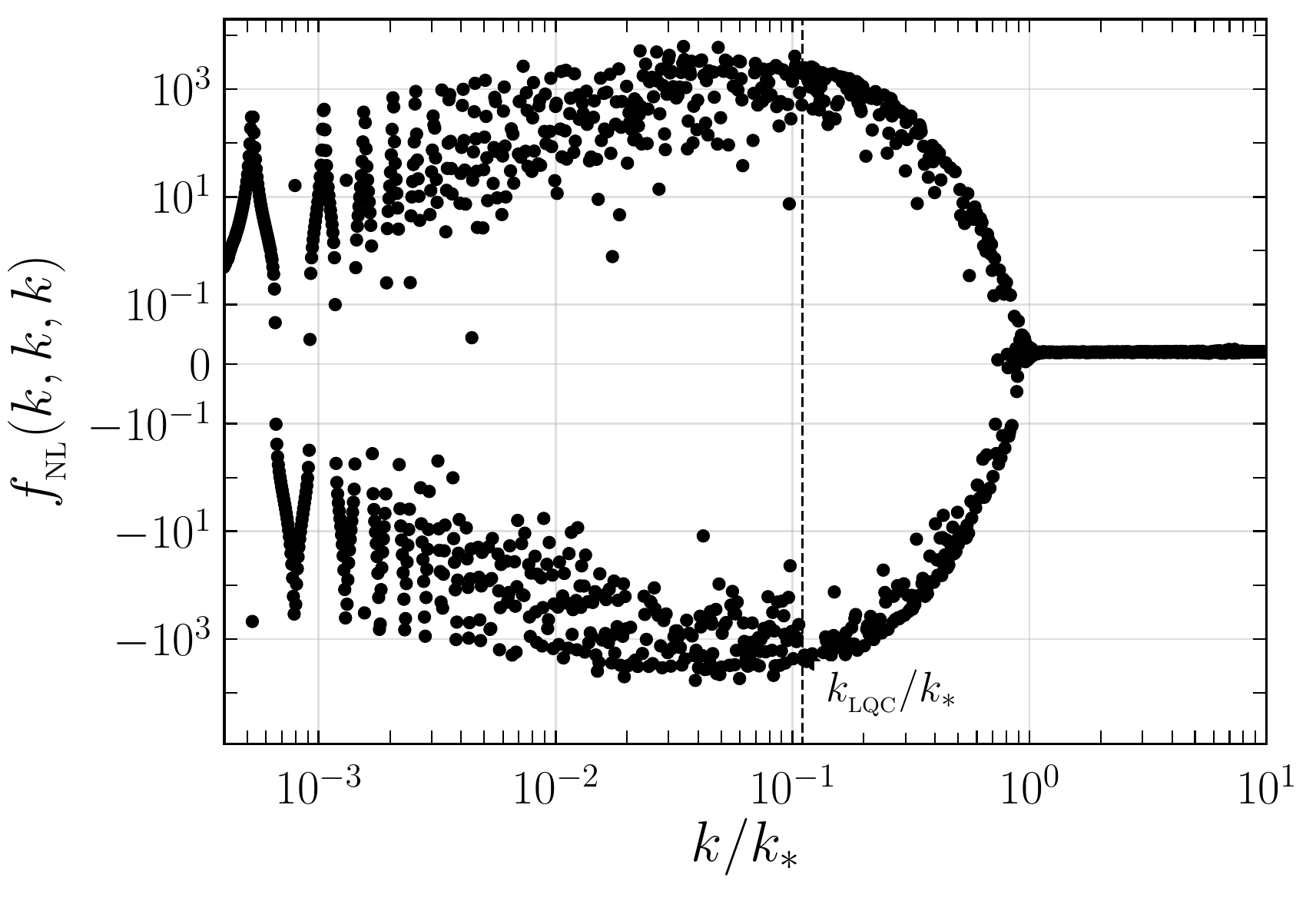}
\caption{
{\em Equilateral configurations}. Plot of $f_{_{\rm NL}}(k,k,k)$ versus $k$. We have used here the same parameter as in the plot of the power spectrum, figure \ref{figPS}, namely $\phi_{\rm B}=7.62\, M_{P\ell}$, and $\rho_{\rm B}=1 \, M_{P\ell}^4$, and Minkowski-like initial data for perturbations at  $\eta_0=-2.8\,10^3 \, T_{P\ell}$  (or equivalently, $t_0=-10^5 T_{P\ell}$ in cosmic time). The plot shows that $f_{_{\rm NL}}(k,k,k)$ is highly oscillatory, and its amplitude is strongly scale dependent. For the value of the $\phi_{\rm B}$, and $\rho_{\rm B}$ chosen in this plot, $f_{_{\rm NL}}$ grows only for the most infrared scales that we can observe in the CMB, which correspond to angular multipoles $\ell\lesssim 30$.}
\label{fnl}
\efig

 \bfig
 \ig[width=0.7\textwidth]{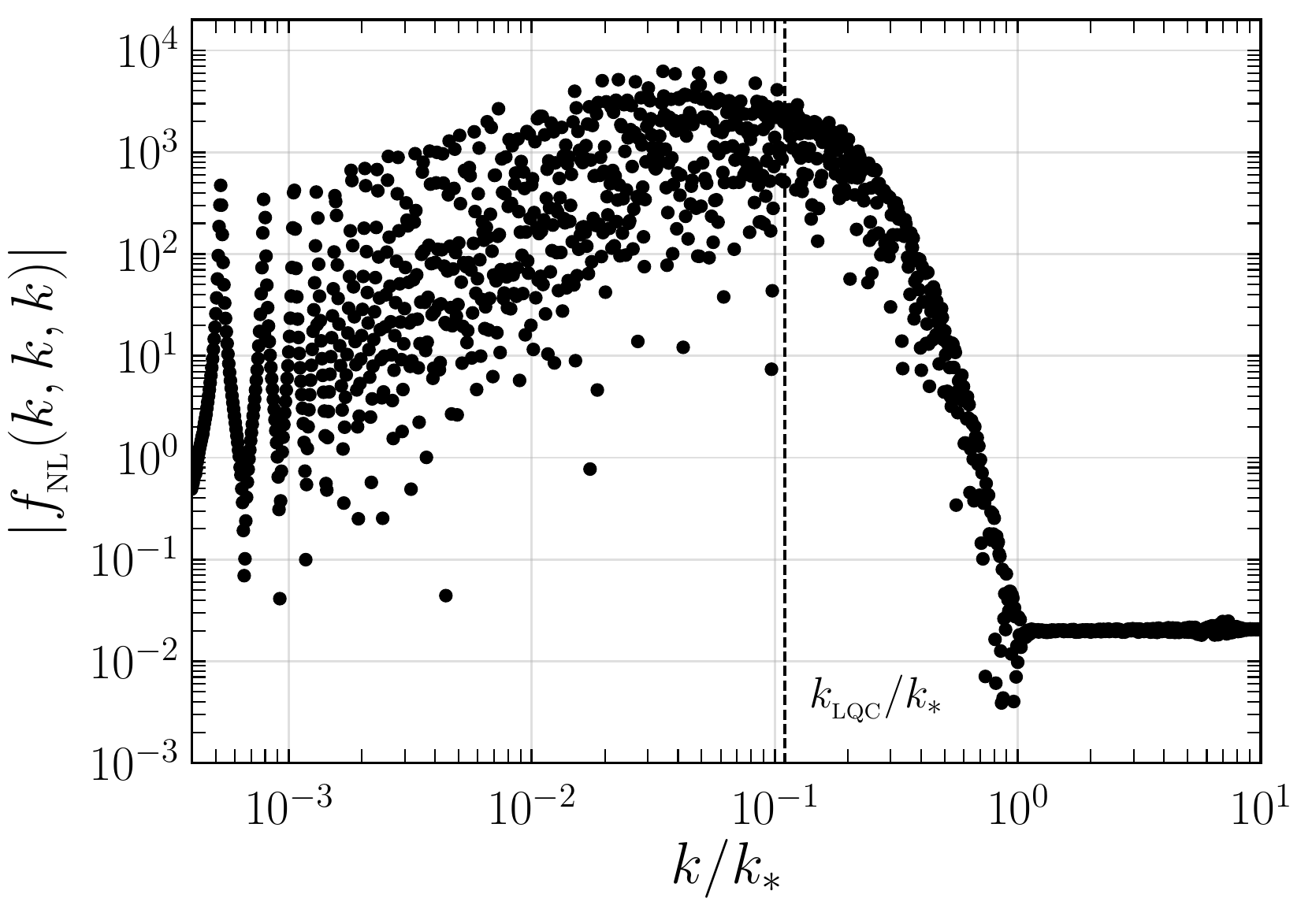}
\caption{{\em Equilateral configurations}. Plot of $|f_{_{\rm NL}}(k,k,k)|$ versus $k$. We have  used the same values of the parameter as in the previous  plot.} \label{fnl2}
\efig

\subsection{The bispectrum}\label{sec:bispectrum}

The numerical evaluation of the bispectrum  requires more effort than what is needed to compute non-Gaussianity {\em during inflation}. 
The first reason is that, in the inflationary era,  only the terms  in the third order Hamiltonian (\ref{eq:H3}) that are leading order in the slow-roll parameters  need to be considered. This provides a significant simplification of the Hamiltonian, which, after integration by parts, reduces to a single term \cite{Maldacena:2002vr}. The second reason is that the background geometry during slow-roll inflation is very close to be described by de Sitter geometry. This makes  an analytical approximation for the modes $\varphi_k(\eta)$ available, which in turn allows for an analytical calculation of the bispectrum. All these simplifications cannot be used in our case because, first of all, before inflation the slow-roll approximation is no longer valid.  And secondly, in our problem the spacetime goes through a contracting phase, followed by a bounce, a pre-inflationary phase on which the kinetic energy of the scalar field is converted to potential energy, and finally an inflationary phase. In each of these phases the scale factor behaves in a quite different manner and, as a consequence, it is difficult to arrive at an analytical approximation for $\varphi_k(\eta)$ valid during the entire evolution.\footnote{There exist efforts to compute non-Gaussianity  in more complicated inflationary scenarios involving deviations from slow-roll, both analytically (see, e.g., \cite{Flauger:2010ja,Martin:2011sn}) and numerically \cite{Hazra:2012yn,Sreenath:2014nca}. However, the pre-inflationary evolution that we are interested in is more complicated than the scenarios previously considered.} 

We present our results for non-Gaussianity in terms of  the function $f_{_{\rm NL}}(k_1,k_2,k_3)$, defined in  (\ref{fNLdef}). We evaluate $f_{_{\rm NL}}(k,\alpha_1\,  k,\alpha_2 \, k)$ as a function of $k$, for different values of $\alpha_1$ and $\alpha_2$. Following standard terminology, we will refer to triads $(k,\alpha_1\,  k,\alpha_2 \, k)$ for which $(\alpha_1 =  \alpha_2=1)$  as \textit{equilateral} configurations of wave-numbers. Similarly, $(\alpha_1\approx 1, \alpha_2\ll \alpha_1)$ and $(\alpha_2\approx 1- \alpha_1)$ are known as \textit{squeezed} and \textit{flattened} configurations, respectively. These names are motivated by the shape of the triangles formed by $\v k_1$, $\v k_2$, and $\v k_3$.

In figure  \ref{fnl} we show $f_{_{\rm NL}}$ in the equilateral configuration as a function of $k/k_*$. In the regime $k\gtrsim k_{\rm LQC}$ the result agrees with the inflationary prediction, i.e., $f_{_{\rm NL}}\sim \epsilon$ where $\epsilon$ is the slow-roll parameter evaluated at horizon exit. For scales that were larger than the curvature radius at the bounce, i.e., $k\lesssim k_{\rm LQC}$, $f_{_{\rm NL}}$ oscillates between positive and negative values with an amplitude of order $10^{3}$. In figure \ref{fnl2} we show the absolute value of $f_{_{\rm NL}}$ in the equilateral configuration in order to analyze the scale dependence of  $f_{_{\rm NL}}$ more carefully. In figure \ref{fnl22} we show $f_{_{\rm NL}}$ in a few  different configurations. In figure \ref{trigfNL} we present  two-dimensional plots  for $f_{_{\rm NL}}$ containing  all configurations, by  fixing  $k_1$ to three different values.

\begin{figure}
\includegraphics[width=0.6\textwidth]{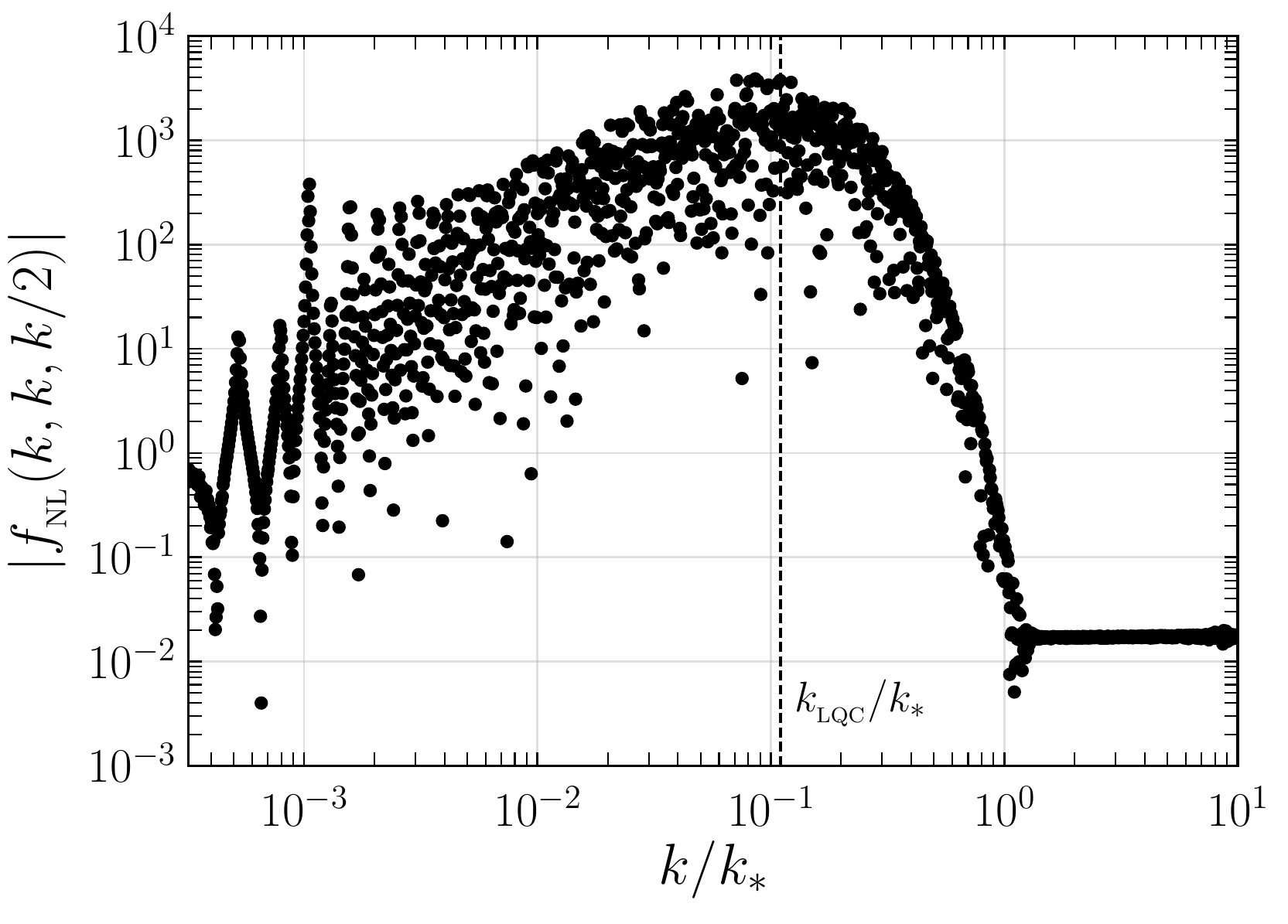}
\includegraphics[width=0.6\textwidth]{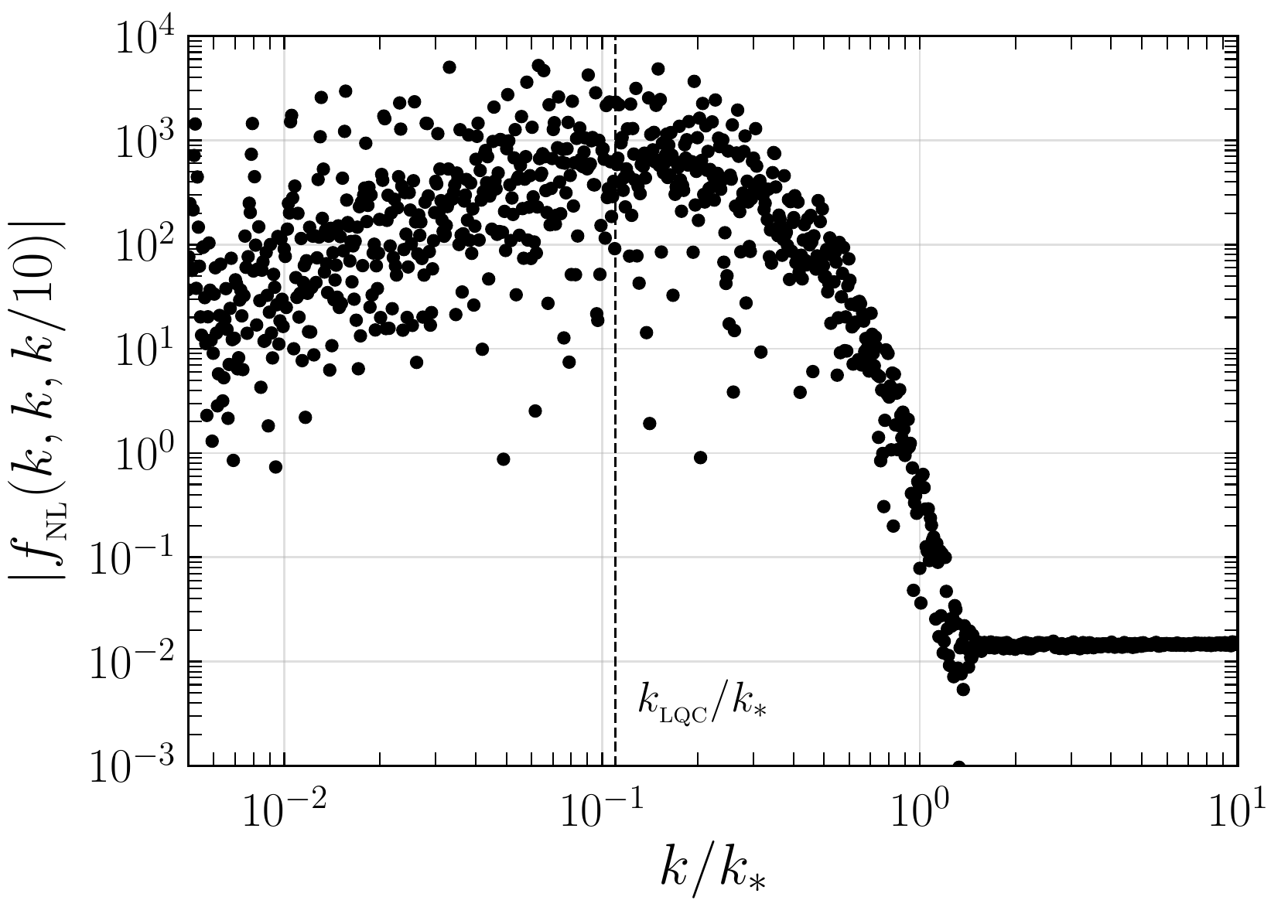}
\includegraphics[width=0.6\textwidth]{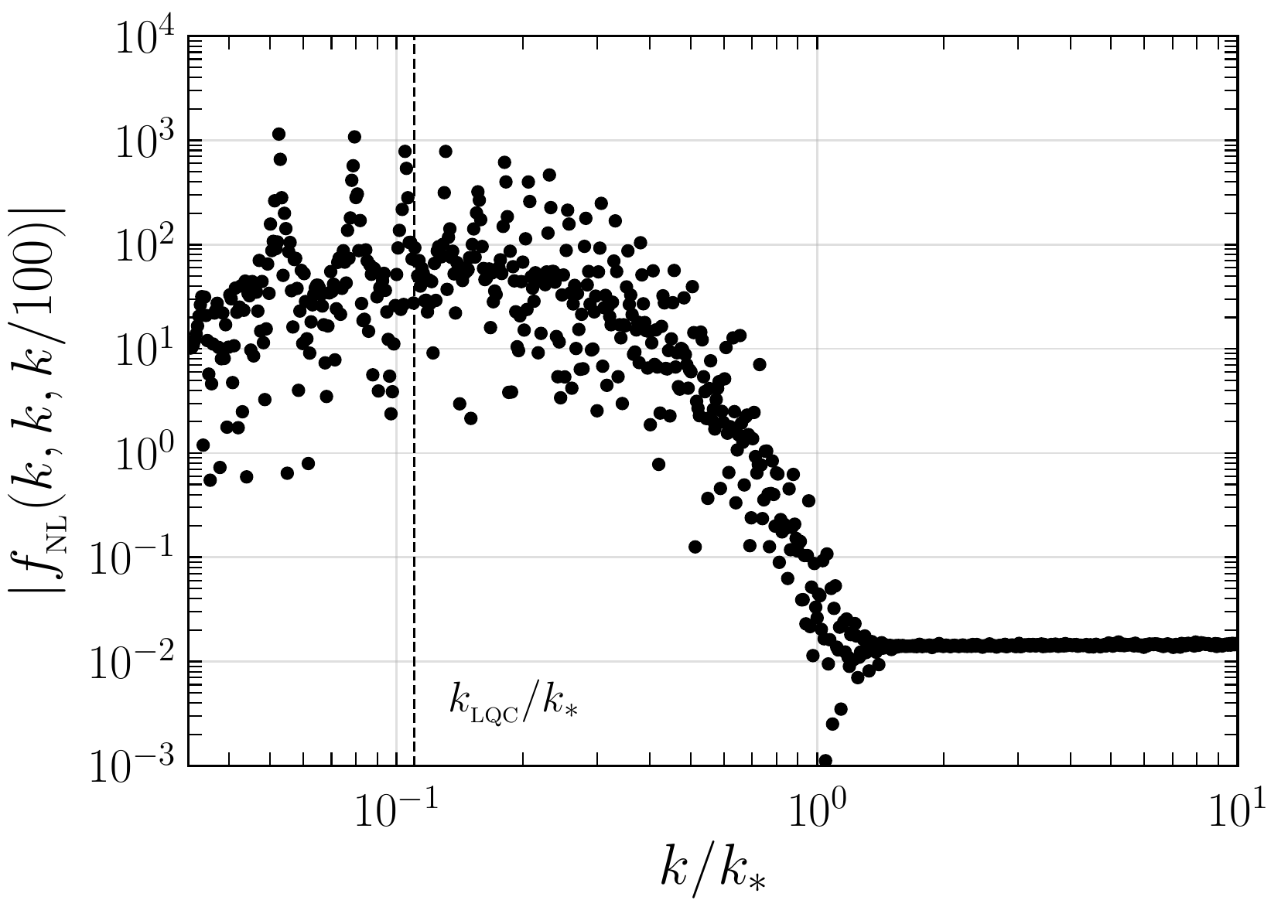}
\caption{Plots of $|f_{_{\rm NL}}(k, k, k/2)|$ (top), $|f_{_{\rm NL}}(k, k, k/10)|$ (middle)  and $|f_{_{\rm NL}}(k, k, k/100)|$ (bottom)  versus $k$. We have  used the same values of the parameter as in the previous  plot.}%
\label{fnl22}
\efig

\bfig
 \ig[width=0.7\textwidth]{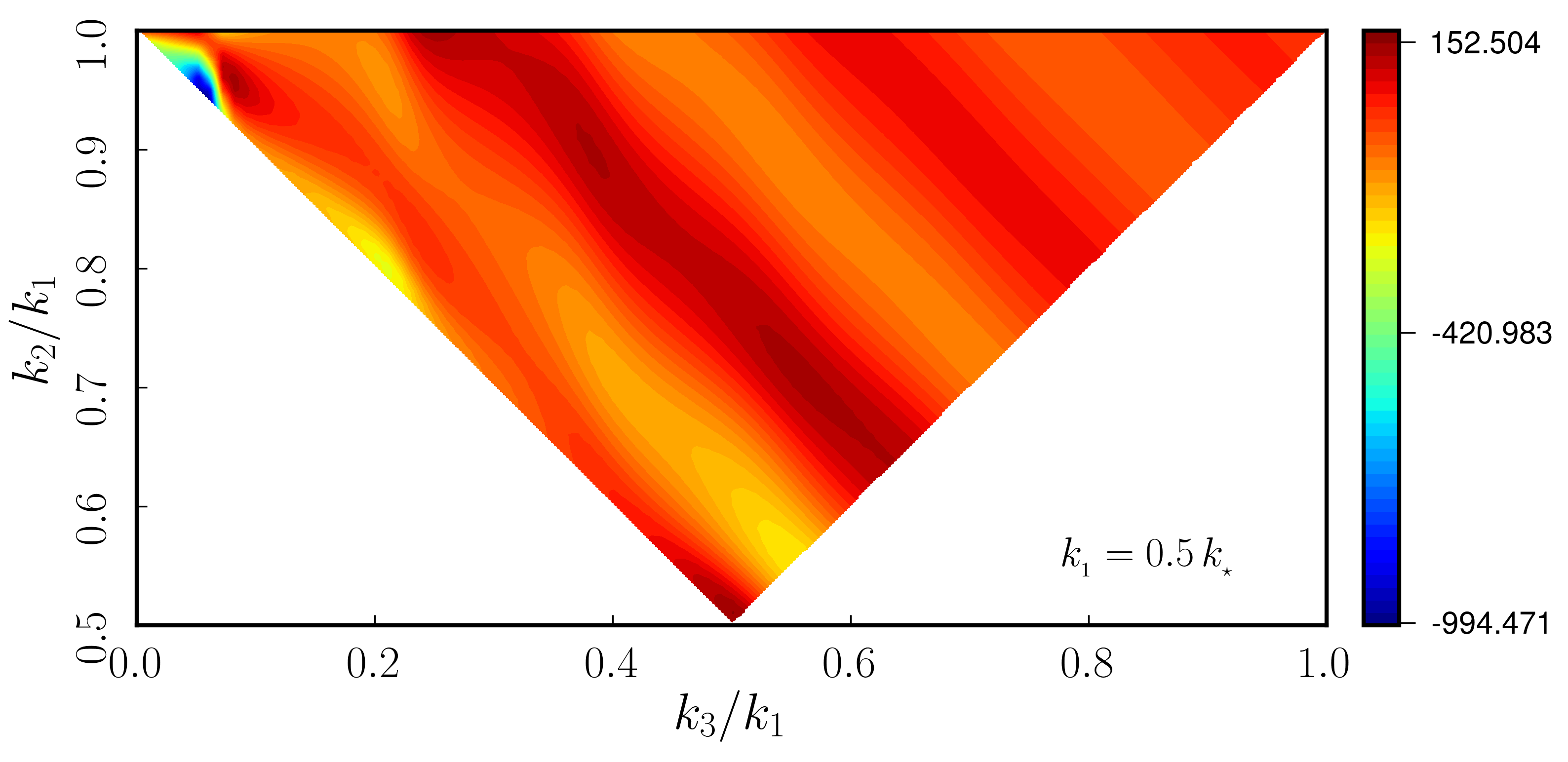}
 \ig[width=0.7\textwidth]{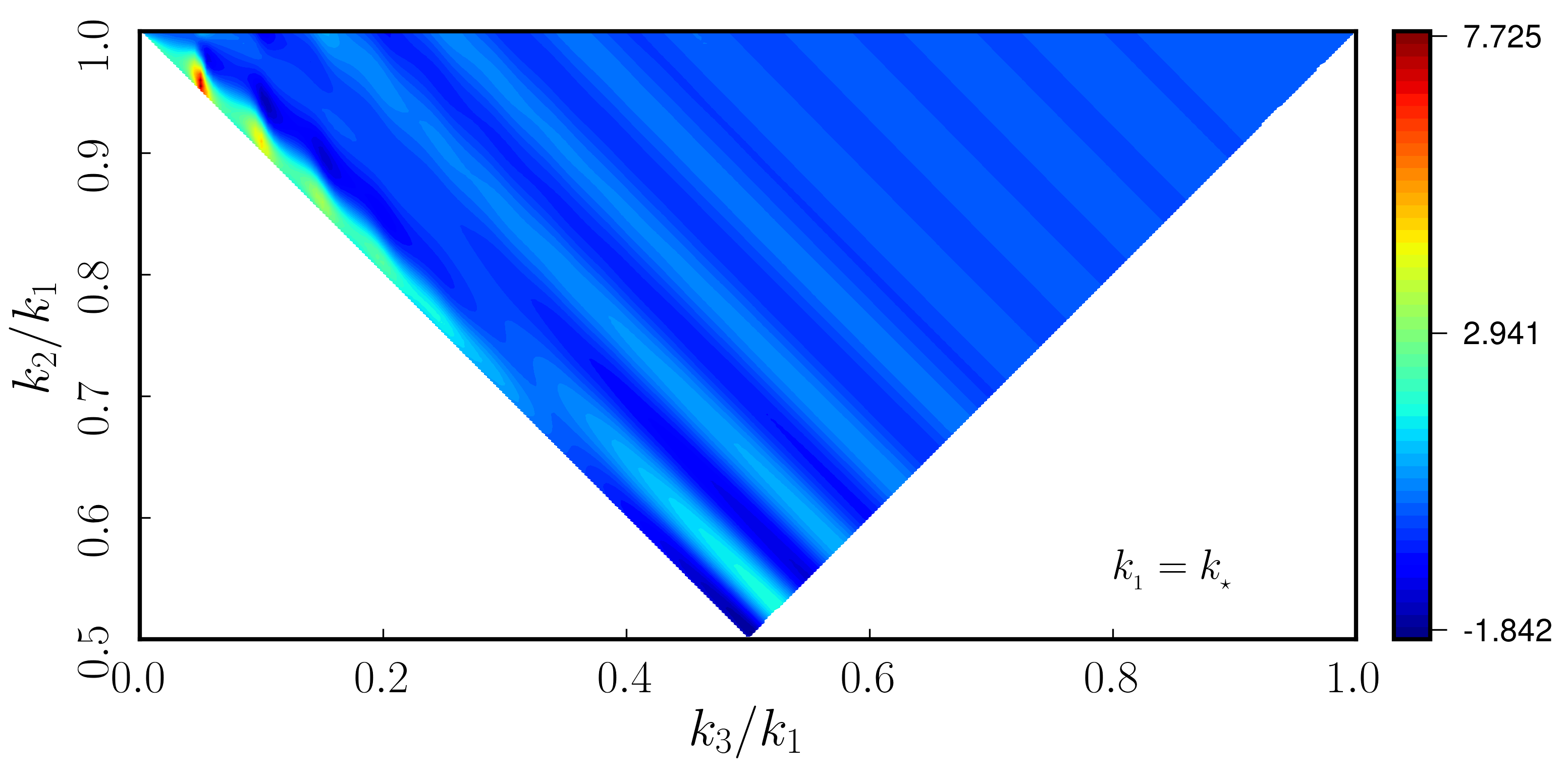}
 \ig[width=0.7\textwidth]{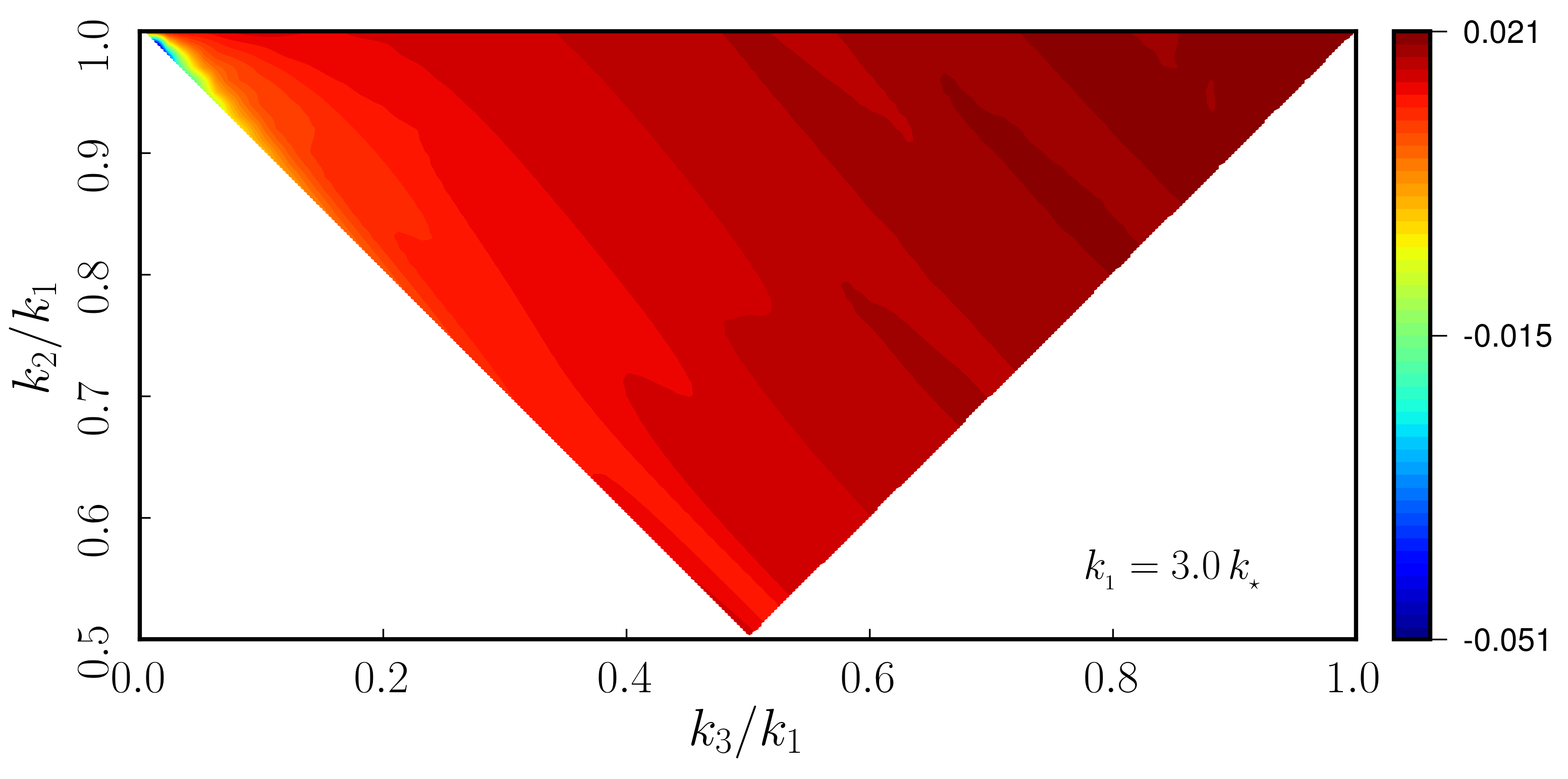}
\caption{
Plots of $f_{_{\rm NL}}(k_1,k_2,k_3)$ versus $x_2\equiv  k_2/k_1$ and $x_3\equiv  k_3/k_1$, for $k_1=0.5\,k_\star$ (top panel), $k_1= k_\star$ (middle panel) and $k_1= 3 k_\star$ (bottom panel). The figure shows configurations allowed by the triangle condition $\v k_1+\v k_2+ \v k_3=0$. Choosing, without loss of generality, $k_1\geq k_2\geq k_3$,   the triangle condition is equivalent to $1\geq x_2\geq 1/2$, $1-x_2\geq x_3\geq x_2$. By comparing the values of $f_{_{\rm NL}}$ among the three plots, we see again its scale dependent character.
These three plots also show  the  oscillatory behavior of  $f_{_{\rm NL}}$, although  this feature is more clearly displayed  in figures  \ref{fnl}-\ref{fnl22}. Furthermore, the plots reveal that  the amplitude of $f_{_{\rm NL}}$ is quite uniform when $k_2$ and $k_3$ are varied while $k_1$ is kept fixed, except for  a small change that makes $f_{_{\rm NL}}$  maximum in the upper left region of the triangle, corresponding to ``squeezed-flattened"  (although not too squeezed) configurations.}
\label{trigfNL}
\efig

These results can be summarized as follows:

\begin{enumerate}

\item $f_{_{\rm NL}}(k_1,k_2,k_3)$ is highly oscillatory. This is a consequence of the oscillatory behavior of the mode functions around the bounce. 

\item As expected, in the regime $k\gtrsim k_{\rm LQC}$, $f_{_{\rm NL}}(k_1,k_2,k_3)$ reduces to standard inflationary prediction ($f_{_{\rm NL}}\sim 10^{-2})$. This is similar to the large-$k$ behavior of the power spectrum (see figure \ref{figPS}). The fact that we recover  the inflationary result for large wave-numbers is a good consistency test of our numerical computations.

\item The amplitude of $f_{_{\rm NL}}(k_1,k_2,k_3)$ is strongly {\em scale dependent}. A scale invariant $f_{_{\rm NL}}$ would not change under simultaneous re-scaling of $k_1$, $k_2$, and $k_3$. The bounce breaks the scale invariance, and makes the amplitude of $f_{_{\rm NL}}(k_1,k_2,k_3)$ to grow for wave-numbers comparable or smaller than $k_{\rm LQC}$. This is a key feature that may allow to contrast this framework with observational data.

\item  By comparing figures \ref{figPS} and \ref{fnl},  we can see that, while the power spectrum deviates from scale invariance for $k\leq k_{\rm LQC}$, $f_{_{\rm NL}}$ does it for $ k\leq 10\, k_{\rm LQC}$. %
This is consistent with the fact that non-Gaussianity  generally provides a {\em better probe of new physics} than the power spectrum \cite{2009astro2010S.158K}.

\item  Consider, without loss of generality, that $k_1\geq k_2\geq k_3$. Now, on the one hand, figure\ \ref{trigfNL} tells that, {\em for fixed} $k_1$, the amplitude of  $f_{_{\rm NL}}$, although quite uniform when we change $k_2$ and $k_3$, attains its maximum value in the upper left region of the triangle. These are configurations for which $k_3 \ll k_2 \approx k_1$,  and  $k_3+k_2 \approx   k_1$, i.e., squeezed-flattened configurations. But note that  $f_{_{\rm NL}}$ becomes small again when $k_3\to 0$ (upper-left corner), that corresponds to very squeezed configurations. Hence, $f_{_{\rm NL}}$ is maximum in the squeezed-flattened, but not too squeezed configurations. A shape of this type was anticipated in more general terms in \cite{Agullo:2010ws,Agullo:2011aa}, and the physical model discussed in this paper provides a concrete example of a single field model in which non-Gaussianity is enhanced in squeezed  configurations.

\end{enumerate}

\begin{figure}
\includegraphics[width=0.7\textwidth]{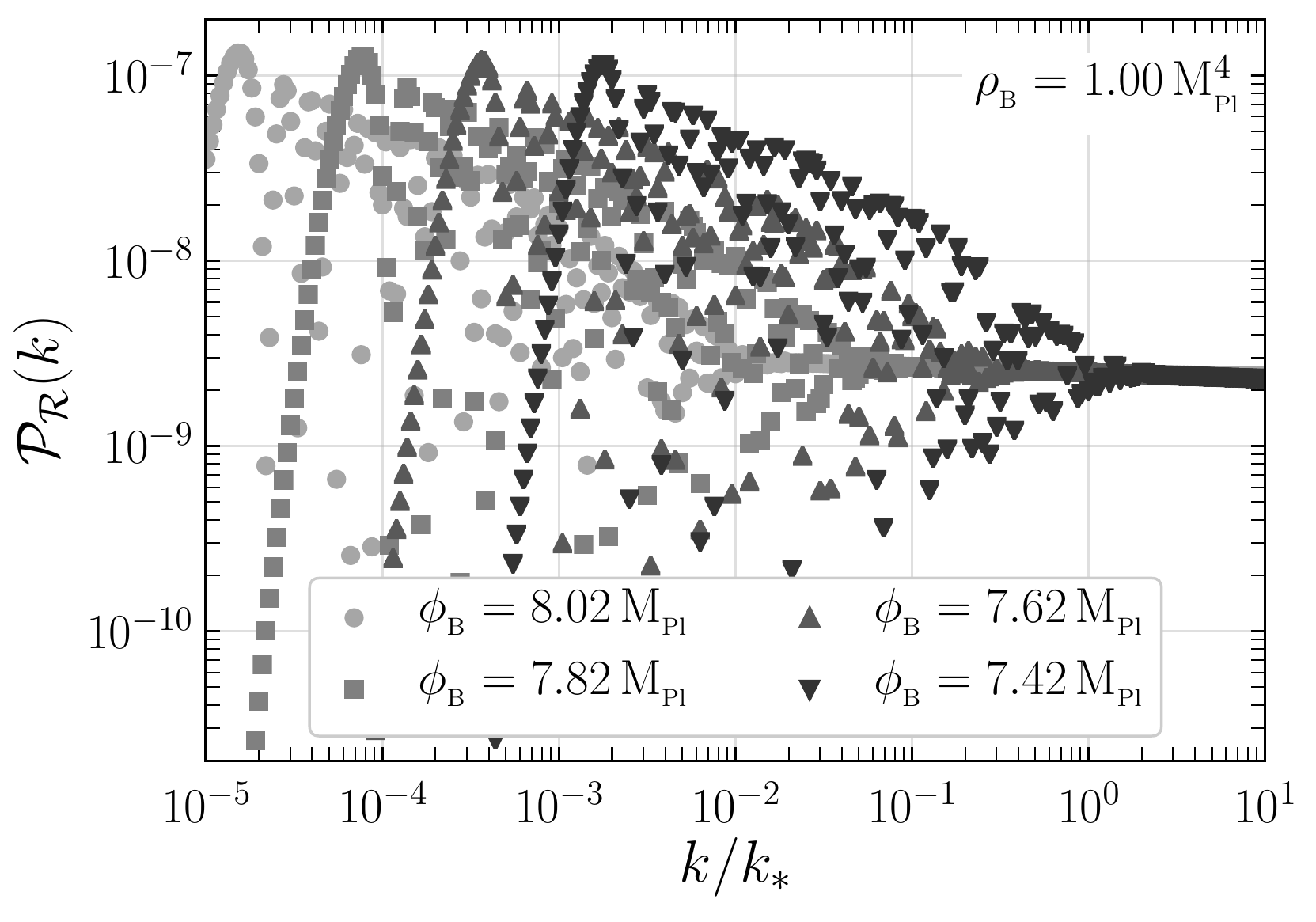}
\includegraphics[width=0.7\textwidth]{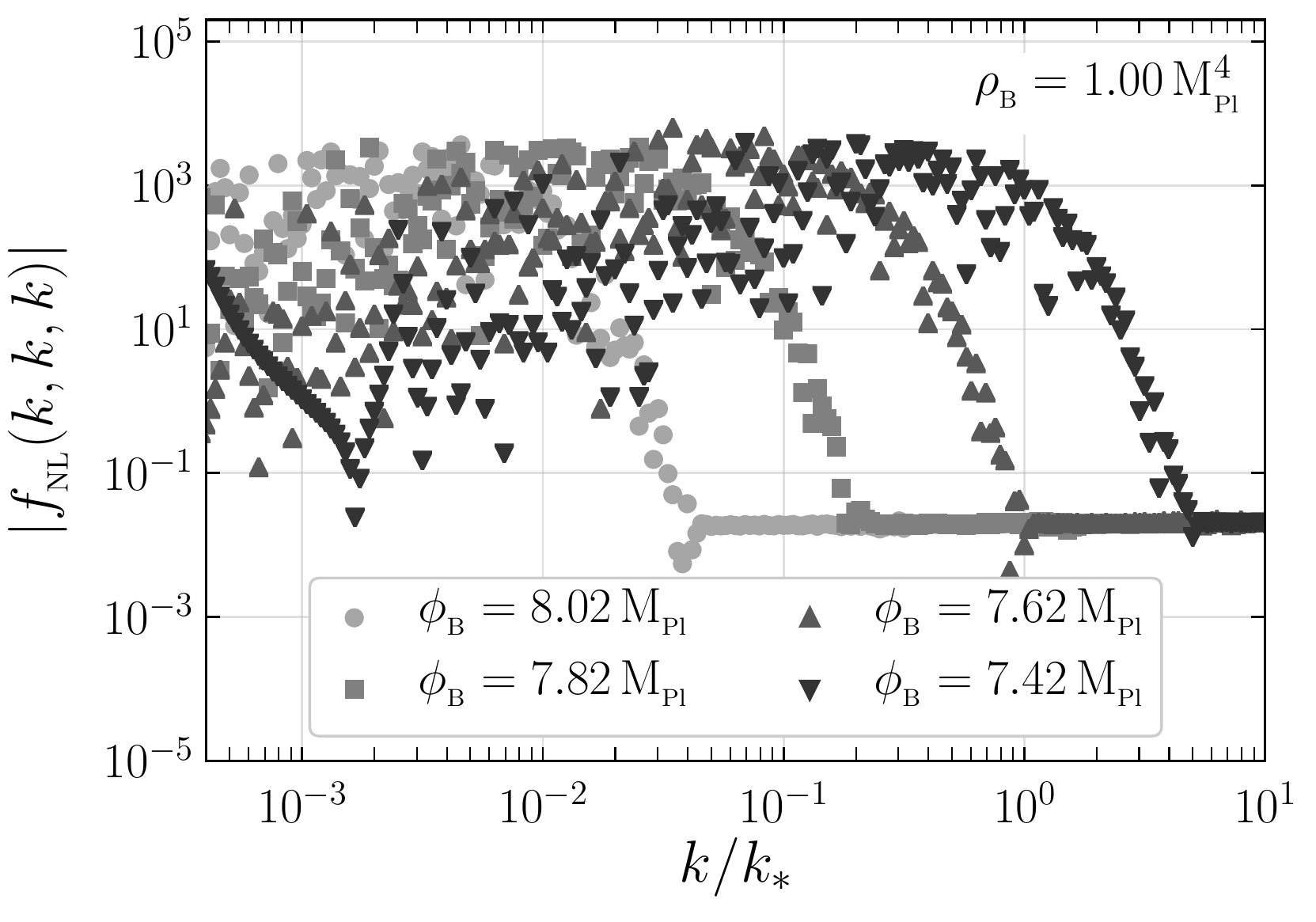}
\hfill
\caption{Power spectrum (upper panel), and $|f_{_{\rm NL}}|$ in the equilateral configuration (bottom panel) for $\rho_{\rm B}=1\, M_{P\ell}^4$, for different values of $\phi_{\rm B}$. The horizontal axis shows wave-number relative to the reference scale $k_*$ that today corresponds to $0.002 \, {\rm Mpc}^{-1}$. Hence the window of observable modes is approximately   $k\in [k_*/10,1000k_*]$. The plot shows that different values of $\phi_{\rm B}$ give rise to power spectra and $f_{_{\rm NL}}$ with exactly the same shape, with the only difference that they are shifted from each other.}\label{fig:ph}
\end{figure}
\subsection{Dependence of $f_{_{\rm NL}}$ on the value of the scalar field at the bounce}\label{sec:phiB}

The value of $\phi_{\rm B}$ determines the number of $e$-folds of expansion between the bounce and the onset of the observable phase of inflation, dubbed $\Nbstar$ \cite{Agullo:2013ai,Agullo:2015tca,Agullo:2016hap,Bolliet:2017czc}.\footnote{\label{fn12} By ``onset" of inflation we refer in this paper to the time $\eta=\eta_*$ at which the reference scale $k_*$ that today has a physical value $k_*/a_{\rm today} =0.002\, M_{Pc}^{-1}$, exits the Hubble radius during inflation. Since inflation lasts  approximately 61 additional  e-folds after $\eta_*$, the number of e-folds from the bounce to the end of inflation is $\Nbstar+61$.} We are interested in effective trajectories for which $\Nbstar \approx 12$. For this value the effects created by the bounce on the power spectrum and non-Gaussianity would appear only in the smallest wave-numbers---or equivalently, the lowest multipoles $\ell$--- that we can observe in the CMB. For  larger values of $\Nbstar$, scales affected by the bounce are red-shifted outside our observable universe, and these effects become unobservable. On the contrary, if $\Nbstar$ is smaller than $12$, the effects of the bounce would appear on all scales in the CMB, and our predictions would be a strongly scale dependent power spectrum with large non-Gaussianity, in clear tension with observations.  $\Nbstar \approx 12$ corresponds to 
$\phi_{B}\approx 7.6 M_{P\ell}$. This small value of the field makes  the kinetic energy to dominate over the  potential energy at the bounce.\footnote{This is the reason why in this paper, as well as in previous analyses \cite{Agullo:2013ai,Agullo:2015tca,Agullo:2016hap}, one focuses on kinetic dominated bounces. In the subsequent evolution, the ratio of the potential energy to the total energy of $\phi$ grows and, at time $\eta=\eta_{\star}$ when slow roll inflation begins, the potential energy dominates over kinetic. } 

What effect should we expect on the observable quantities if we change $\phi_{\rm B}$? Since a change in $\phi_{\rm B}$ modifies the amount of expansion $\Nbstar$, 
we expect that changing $\phi_{\rm B}$ will {\em shift $\mathcal{P}_{\mathcal{R}}(k)$ and $f_{_{\rm NL}}$ with respect to the set of wave-numbers that we can directly observe.} However, the shape of $\mathcal{P}_{\mathcal{R}}(k)$ and $f_{_{\rm NL}}$ is not expected to change, since the  bounce itself is not modified by changing $\phi_{\rm B}$.\footnote{The bounce is  dominated by quantum gravity effects, rather than by matter, and therefore a small change on $\phi_{\rm B}$ does not modify the spacetime geometry around the time of the bounce  in any significant amount.}

Figure \ref{fig:ph} shows the power spectrum and $f_{_{\rm NL}}$ in the equilateral  configuration for different values of $\phi_{\rm B}$, and for $\rho_{\rm B}=1\, M_{P\ell}^4$. The results are qualitatively the same for other configurations.  As expected, the only effect of changing $\phi_{\rm B}$ is a shift of $\mathcal{P}_{\mathcal{R}}(k)$ and $f_{_{\rm NL}}$ relative to $k_*$. We  see, for instance,  that for $\phi_{\rm B}=8.02\, M_{P\ell}$ both the power spectrum  and $f_{_{\rm NL}}$ are indistinguishable from the standard results of slow-roll inflation for observable modes $k\in [k_*/10,1000k_*]$. All the effects from the bounce are red-shifted to super-Hubble scales for this value of $\phi_{\rm B}$. On the contrary, for $\phi_{\rm B}=7.42\, M_{P\ell}$ the bounce affects both the power spectrum and non-Gaussianity, although only for infra-red scales in the CMB. 

In summary, the scalar field at the bounce $\phi_{\rm B}$ determines the amount of cosmic expansion accumulated after the bounce, and  changing it produces a shift of the power spectrum and non-Gaussianity with respect to the scales that are directly observable in the CMB, without  modifying their shape.

\begin{figure}
\includegraphics[width=0.7\textwidth]{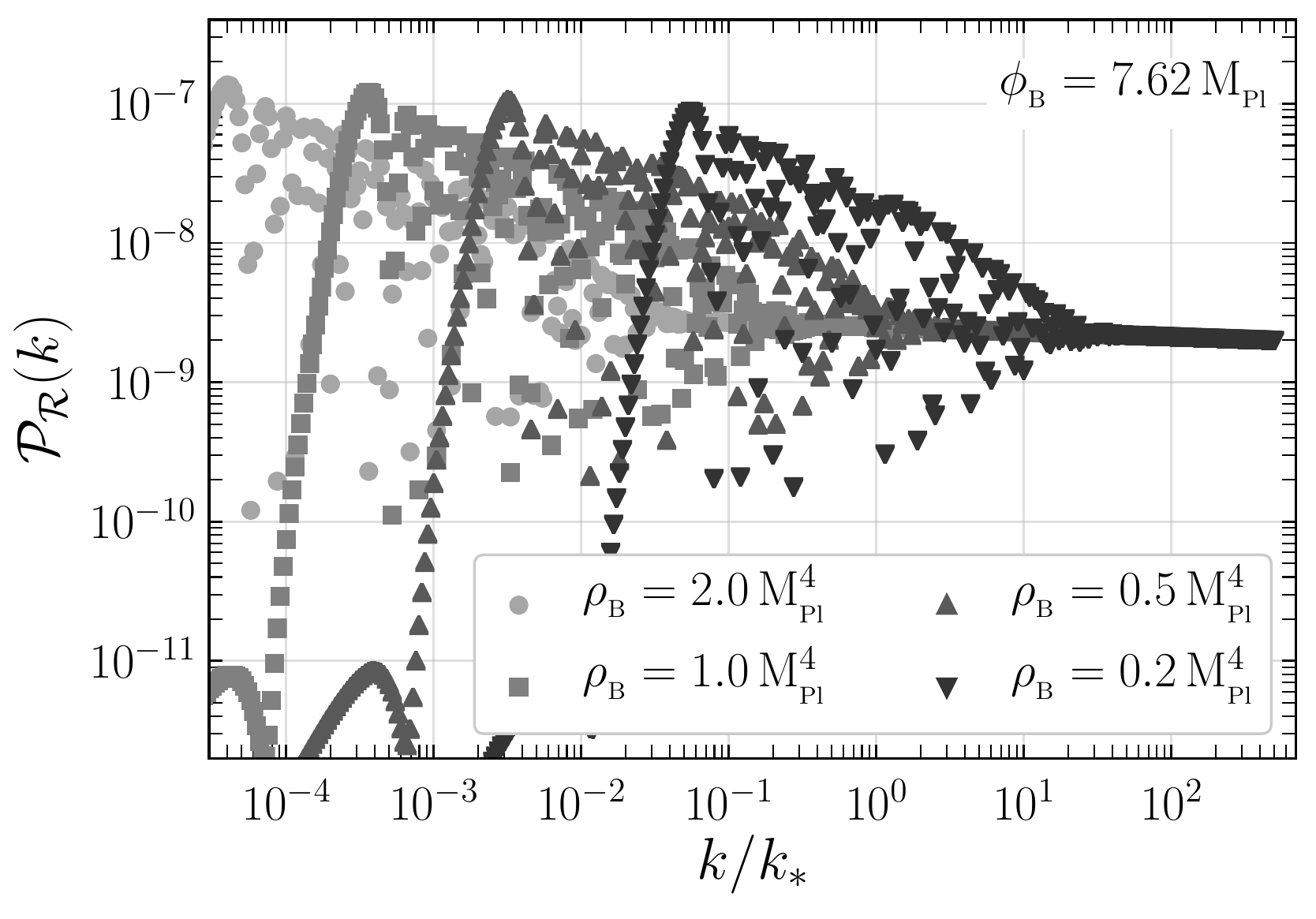}
\includegraphics[width=0.7\textwidth]{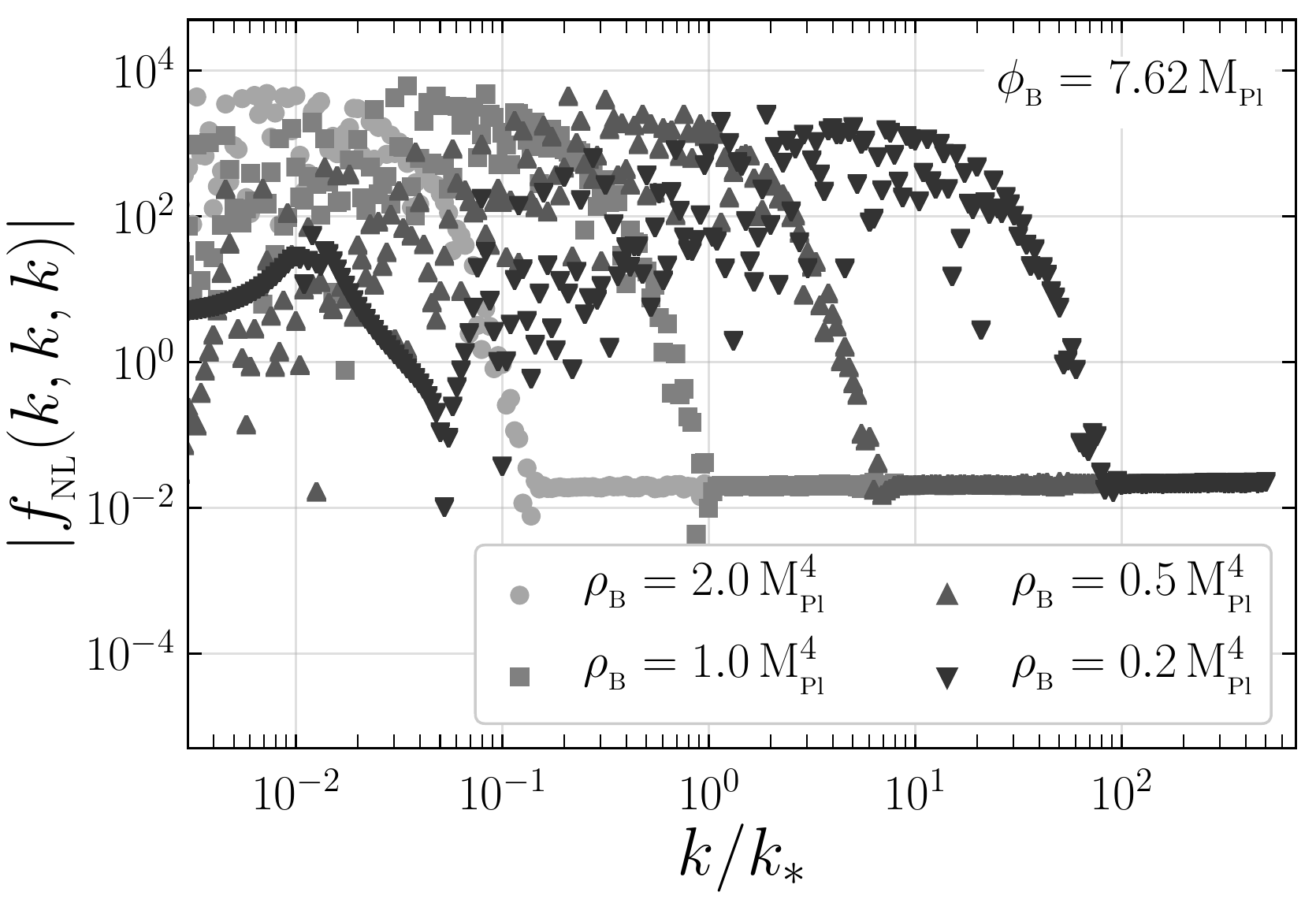}
\caption{Power spectrum (upper panel), and $|f_{_{\rm NL}}|$ in the equilateral configuration (bottom panel) for $\phi_{\rm B}=7.62\, M_{P\ell}$, for different values of $\rho_{\rm B}$. The plot shows that different values of $\rho_{\rm B}$ change the maximum value of $f_{_{\rm NL}}$. We also see that both, the power spectrum and $f_{_{\rm NL}}$ are shifted towards more infrared scales relative to $k_*$ for large values of $\rho_{\rm B}$.\\
}\label{fig:rhoB}
\end{figure}

\subsection{Dependence of  $f_{_{\rm NL}}$ on the   energy density at the bounce}\label{sec:rhoB}

Changing the energy density at the bounce also changes the amount of expansion from the bounce to the onset of inflation. This is because larger the value of $\rho_{\rm B}$, larger would be the expansion needed for the energy density to decrease and reach the inflationary value.  Therefore, we  expect  $f_{_{\rm NL}}$, as well as the power spectrum, to shift its position in relation to observables scales, in a way similar to the effect of changing $\phi_{\rm B}$. 

There two different factors that could change the energy density at the bounce: (i) a change in the value of the are gap $\Delta_0$, (iii) a change in the quantum state $\Psi(v,\phi)$ that describes the background quantum geometry. The analysis of this section is, therefore, well-motivated. 

Figure \ref{fig:rhoB} shows the power spectrum and $f_{_{\rm NL}}$ in the equilateral configuration (the result is  similar for other configurations) for different values of $\rho_{\rm B}$, with $\phi_{\rm B}=7.62\, M_{P\ell}$. As expected, both  $\mathcal{P}_{\mathcal{R}}$ and $f_{_{\rm NL}}$ are redshifted towards  infra-red scales for larger values of $\rho_{\rm B}$. We  also observe  that  $\mathcal{P}_{\mathcal{R}}$ and $f_{_{\rm NL}}$ are {\em more enhanced} for large values of $\rho_{\rm B}$.  For the power spectrum, the change in the amplitude produced by changing $\rho_{\rm B}$ is very small, and therefore the dominant effect is simply a shift relative to $k_*$. Therefore, regarding $\mathcal{P}_{\mathcal{R}}(k)$, changing $\rho_{\rm B}$ and $\phi_{\rm B}$ produces the  same results. This fact was analyzed  in \cite{Agullo:2016hap}, and it was pointed out that, if one restricts to observable scales and takes into account observational error bars, the effect produced by a change in $\rho_{\rm B}$ in the power spectrum $\mathcal{P}_{\mathcal{R}}(k)$ can be compensated by a change in $\phi_{\rm B}$. Hence, observations of the power spectrum alone can only provide information about a combination of $\phi_{\rm B}$ and $\rho_{\rm B}$, and not about their individual values.  We find that this does not happen for $f_{_{\rm NL}}$. Hence {\em the degeneracy between the observable effects of $\phi_{\rm B}$ and $\rho_{\rm B}$ disappears} for non-Gaussianity. Consequently, an observation of the power spectrum and non-Gaussianity generated by the bounce would provide  information about the energy (or curvature) scale of the bounce. 

The results of this section can be interpreted in more general terms. Recall that, as discussed in \cite{Ashtekar:2015iza} and \cite{Agullo:2016hap} and summarized in section \ref{sec:EE},  a change in the quantum state $\Psi_0(v,\phi)$ that describes  the background geometry has effects on observable quantities that, with great accuracy, can be mimicked by  a change in $\rho_{\rm B}$. Therefore, the content of this section can  be also understood as an investigation of the sensitivity of observable quantities to the  choice of $\Psi_0(v,\phi)$.

\subsection{Influence of the scalar field's potential}\label{sec:Starobinsky}
In this section, we investigate the sensitivity of the results for non-Gaussianity in LQC under a change of the scalar field's potential. In LQC the bounce is generated by quantum gravity effects, and the contribution of $V(\phi)$ is subdominant. Therefore, we expect that the results for $f_{_{\rm NL}}(k_1,k_2,k_3)$ obtained in the previous sections by using the quadratic potential will remain largely unaltered for other choices of $V(\phi)$. We compute $f_{_{\rm NL}}(k_1,k_2,k_3)$ in this section for the so-called Starobinsky potential \cite{Barrow:1988xi,Barrow:1988xh,PhysRevD.39.3159,Starobinsky:2001xq},
\begin{equation} \label{SV}
 V(\phi)\, =\, \f{3\,M^2}{4\,\kappa}\,\l(\, 1\, -\, e^{-\sqrt{\f{2\,\kappa}{3}}\phi}\r)^2 \, .
\end{equation}
The power spectrum in LQC has been analyzed in detail in \cite{Bonga:2015kaa, Bonga:2015xna}, and the results are qualitatively similar to the quadratic potential.

\bfig
\ig[width=0.7\textwidth]{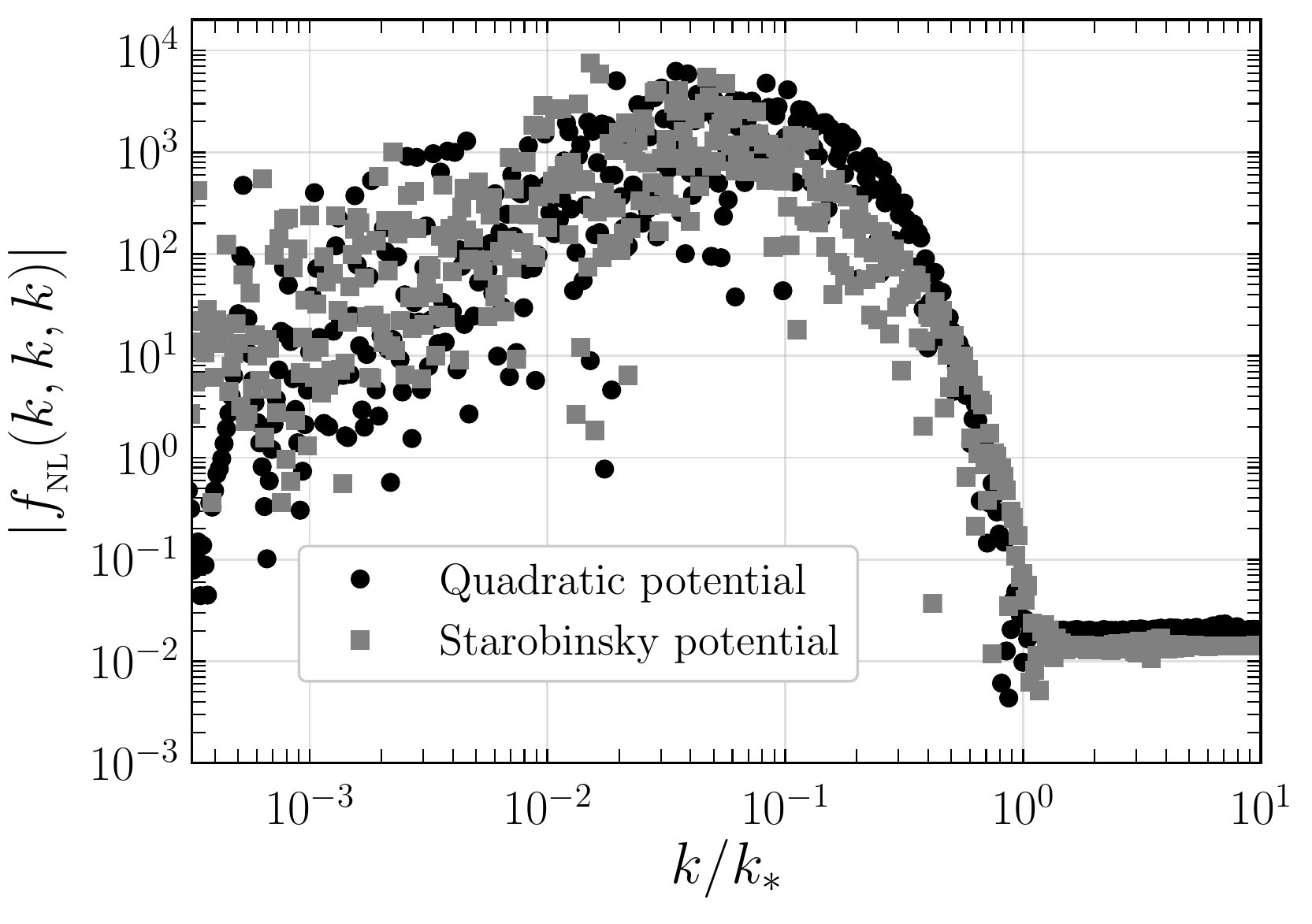}
\ig[width=0.7\textwidth]{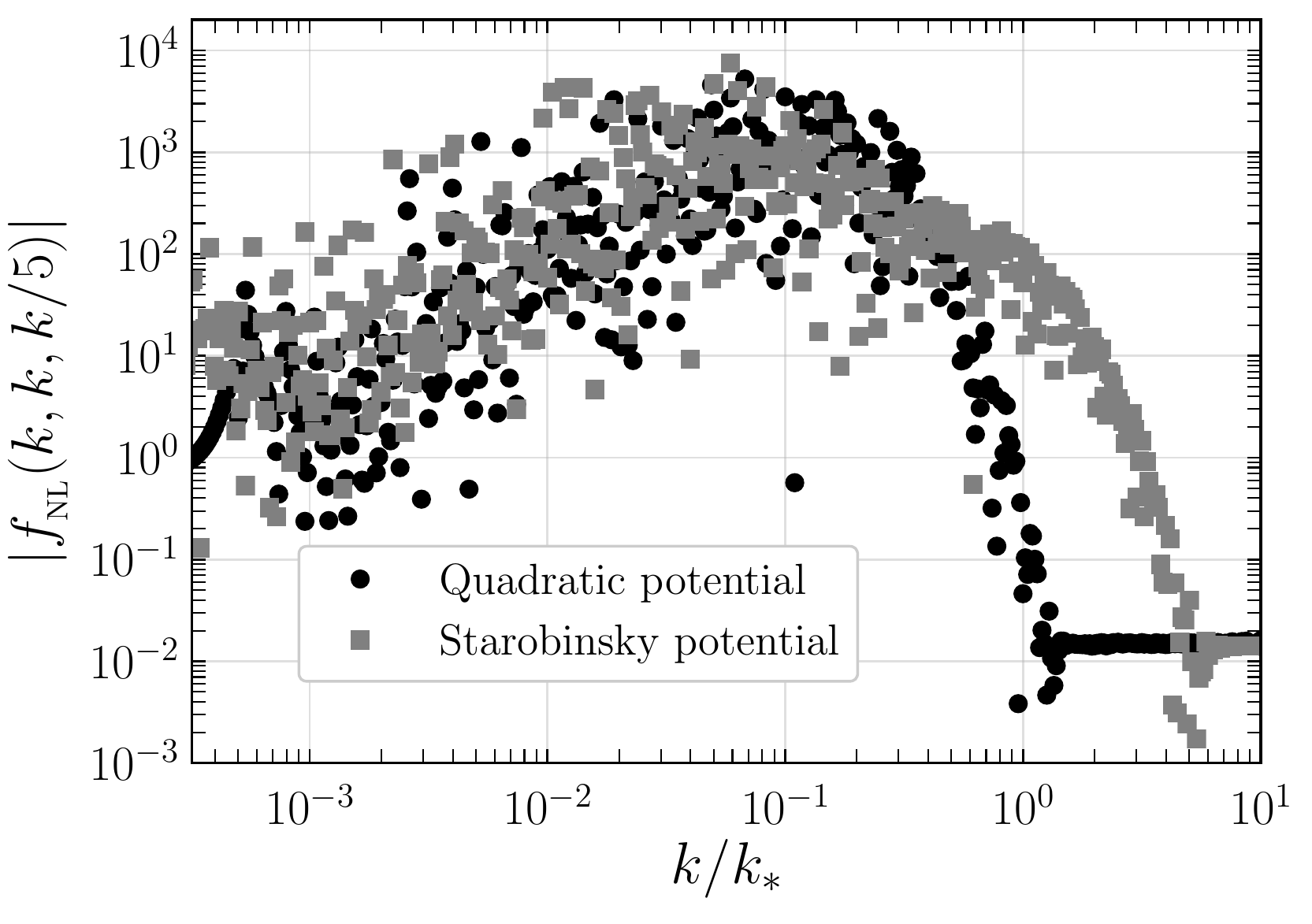}
\caption{Comparison of $|f_{_{\rm NL}}(k,k,k)|$ (upper panel) and $|f_{_{\rm NL}}(k,k,k/5)|$ (bottom) evaluated at the end of inflation for the quadratic  and the Starobinsky potential. 
The figure illustrates that the spectral shape is very similar regardless of the potential. The differences, more evident in the bottom panel, arise from contributions generated far from the bounce.}\label{fig:fnl-starobinsky}
\efig

We compute $f_{_{\rm NL}}(k_1,k_2,k_3)$ by using (\ref{SV}) for the value of $M$ obtained from  the Planck normalization, $M = 2.51\times 10^{-6} M_{P\ell}$. Figure \ref{fig:fnl-starobinsky} shows the results for two different configurations, and for $\phi_{\rm B} = -4.88 \, M_{P\ell}$ and $\rho_{\rm B} = 1\, M_{P\ell}^4$. The initial state of perturbations has been chosen to be the Minkowski-like vacuum at $\eta_0=-281.5 \, T_{P\ell}$ (equivalently, $t_0 = -2.32 \times 10^5\, T_{P\ell}$). At this time all modes of interest  are in the adiabatic regime.  Our analysis indicate that the conclusion reached in all previous section remain true, as expected, since most of these features are due to the bounce. 

At the quantitative level, the results also agree, although  some  small difference appear both for large and small wave-numbers. The value of $|f_{_{\rm NL}}(k_1,k_2,k_3)|$  for large $k_i$  is proportional to the slow-roll parameter $\epsilon$ evaluated at horizon exit during inflation. This parameter is smaller for the Starobinsky potential (grey squares) than for the quadratic potential  (black circles), and  explains the small difference in amplitude  in figure \ref{fig:fnl-starobinsky}. The differences in the bottom panel of figure \ref{fig:fnl-starobinsky} are larger, and they originate from the discrepancies in the background dynamics at early and late times, far from the bounce. These differences can  be reduced by adjusting appropriately  the value of the free parameters $\phi_{\rm B}$ and $\rho_{\rm B}$. %

\bfig
 \ig[width=.7\textwidth]{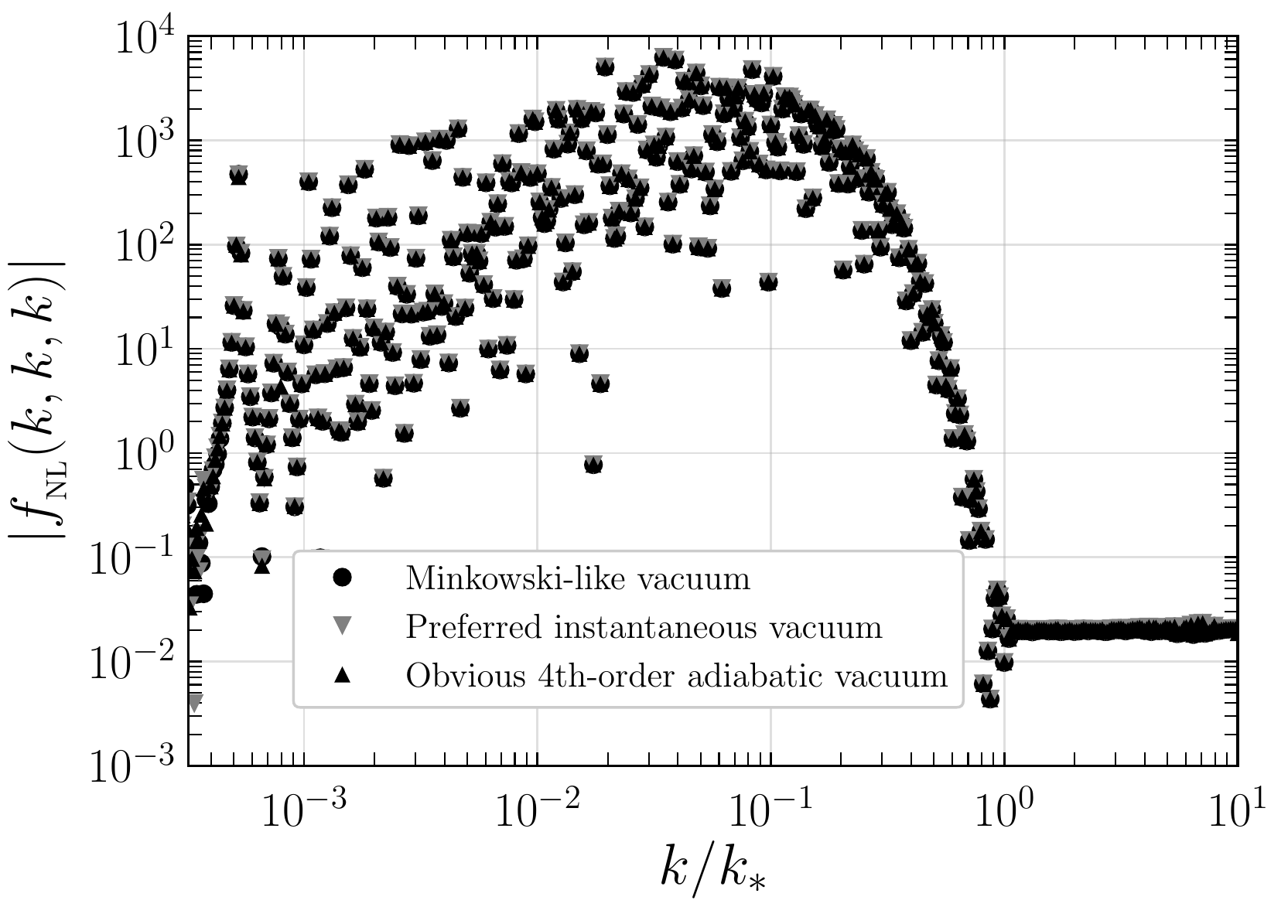}
\caption{Plot of $|f_{_{\rm NL}}(k_1,k_2,k_3)|$ in the equilateral configuration ($k_1=k_2=k_3$) for different choices of initial quantum state for perturbations. The plot shows that the three choices considered in this paper produce results that are all very similar. Differences only appear for the most infrared part of the spectrum, that corresponds to unobservable scales. %
}
\label{fig:states}
\efig

\subsection{Dependence of  $f_{_{\rm NL}}$ on the initial states for  perturbations}\label{sec:inistate}

We explore in this section the sensitivity of non-Gaussianity to different choices of initial state for the quantum scalar perturbations. This question is relevant because in spacetimes with no time-like isometries, such as the spatially flat FLRW spacetime considered in this paper, the notion of quantum vacuum for a test field is ambiguous: there are infinitely many candidates for Fock vacua, and none are preferred with respect to the other  \cite{1994qftc.book.....W} (see \cite{Agullo:2015qqa} for further discussions).  In FLRW, one can narrow the freedom by restricting to homogenous and isotropic states  that are adiabatic of, at least, fourth order---so that the energy-momentum tensor is well-defined for these states  \cite{Parker:1974qw}---but the mathematical freedom is still  large.  Consequently,  one could in principle obtain very different results by appropriately tuning the initial state. 

Notice that this freedom is not specific to LQC. It is common to any cosmological model dealing with quantum perturbations, including the inflationary framework. A way to make progress is to add physical principles  to select appropriate initial data for perturbations. For instance, if evolution begins at a time at which all wavelengths of interest for observations are  small compared to the curvature scale,  then the adiabatic analysis \cite{Parker:1974qw} provides guidance. This is the strategy that one follows in standard inflation and  we adopt it here as well. We use three different proposals for initial state, all based on reasonable  criteria, and compute $f_{_{\rm NL}}$ in each case. A similar exploration using these three different initial states, has been done for the power spectrum in LQC in   \cite{Agullo:2013ai,Agullo:2015tca}. The outcome of these analyses was that  the power spectrum is very similar for observable scales in all three cases considered. Here, we  reach the same conclusions for non-Gaussianity. Therefore, we argue that the results of this paper do not rely on a fine-tuned choice of initial conditions for perturbations, and are therefore generic, within the mathematical limitations mentioned  above.

More precisely, the three types of initial state that we choose are:
\begin{itemize}

\item
Minkowski-like initial state. This state was introduced at the beginning of section \ref{PS}. This state is not a forth-order adiabatic state (it is only of adiabatic order zero).

\item  Obvious adiabatic vacuum. This state was introduced in  \cite{Agullo:2012fc}. It is the state obtained by using initial data for the mode functions given by the  first four terms of the adiabatic expansion of $\varphi_k(\eta)$. This state is therefore of fourth adiabatic order. This prescription, however, cannot be specified for very infrared modes, since it produces modes with the incorrect normalization. Nevertheless, the  ambiguity will only modify  the most infra-red part of our results that correspond to modes that are not directly observable, and therefore we use this state for the purpose of this section. 
\item  Preferred instantaneous vacuum. This state was introduced in \cite{Agullo:2015qqa}, and it is defined as the only state that has zero expectation value of the adiabatically renormalized energy-momentum tensor at the initial time $\eta_0$. In this sense, this is a  generalization of the Minkowski vacuum to cosmological spacetimes. It is also a state of fourth adiabatic order.

\end{itemize}

Figure \ref{fig:states} shows the function $f_{_{\rm NL}}$ for equilateral configurations computed using these three different initial states, specified at  $\eta_0= 2.842 \times 10^3\, T_{P\ell}$. As anticipated, the results are essentially the same.\\

We have also explored the sensitivity of $f_{_{\rm NL}}$ to the  time at which the initial conditions are imposed.  We found that as long as  $\eta_0$ is chosen such that all modes of interest are ultra-violet compare to the curvature-scale, $k^2\gg a''/a$, the results for  $f_{_{\rm NL}}(k_1,k_2,k_3)$ are insensitive to the choice of $\eta_0$.

Another physically motivated instant to specify initial data is the bounce. At that time, however, the condition $k^2\gg a''/a$ is not satisfied for all modes of interest, and therefore the adiabatic condition is not sufficient to choose an initial state. We found that $f_{_{\rm NL}}$ {\em is very sensitive} to the ambiguity in the choice of initial data for perturbations at the bounce. Different choices produce results that differ significantly from each other, and therefore we were unable to make  any generic statement about the value of $f_{_{\rm NL}}$ when the evolution begins at the bounce. Unless one adds new principles that enables us to select preferred initial data for perturbations at the bounce (see \cite{Ashtekar:2016wpi,Ashtekar:2016pqn,deBlas:2016puz} for interesting examples within LQC) it seems difficult to reach  any conclusion. In absence of such principles, the far past well before the bounce appears as the most natural place to specify the initial state of perturbations.

\bfig
 \ig[width=0.7\textwidth]{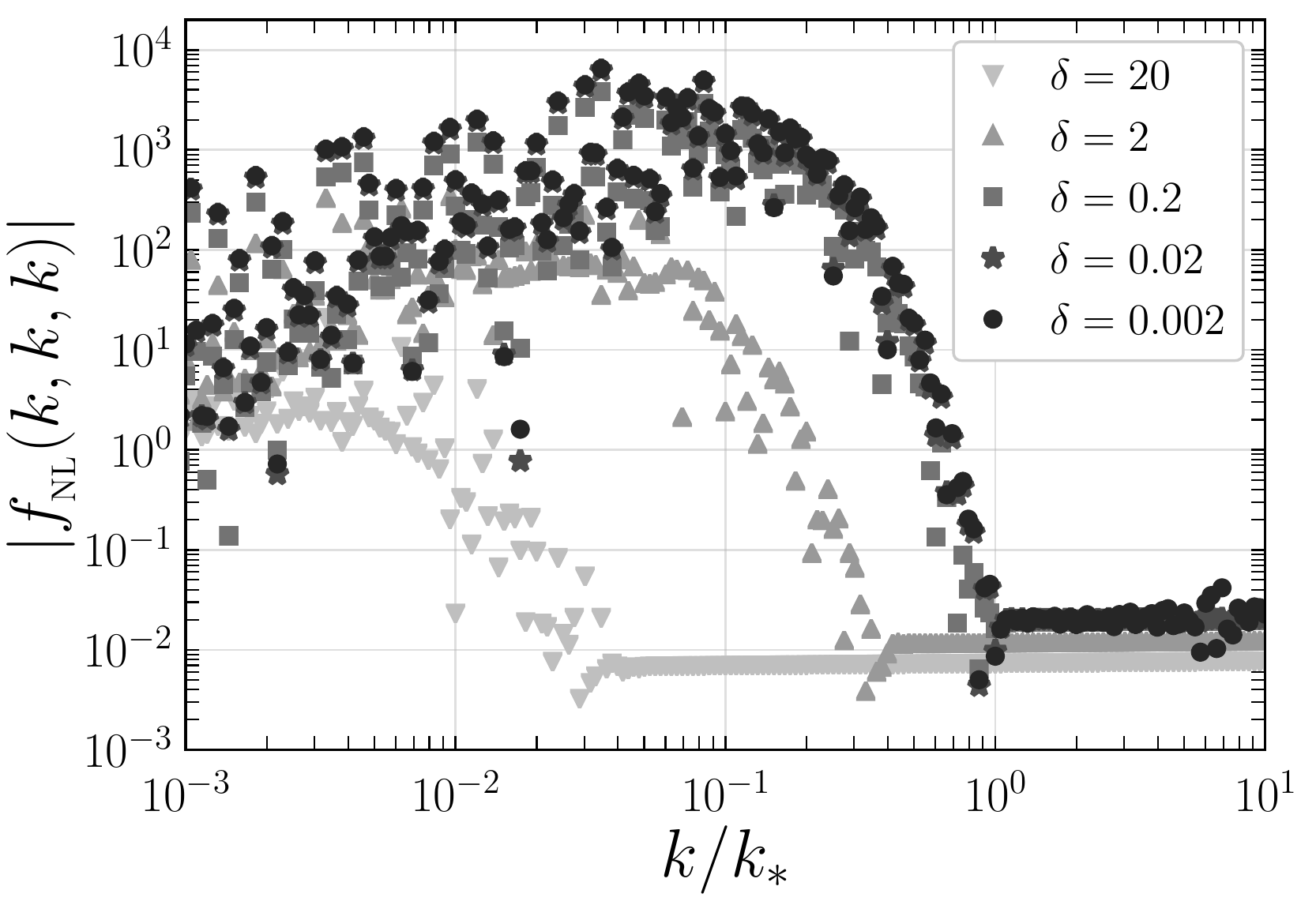}
\caption{$|f_{_{\rm NL}}(k_1,k_2,k_3)|$ at the end of inflation computed by adding a damping factor $\exp{[-\delta \, (k_1+k_2+k_3)/(a \sqrt{\kappa \rho/3})}]$ to the numerical integrals. The plot shows the equilateral configuration, $k_1=k_2=k_3$. For large values of $\delta$ ($\delta=20,2,0.2$) the computation underestimate the real value of $f_{_{\rm NL}}$. For smaller values of $\delta$, the actual value of $f_{_{\rm NL}}$ is no longer suppressed, but then numerical artifacts appear for large $k$ if $\delta$ is chosen too small, as can be seen in the plot for $\delta=0.002$. These instabilities originate in the highly oscillatory nature of these modes.  This analysis indicates that the optimal value of $\delta$  is around $0.02$.}\label{cutoff}
\efig
\subsection{Tests of the numerics}\label{sec:tests}

In this subsection we provide further tests of the numerical computations, with the goal of increasing our confidence on the results and rule out  potential numerical  artifacts.

The main challenge of the numerical evaluation of the bispectrum is that it involves integrals of highly oscillatory functions. These  integrands  include products of three mode functions $\varphi_k(\eta)$ (see equation\ \eqref{Bphi}). As discussed in section \ref{PS}, these functions transition from being slowly evolving when $k \lesssim \sqrt{|f(\eta)|} = \sqrt{|a^2(\u-\f{R}{6})|}$, to highly oscillatory functions when $k \gg \sqrt{|f(\eta)|}$. It is the latter case that produces numerical instabilities. 

However, because the main  contribution to the integrals comes from times when at least one of the modes satisfies $k \lesssim \sqrt{|f(\eta)}|$, a convenient strategy to reduce numerical instabilities, and also to reduce the computation time without affecting the result, is to remove from the integration time intervals for which  {\em all the three modes} are highly oscillatory. This can be easily done by including a damping factor to the integrand in equation (\ref{Bphi}) of the form $\exp{[-\delta \, (k_1+k_2+k_3)/\sqrt{|f(\eta)|}}]$, with $\delta < 1 $, similar to the strategy followed in other approaches \cite{Maldacena:2002vr,Hazra:2012yn,Sreenath:2013xra}. However, because the function $f(\eta)$ has a complicated behavior close to the bounce, at the practical level it is more convenient to work with a smoother damping factor of the form $\exp{[-\delta\, (k_1+k_2+k_3)/(a \sqrt{\kappa \rho/3})]}$. Figure  \ref{cutoff} shows the result for $f_{_{\rm NL}}(k_1,k_2,k_3)$ evaluated at the end of inflation for different values of the cut-off $\delta$. As expected, for large values of $\delta$  the integral is artificially suppressed, and the result underestimates the value of $f_{_{\rm NL}}$. On the contrary, when $\delta$ is very small, numerical instabilities appear for large  wave-numbers. Our analysis shows that there is an optimal value, around $\delta=0.02$ for which the numerical calculation is fast and reliable. This is the value that we have  used to produce the figures in section \ref{sec:bispectrum}. %

\bfig
 \ig[width=0.7\textwidth]{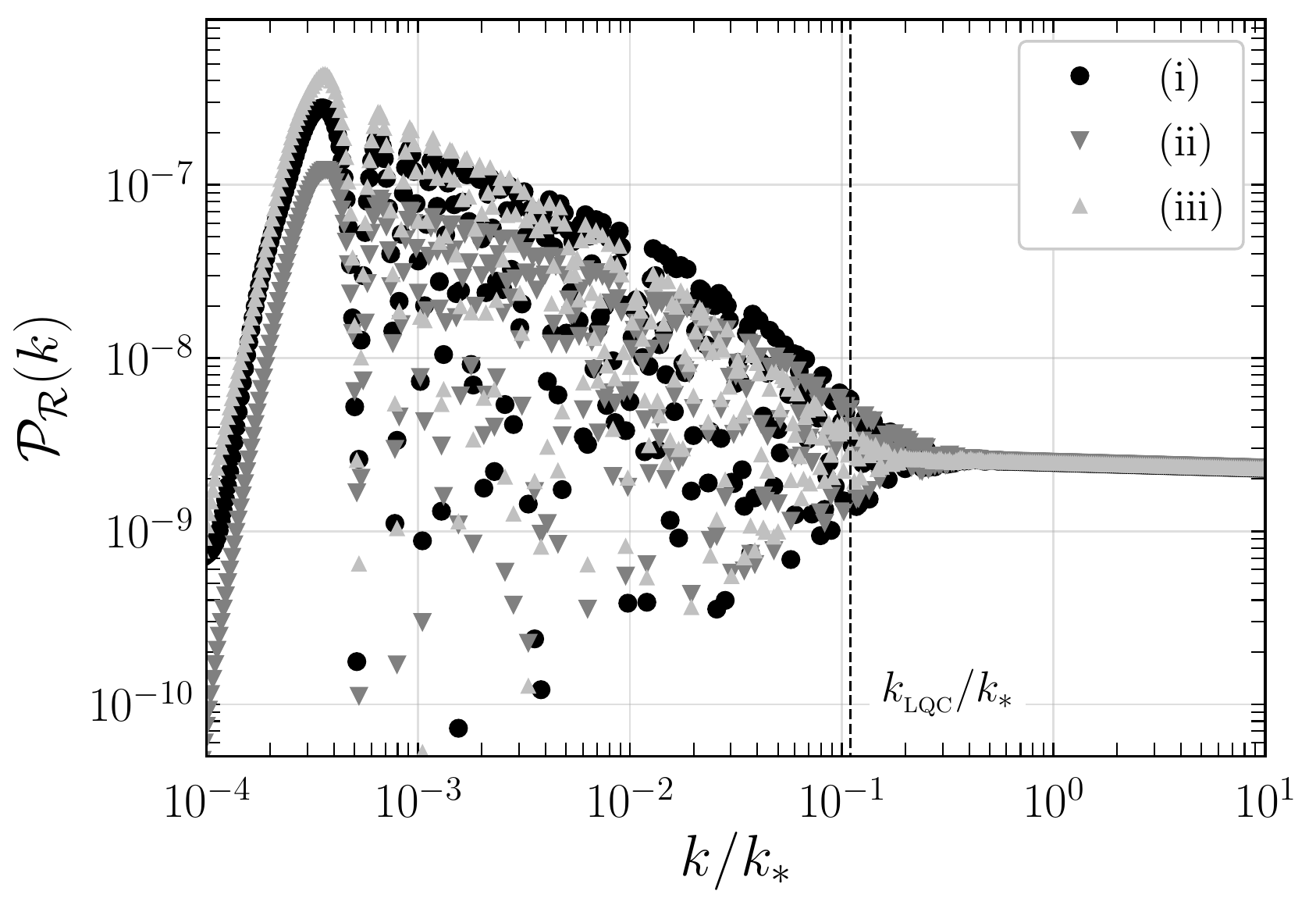}
\caption{{ The scalar power spectrum ${\mathcal P}_s(k)$ evaluated at the end of inflation for the three different strategies  for evaluating $\pi_a$ described at the end of \ref{sec:freeev}. The power spectrum is very similar in the three cases, and important differences  appear only for the very infra-red part of the spectrum, that corresponds to wave-lengths that are several orders of magnitude larger than today's Hubble radius.}}\label{fig:Ps-strategies}
\efig
\bfig
 \ig[width=0.7\textwidth]{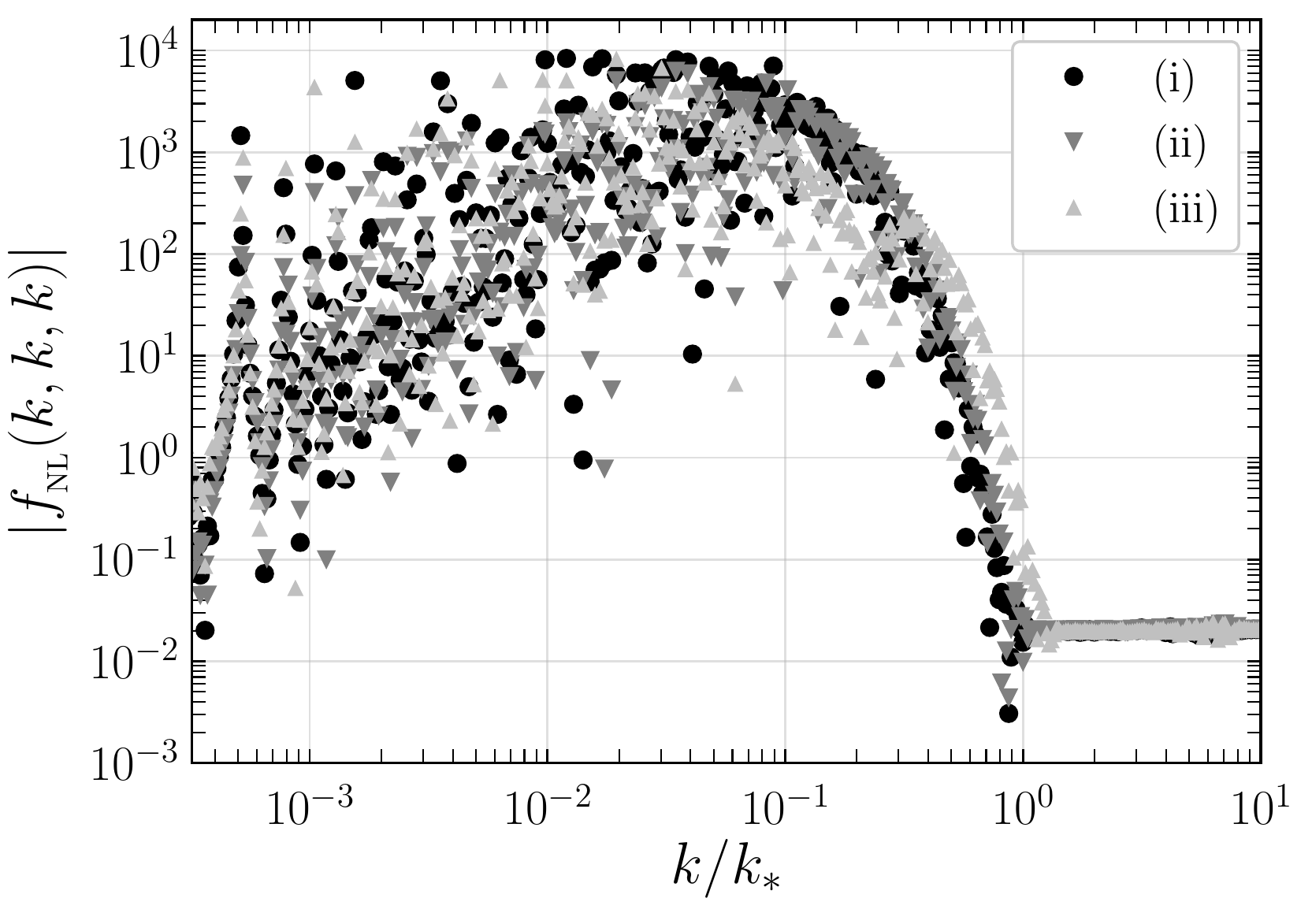}
\caption{{$|f_{_{\rm NL}}(k, k, k)|$ evaluated at the end of inflation for the three different strategies for evaluating $\pi_a$ described at the end of \ref{sec:freeev}. The results are very similar in all three cases, and the differences between them are small compared  to current observational error bars.}}\label{fig:fnl-strategies}
\efig  
The second  test that  we perform in this section concerns the ambiguity regarding the value of $\pi_a$ in LQC, discussed at the end of  section \ref{sec:freeev}. There, we proposed  three different strategies for evaluating $\pi_a$ and the various powers of it that appear in the classical Hamiltonian for perturbations. We will now show that the results obtained  for the power spectrum and non-Gaussianity are very similar in all three cases. In order to do this, we compare the power spectrum in figure \ref{fig:Ps-strategies}, and $|f_{_{\rm NL}}|$ in figure \ref{fig:fnl-strategies}, obtained by using the three proposed strategies. Although some small differences appear, they are either smaller than observational error bars, or they appear for very infrared modes that cannot be observed in our Hubble patch of the universe. Note also that the freedom that we have in changing the free parameters of the theory, and that we explore in  previous sections, make these differences even less relevant, since, as we saw, a small change in the value of some of these parameters would compensate the effects in the power spectrum and non-Gaussianity.

\section{Analytical understanding of the evolution of non-Gaussianity across the bounce}\label{sec:analytical}

A characteristic feature of the non-Gaussianity produced by the LQC bounce is an enhancement of $f_{_{\rm NL}}$ for wave-numbers comparable to the scale $k_{\rm LQC}$ set by the bounce. The goal of this section is to complement the previous numerical analysis with an analytical understating of the origin of this feature. By doing so we will, on the one hand, increase our confidence on the numerical results and, on the other, understand better  the physical origin of such behavior. 

We will use standard techniques from asymptotic analysis of integrals to find approximate expressions for the way the amplitude of $f_{_{\rm NL}}$ behaves. Although our arguments are quite simple, the result  captures the  physics of the problem remarkably well, both qualitatively and  quantitatively.

First of all, we want to isolate the contribution to $f_{_{\rm NL}}$ that comes exclusively from the bounce. For this, we go back to the definition of $f_{_{\rm NL}}$ in section \ref{sec:bisp}, and find that this contribution is given by
\bea \label{1int} I(k_1,k_2,k_3)&=&\int_{-\Delta \eta}^{\Delta \eta} \d\eta\, \Big( f_1(\eta)\, \varphi^*_{{k}_1}( \eta)\varphi^*_{{k}_2}(\eta)\varphi^*_{{k}_3}( \eta) 
+f_2( \eta)\, \varphi^*_{{k}_1}( \eta)\varphi^*_{{k}_2}( \eta){\varphi'}_{{k}_3}^*(\eta) \nonumber \\
&+& f_3( \eta)\, \varphi_{{k}_1}(\eta){\varphi'}_{{k}_2}^*( \eta){\varphi'}_{{k}_3}^*( \eta)
+(\v k_1 \leftrightarrow \v k_3)+ (\v k_2 \leftrightarrow \v k_3) \Big) \, ,
 \eea
where $f_1( \eta)$, $f_2(\eta)$ and $f_3(\eta)$ are  background functions, given in Appendix B. We use $\Delta \eta=1000\, T_{P\ell}$ (recall the bounce happens at $\eta=0$). %
 For $k\gtrsim k_{\rm LQC}$ the mode function can be approximated by $\varphi_{{k}}\sim e^{-i k\eta}$. With this we have

\be \label{eqnr} I(k_1,k_2,k_3)\sim\int_{-\Delta \eta}^{\Delta \eta} \d\eta \, g(\eta)\, e^{i (k_1+k_2+k_3)\, \eta}\approx  \int_{-\infty}^{\infty} \d\eta \, g(\eta)\, e^{i k_t\, \eta} \, W(\eta,\Delta), \ee
where $k_t \equiv  k_1+k_2+k_3$; $g(\eta)$ is a combination of the functions $f_i$'s in (\ref{1int}); and $W(\eta,\Delta \eta)$ is a  window function that is equal  to zero for $|\eta|>\Delta \eta$, equal to one for $|\eta|<\Delta \eta$, and smoothly interpolates between both values. The function  $W(\eta,\Delta \eta)$ allows us to extend the integration limits to $-\infty$ and $+\infty$, without modifying the value of the integral, and its concrete form  will be unimportant for our purposes. 

Now, Cauchy's integral theorem tells us that the right hand side of \eqref{eqnr} is equal to  $2\pi i$ times  the sum of the residues of the poles of $g(\eta)$ with {\em positive imaginary part}. The real part of each pole contributes to the oscillatory behavior of the integral as a function of $k_t$, while the imaginary part adds an exponentially decreasing factor.  Hence,  the asymptotic behavior of the amplitude of the integral $I$ as a function of $k_t$ is given by {\em the pole of $g(\eta)$ with the largest imaginary part}.

To find this pole, it is sufficient to realize that, out of the four background functions $a(\eta)$, $\pi_a(\eta)$, $\phi(\eta)$, and $p_{\phi}(\eta)$ that appear in $g(\eta)$, the scale factor is the only one having a {\em minimum} at the bounce. From this, we know that the pole we are looking for comes from factors $\frac{1}{a^n(\eta)}$ contained in $g(\eta)$. To compute this pole, we use an analytical approximation for the scale factor, valid close to the bounce, that in cosmic time reads (see, e.g., \cite{Bolliet:2015bka})
\be a(t)= a_B\, \left(1+3\, \kappa\rho_{\rm B} \, t^2\right)^{1/6},\ee
where we have chosen the bounce to take place at $t=0$. The pole of $a(t)^{-1}$ is at {$t_p=i\, /\sqrt{3\, \kappa\, \rho_{\rm B}}$} and, in conformal time, at\footnote{The relation between $t$ and $\eta$ close to the bounce can be written in terms of a hypergeometric function as $\eta=\int_0^t  a(t')^{-1}\, dt'=t\, a_B^{-1} \, \, _2F_1[\f{1}{6},\f{1}{2},\f{3}{2},-3 \, \kappa\, \rho_{\rm B}\, t^2]$.}
\be \eta_p=i\,\sqrt{\pi/3}\, \frac{\Gamma[5/6]}{2\Gamma[4/3]}\, \f{1}{a_B\, \sqrt{\kappa\, \rho_{\rm B}}}  =i\,  \, \f{\alpha}{k_{\rm LQC}}\, ,\ee
where $\Gamma[x]$ is the Gamma function, $\alpha\simeq0.64677$, and we have used   $k_{\rm LQC}=a_B\, \sqrt{\kappa \rho_{\rm B}}$. Therefore, this  argument tells us that  the  bounce produces a contribution to $f_{_{\rm NL}}(k_1,k_2,k_3)$ whose amplitude  changes with  $k_i$ according to $e^{- \alpha  (k_1+k_2+k_3)/k_{\rm LQC}}$, when $(k_1+k_2+k_3)\gtrsim k_{\rm LQC}$. In figure \ref{analytical} we compare this analytical approximation  with the numerical result, for three different  configurations finding a good agreement. 

To summarize, the analysis of this section confirms that  the scale dependent enhancement of $f_{_{\rm NL}}$  originates from the bounce, and it is the scale $k_{\rm LQC}$  that dictates how pronounced this enhancement is. Furthermore, since it is only the complex pole of the scale factor at the bounce that accounts for the main features of $f_{_{\rm NL}}$, it is expected that bounces in other cosmological models different from LQC will produce similar non-Gasussianity.

\bfig
 \ig[width=0.7\textwidth]{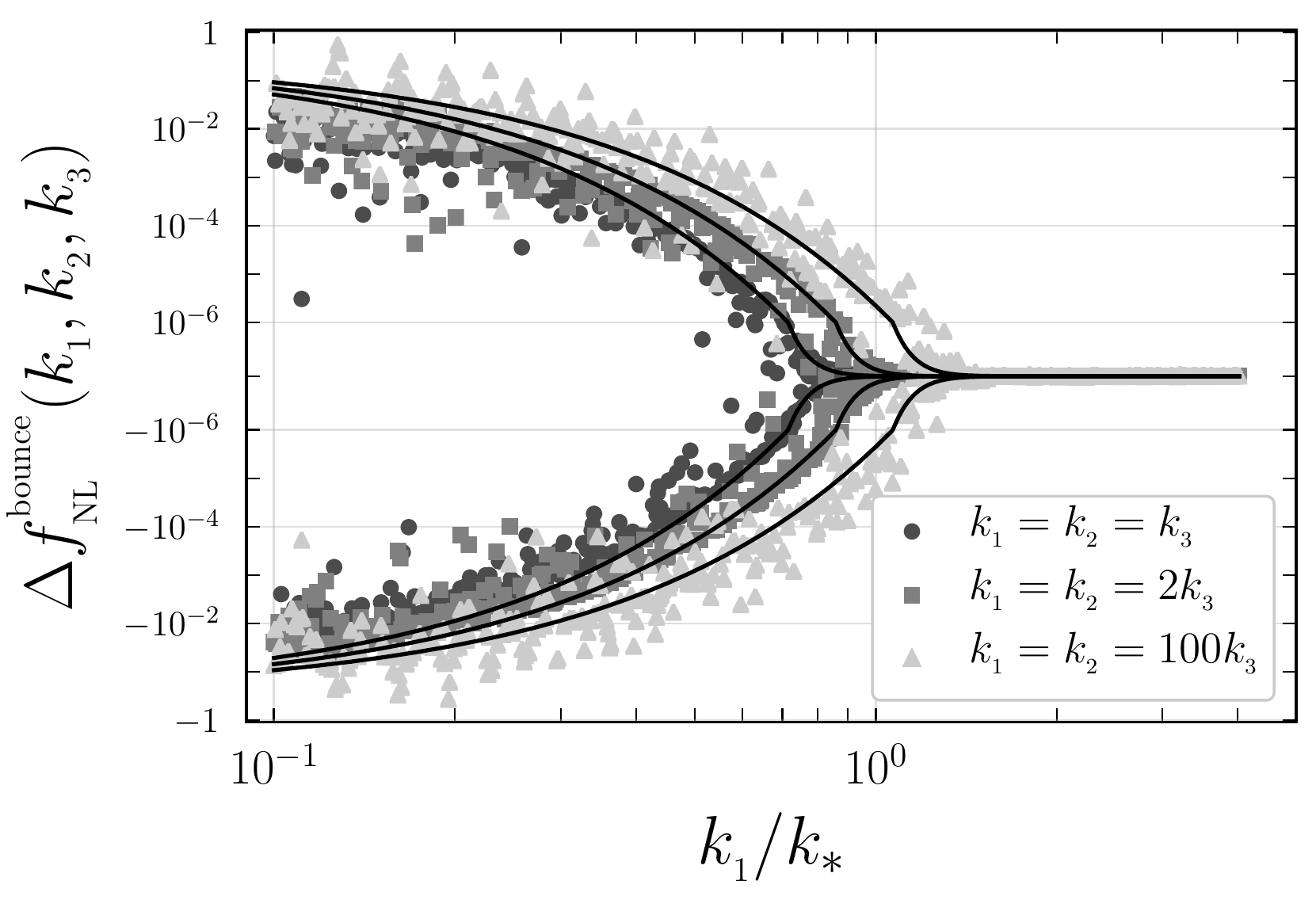}
\caption{{Comparison of the  numerically-computed contribution from the bounce to $f_{_{\rm NL}}$  (gray points),  called $\Delta f^{\rm bounce}_{NL}$ in the figure, and the analytical approximation  $e^{- \alpha  (k_1+k_2+k_3)/k_{\rm LQC}}$ (black line). The comparison is made for three different configurations of wave-numbers. The agreement is very good for all of them.   $\Delta f^{\rm bounce}_{NL}$ is defined as the value of $f_{_{\rm NL}}$ given only by the first term in equation (\ref{BR}), and evaluating the integral in  (\ref{Bphi}) just before the onset of inflation.}}
\label{analytical}
\efig

\section{Stability of  perturbation theory}\label{sec:stpertb}

We found that a cosmic bounce taking place close to the Planck scale produces large values of  $f_{_{\rm NL}}$, of order $10^3$. This result is in agreement with the extended intuition that, near the Planck regime,  self-interactions of scalar perturbations  with purely gravitational origin---i.e.,\ described by  terms in the third order interaction Hamiltonian (\ref{eq:H3}) that are independent of the potential $V(\phi)$---become strong.   This large value of $f_{_{\rm NL}}$ raises concerns about the validity of the perturbative expansion, on which the entire analysis rests. 

To evaluate  the validity of the perturbative series, we need to compute the corrections that  $f_{_{\rm NL}}$ (the three-point functions) introduces in the power spectrum (the two-point function). If this correction is similar or larger than  the leading order contribution, then the perturbative expansion would break down. As we show in this section, this is not the case.

The two-point function of comoving curvature perturbations at the end of inflation at next-to-leading order, is obtained from the correlation function of $\delta\phi$ by keeping the first correction arising from (\ref{Rtodph}). We get
\bea \label{2pfno} \langle 0|\h{\mathcal{R}}_{\vec k_1} \h{\mathcal{R}}_{\vec k_2}|0\rangle &=&
\l(-\f{a}{z}\r)^2 \langle 0|\h{\dph}_{\vec k_1} \h{\dph}_{\vec k_2}|0\rangle \nonumber \\ 
&+&2 \l(-\f{a}{z}\r)^3\, \left[-\f{3}{2}+3\f{V_{\phi}\, a^5}{\kappa\, \pp\, \pi_a}+\f{\kappa}{4}\f{z^2}{a^2}\right] \int\f{\d^3p}{(2\pi)^3} \, \langle 0|\h{\dph}_{\vec k_1} \h{\dph}_{\vec p}\, \h{\dph}_{\v k_2- \vec p}|0\rangle \nonumber \\ 
&+& \l(-\f{z}{a}\r)^4\,\left[-\f{3}{2}+3\f{V_{\phi}\, a^5}{\kappa\, \pp\, \pi_a}+\f{\kappa}{4}\f{z^2}{a^2}\right]^2 \int\f{\d^3p}{(2\pi)^3}\f{\d^3q}{(2\pi)^3} \, \langle 0|\h{\dph}_{\vec p}\, \h{\dph}_{\vec k_1-\v p}\,  \h{\dph}_{\vec q}\, \h{\dph}_{\v k_2- \vec q}|0\rangle \nonumber \\&+& \cdots
\eea
The power spectrum computed in previous sections was obtained by considering only the first line of this equation and, additionally, by ignoring  corrections from the interaction Hamiltonian when computing it. Now, we go to the next order in perturbations.

For the two-point function in the first line of (\ref{2pfno}), we have
\bea \langle 0|\h{\dph}_{\vec k_1} \h{\dph}_{\vec k_2}|0\rangle = \langle 0|\h{\dph}_{\vec k_1}\h{\dph}_{\vec k_2}|0\rangle-i/\hbar \int_{\eta_0}^{\eta}\d\eta'\, \langle 0|\left[\h{\dph}^I_{\vec k_1} \h{\dph}^I_{\vec k_2} , \h{\mathcal{H}}^{\rm I}_{\rm int}( \eta')\right]|0\rangle +\,  \mathcal{O}(\mathcal{H}^2_{\rm int}) \, . \eea
The first term in the right hand side was the one computed in  equation (\ref{2pleadord}). The second term in the right hand side vanishes, since it involves expectation values of an odd number of  fields in the interaction picture, which are Gaussian. Therefore, there is no correction linear in the third order Hamiltonian to this term.  
Hence, the leading order correction to the two-point function comes from the second and third line of (\ref{2pfno}). 

The  three-point function in the second line contributes with terms linear in the third order Hamiltonian. In contrast, the leading order term in the four-point function is independent of the interaction Hamiltonian. By using (\ref{4p2}) and the definition of the bispectrum of $\delta \phi$ given in (\ref{bispdph}), we obtain  the first perturbative correction to the power spectrum:
\bea 
\langle 0|\h{\mathcal{R}}_{\vec k_1} \h{\mathcal{R}}_{\vec k_2}|0\rangle &=& (2\pi)^3 \delta^{(3)}(\vec k_1+\vec k_2)\, \f{2\pi^2}{k_1^3} \, \hbar \,\left[ \mathcal{P}_{\mathcal{R}}(k_1)+ \, \Delta \mathcal{P}_{\mathcal{R}}(k_1)\right] \, ,\eea
where
\bea \label{DeltaP} \Delta \mathcal{P}_{\mathcal{R}}(k_1)&=&\hbar\, \f{k_1^3}{\pi^2}\, \Bigg[  \l(-\f{a}{z}\r)^3\, \left[-\f{3}{2}+3\f{V_{\phi}\, a^5}{\kappa\, \pp\, \pi_a}+\f{\kappa}{4}\f{z^2}{a^2}\right] \int\f{\d^3p}{(2\pi)^3}  \, B_{\delta\phi}(\vec{k}_1,\v p, -\v k_1-\v p)\, , \nonumber \\ &+&  \l(-\f{a}{z}\r)^4\, \left[-\f{3}{2}+3\f{V_{\phi}\, a^5}{\kappa\, \pp\, \pi_a}+\f{\kappa}{4}\f{z^2}{a^2}\right]^2 \,   \int\f{\d^3p}{(2\pi)^3}\,  |\varphi_p|^2 \, |\varphi_{| \v k_1-\v p|} |^2\Bigg],
\eea
where all quantities are evaluated at the end of inflation. Note from this expression that the next-to-leading order correction to the power spectrum for a wave-number $k_1$, gets contributions from other wave numbers, as a result of the correlations arising from the three-point function.

\bfig
 \ig[width=.7\textwidth]{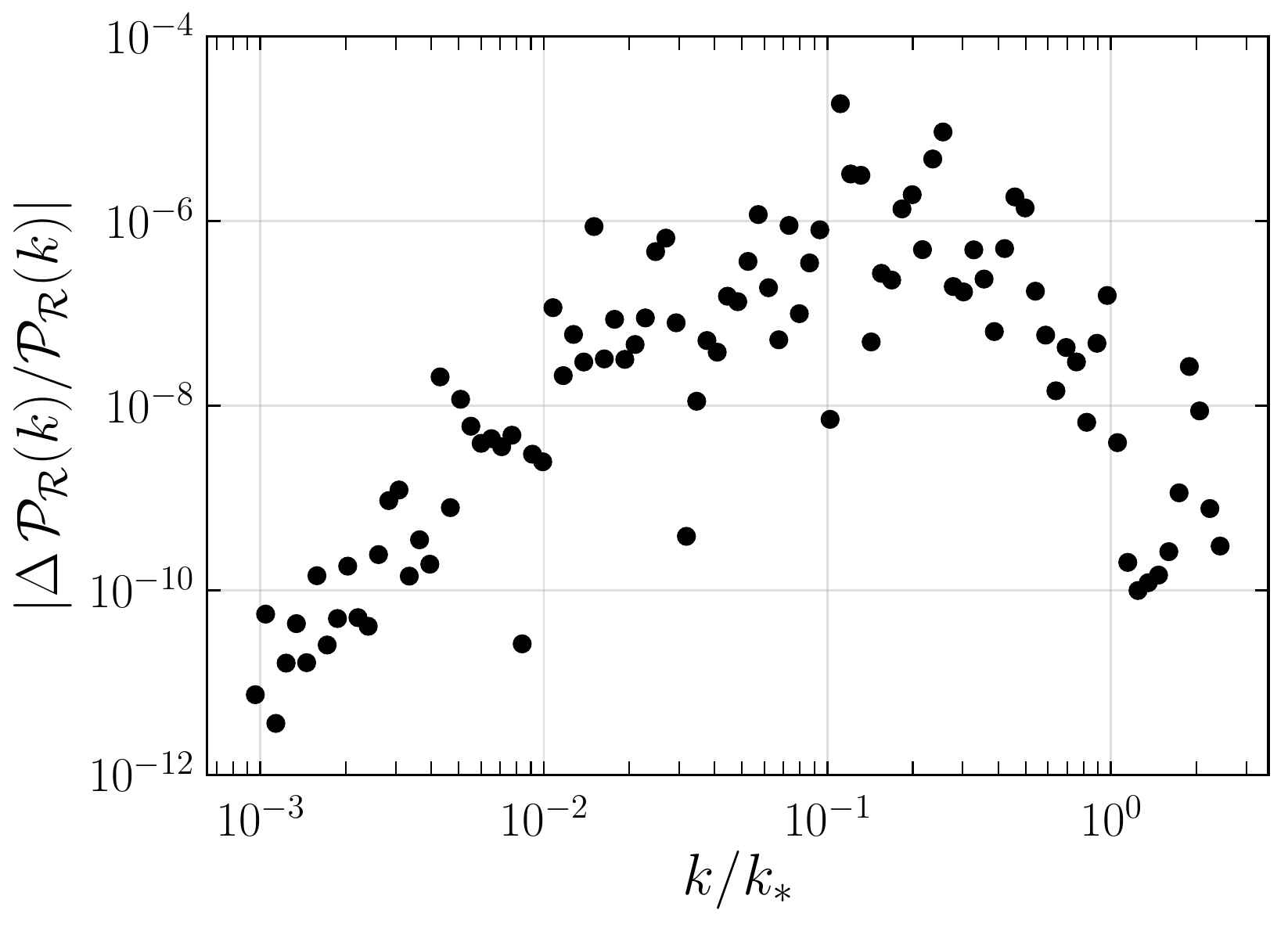}
\caption{Plot of the relative size of the first order correction to the power spectrum, $|\Delta \mathcal{P}_{\mathcal{R}}/\mathcal{P}_{\mathcal{R}}|$. The plot shows the numerically computed value as a function of the  wave-numbers $k$. The result shows that, indeed, $|\Delta \mathcal{P}_{\mathcal{R}}/\mathcal{P}_{\mathcal{R}}|\ll 1$, confirming that we are well inside the perturbative regime.  This plot is obtained by using the same values for the free parameters as in section \ref{sec:bispectrum}, and the conclusions remain unchanged for other choices.} \label{fig:DeltaP}
\efig

An order of magnitude estimate of (\ref{DeltaP}) can be obtained as follows. In the first line, the background function between square brackets is of order $\epsilon$ ($\epsilon$ symbolizes here any of the slow-roll parameters). The bispectrum $B_{\delta\phi}$, is of order $f_{_{\rm NL}}\, \mathcal{P}_{\mathcal{R}}^2$ and, therefore, the term in the  first line of  (\ref{DeltaP}) is of order $\epsilon \, f_{_{\rm NL}}\, \mathcal{P}_{\mathcal{R}}^2$ ($\epsilon$ symbolizes a slow-roll parameter). Similarly, the second line of  (\ref{DeltaP}) is of  order $\epsilon^2 \, \mathcal{P}_{\mathcal{R}}^2$. Since $f_{_{\rm NL}} \lesssim 10^4$, and  $\epsilon\sim10^{-2}$, the first line of  (\ref{DeltaP}) is much larger than the second one. Then, we expect $\Delta \mathcal{P}_{\mathcal{R}}/\mathcal{P}_{\mathcal{R}}\sim \epsilon\, f_{_{\rm NL}}\, \mathcal{P}_{\mathcal{R}}\lesssim 10^{-4}$.

 We have  numerically  evaluated expression (\ref{DeltaP}), and the results appear in figure\ \ref{fig:DeltaP}. The figure shows that $\Delta \mathcal{P}_{\mathcal{R}}/\mathcal{P}_{\mathcal{R}}$ is smaller than $10^{-4}$, confirming that  the next to leading order corrections to the power spectrum are indeed negligible. Therefore, we find that  although $f_{_{\rm NL}}$ experiences an enhancement of several orders of magnitude, the perturbative expansion remains valid. The reason is found in the smallness of the leading order power spectrum $\mathcal{P}_{\mathcal{R}}(k)\lesssim 10^{-7}$. From the expressions above, we see that the leading order correction contains, in addition to $f_{_{\rm NL}}$, an additional power of $\mathcal{P}_{\mathcal{R}}(k)$. The smallness of $\mathcal{P}_{\mathcal{R}}$ compensated for the enhancement of $f_{_{\rm NL}}$. Higher order corrections contain even higher powers of $\mathcal{P}_{\mathcal{R}}(k)$. In this sense, one can intuitively think about $\mathcal{P}_{\mathcal{R}}(k)$ as the small `parameter' in terms of which the perturbative expansion is defined.\\

\section{Discussion and conclusions}\label{sec:sum}

The goal of this section is to provide a summary of the main results of this paper, contrast them with observational data, and discuss the main consequences. The main take-home messages from our analysis are the following: \\

(1) The evolution of scalar perturbations across the LQC bounce, starting from an adiabatic vacuum state before the bounce when all the Fourier modes of interest have  wavelengths much smaller than the (spacetime) curvature radius,  produces a state that at the onset of inflation   is both excited and non-Gaussian, relative to the Bunch-Davies vacuum. In other words, both the two- and three-point correlation functions of scalar perturbations deviate significantly from their Bunch-Davies counterparts at the onset of inflation. Consequently, the predictions for the  primordial power spectrum and non-Gaussianity  are modified as a result of the pre-inflationary evolution. (See section \ref{sec:EXT} and \ref{sec:numres}.)\\

(2) The bounce of LQC produces a strong enhancement of the non-Gaussianity as compared to that generated by inflation alone, producing  values for the  function $f_{_{\rm NL}}(k_1,k_2,k_3)$ as large as $10^4$ for some wave-numbers and for some choices of the free parameters of the model. Recall that inflation alone produces  $f_{_{\rm NL}}$ of order of $10^{-2}$. (See section \ref{sec:numres}.)\\

(3) The large enhancement of non-Gaussianty raises concerns about the validity of perturbation theory. We have computed higher order contributions to correlation functions and found that they are small compared to the leading order result. Hence, perturbation theory remains a valid tool to compute the primordial power spectrum and bispectrum of cosmological perturbations in LQC. (See section \ref{sec:stpertb}.)  \\

(4) The non-Gaussianity produced by the LQC bounce is {\em strongly scale dependent}. The  bounce introduces a new scale, determined by the Ricci spacetime  curvature scalar at the bounce, $R_{\rm B}$. For perturbations,  this new scale can be written as $k_{\rm LQC}\equiv  a_{\rm B}\sqrt{R_{\rm B}/6}$---or, equivalently, in terms of the energy density at the bounce, $\rho_{\rm B}$, as $k_{\rm LQC}\equiv  a_{\rm B}\sqrt{\kappa\, \rho_{\rm B}}$. Fourier modes with comoving wave-numbers $k\gg k_{\rm LQC}$ are not affected by the bounce, and their primordial non-Gaussianity  originate entirely from the inflationary phase and are small. On the contrary, for Fourier modes that are infra-red enough to ``feel'' the bounce, i.e.,\  $k\lesssim k_{\rm LQC}$, the bounce contributes significantly to their non-Gaussianty. We have provided an analytical argument to understand  the enhancement observed in our numerical computations, and concluded that it is given by $|f_{_{\rm NL}}(k_1,k_2,k_3)|\propto e^{-\alpha\,(k_1+k_2+k_3)/k_{\rm LQC}}$, with $\alpha\approx 0.65$. (See section \ref{sec:analytical}.)\\

(5) The non-Gaussianty generated by the LQC bounce has a very particular ``shape'', discussed in section \ref{sec:bispectrum}, that can be used to differentiate the results for  LQC from other models of the early universe. Namely,  in addition to  the scale-dependence mentioned above, $f_{_{\rm NL}}(k_1,k_2,k_3)$ peaks in the flattened-squeezed configurations. ( See section \ref{sec:bispectrum}).\\

(6) The function $f_{_{\rm NL}}(k_1,k_2,k_3)$  is {\em highly oscillatory} with respect to the wave numbers $k_1,k_2,k_3$.\\ 

(7) Non-Gaussianity is {\em more sensitive} to the bounce than the power spectrum. For both the power spectrum and non-Gaussianity, the relative size of the modifications  that the bounce introduces  decreases for large wave-numbers $k$, and becomes negligible for $k\gg k_{\rm LQC}$. However, the effects  on the power spectrum disappear `faster'  than  on $f_{_{\rm NL}}$, when we move towards larger $k$. As a consequence, there is an interval of wave-numbers, given approximately by $k\in(2\, k_{\rm LQC}, 10 \, k_{\rm LQC})$ for which the modifications in the power spectrum are already negligible, but they are still important in non-Gaussianity.\\%

(8) Impact of different choices of the free parameters in the model.

\begin{itemize}
\item A change in the value of the scalar field at the bounce $\phi_{\rm B}$ increases the number of e-folds of expansion between the bounce and the beginning of inflation, and this produces a  {\em shift} of the function $f_{_{\rm NL}}(k_1,k_2,k_3)$ relative to the interval of wave-numbers that  we can directly observe in the CMB. Increasing $\phi_{\rm B}$ produces a shift of $f_{_{\rm NL}}(k_1,k_2,k_3)$ towards infra-red scales with respect to the observable window. This effect was known to happen for the power spectrum (see, e.g.,\ \cite{Agullo:2012fc}), and we have shown that it also occurs for non-Gaussianty. (See section \ref{sec:phiB}.)  

\item A change in the value of the energy density of the scalar field at the bounce,  $\rho_{\rm B}$, produces  also a {\em shift} on the function $f_{_{\rm NL}}(k_1,k_2,k_3)$, {\em together with a change in its amplitude}. For the power spectrum, the effects of changing $\phi_{\rm B}$ and $\rho_{\rm B}$ compensate each other (except for extreme infra-red scales), and therefore their consequences cannot be individually distinguished. This degeneracy is broken for the bispectrum. (See section \ref{sec:rhoB}.)

\item The contribution from the bounce to $f_{_{\rm NL}}(k_1,k_2,k_3)$ is largely insensitive to the choice of the scalar field's potential. We have checked this by comparing the result for $f_{_{\rm NL}}(k_1,k_2,k_3)$ obtained with two commonly used potentials: the quadratic and the Starobinsky potential.  (See section \ref{sec:Starobinsky}.)

\item The predictions for $f_{_{\rm NL}}(k_1,k_2,k_3)$ are unchanged for several different choices of initial quantum vacuum states for scalar perturbations, provided this initial state  is specified at a time well before the bounce, when all modes of interest are in the adiabatic regime (see section \ref{sec:inistate}). On the contrary, we find that the result for $f_{_{\rm NL}}(k_1,k_2,k_3)$ is sensitive to the choice of initial data for perturbations {\em if they are specified at or close to the bounce}. This  does not happen for the power spectrum and shows again that non-Gaussianity is more sensitive to the physics of the bounce than the power spectrum. (See section \ref{sec:inistate}).

\end{itemize}

Finally, we discuss the observational perspective of our analysis in regard of the current and forthcoming constraints on primordial non-Gaussianity.

The Planck Collaboration reported results on their search for non-Gaussianty in the CMB in \cite{Ade:2015ava}. They were unable to confirm any primordial non-Gaussianity, and provided  tight constraint on different models of the early universe. These constraints are rather strong for models producing {\em scale-invariant}\footnote{These are models for which  $f_{_{\rm NL}}(k_1,k_2,k_3)$ does not change when the three wave-numbers are simultaneously re-scaled.} non-Gaussianity. They found $f_{_{\rm NL}}^{\rm local}=0.8\pm 5.0$ for the  local template, $f_{_{\rm NL}}^{\rm equil}=-16\pm 70$ for the equilateral template, and $f_{_{\rm NL}}^{\rm ortho}=-34\pm 33$ for the orthogonal one, at $68\%$ confidence level \cite{Ade:2015ava}. These results provide little information about  models  with scale-dependent non-Gaussianity, especially on large angular scales. In those cases the comparison with observational data must be done individually for each model. Recall that due to the sampling variance observational error bars  at low multipole scale approximately as $1/\sqrt{\ell}$, where $\ell$ is the angular multipole. Planck observational error bars are smaller for large multipoles, attaining uncertainties $\Delta f_{_{\rm NL}} \approx \pm 10$ for $\ell \gtrsim 1000$. If $f_{_{\rm NL}}$ is assumed to be scale-invariant, then the  precision at large multipoles suffices to constrain $f_{_{\rm NL}}$ with great accuracy at all scales. The situation is different for scale-dependent $f_{_{\rm NL}}$, as the one we obtained. Nevertheless, we can still find  estimates for the constraints that Planck data implies for the parameters of our model. We found that $f_{_{\rm NL}}$ is of order $10^{-2}$ for large wave-numbers, and then it increases for small wave-numbers, reaching values of order $10^{3}$. In order to respect observational constraints, the enhancement of $f_{_{\rm NL}}$ may only occur for the largest scales probed by the CMB data, for which  error bars are large.  It is important to emphasize that, {\em the non-Gaussianity generated by the bounce has a shape that allows having large non-Gaussianty at low multipole, while being consistent with observational constraints at large multipoles of the CMB.}
 
Taking a conservative viewpoint, we require that the non-Gaussianty generated by the bounce shall only appear for multipoles $\ell \lesssim 50$ (which corresponds  to $k\lesssim 2k_*$, for $k_*=0.002{\rm Mpc}^{-1}$). Recall that the values of  $\phi_{\rm B}$ and $\rho_{\rm B}$ control the scales  at which  the effects from the bounce would manifest themselves in the CMB. Therefore, observational constraints on non-Gaussianity translate into a restriction for the permissible values of  $\phi_{\rm B}$ and $\rho_{\rm B}$; see Table 1.

  \begin{table}
\centering
    \begin{tabular*}{6.5cm}{@{\extracolsep{\fill}}ccc}
    \hline
    \hline
      $\rho_{\rm B}$ & $\phi_{\rm B}(\rm min)$ & $\phi_{\rm B}(\rm max)$ \\
    \hline
$0.2\, M_{P\ell}^4$ & $8.05 \, M_{P\ell}$& $8.41 \, M_{P\ell}$\\
$0.5\, M_{P\ell}^4$ & $7.70 \, M_{P\ell}$ & $8.08 \, M_{P\ell}$\\
$1\, M_{P\ell}^4$ & $7.46\, M_{P\ell}$ & $7.82 \, M_{P\ell}$\\
$2\, M_{P\ell}^4$ & $7.19\, M_{P\ell} $ & $7.58 \, M_{P\ell}$ \\
$5\, M_{P\ell}^4$ & $6.88 \, M_{P\ell}$ & $7.24 \, M_{P\ell}$\\ 
    \hline
    \hline
    \end{tabular*}
\caption{In this table $\phi_{\rm B}(\rm min)$ represents the minimum value of $\phi_{\rm B}$ for different values of $\rho_{\rm B}$ obtained from a conservative application of observable constraints on non-Gaussianity. On the other hand, $\phi_{\rm B}(\rm max)$ is the maximum value of $\phi_{\rm B}$ for which the enhancement of non-Gaussianity produced by the bounce appears in observable scales. We emphasize that values of $\phi_{\rm B}$ larger than $\phi_{\rm B}(\rm max)$ are allowed, but for them the bounce does not produce any direct effect  in the CMB, neither in the power spectrum nor in  non-Gaussianity, and hence the results agree with those obtained from standard inflation. The numbers in this table are obtained by using the quadratic potential with the mass parameter fixed by the Planck normalization, $m=6.4\times10^{-6}M_{P\ell}$. \label{valuetable}}
\end{table}
 
As mentioned earlier, the enhancement that  the bounce produces in the power spectrum  appears for smaller wave-numbers than the enhancement in non-Gaussianty. This implies that if $\phi_{\rm B}$ is chosen to be equal or larger than $\phi_{\rm B}(\rm min)$, in such a way that the LQC-effects on non-Gaussianity appear only for $\ell \lesssim 50$,  then the LQC-effects in the power spectrum would appear only for the first few multipoles $\ell \lesssim 5$, and would be difficult to observe.

However, one should keep in mind this analysis corresponds to the most conservative application of observational constraints. It is likely that the oscillatory character of the non-Gaussianity found in this paper may partially attenuate some of its effects in the CMB, and such attenuation would relax the restrictions on $\phi_{\rm B}$. For this reason, the numbers given above, and the conclusions extracted from them, are meant  to be taken as  ``worse-case scenario'', rather than a strict constraint. \\

 Regarding observational consequences of the non-Gaussianity generated by the bounce, we point out two possibilities. On the one hand, although the CMB  has been the main  source of information about primordial perturbations, the large scale structure will take this role  in the near future \cite{Alvarez:2014vva}. The characteristic shape of the non-Gaussianity produced by a bounce obtained in this paper could then be used as the smoking gun to contrast our findings with future observations of the large scale structure. 

On the other hand, even though error bars for non-Gaussianity in CMB observations are too large to directly observe the predictions obtained in this paper, it was recently emphasized in \cite{Schmidt:2012ky,Agullo:2015aba} that this non-Gaussianity can modify the power spectrum at low multipoles, via higher order effects known as non-Gaussian modulation of the power spectrum. A detailed analysis shows that these effects can be large enough to be observable for multipoles $\ell\lesssim 30$ in the power spectrum, and that they are expected to produce effects very similar to the anomalous features that the Planck and WMAP missions have observed at low angular multipoles in the CMB, and that remain unexplained at the present time (see \cite{Ade:2015hxq,Dai:2013kfa}  for a detailed account of the observational aspects of these anomalies, and their statistical significance). The possibility that these  features could originate from a bounce that takes place before inflation, as the one predicted by LQC, is exciting, and the quantitative details are worth to be explored.\\

\section*{Acknowledgments}
We thank Abhay Ashtekar, Aur\'elien Barrau, Eugenio Bianchi, Martin Bojowald, Beatrice Bonga, Robert Brandenberger,  Jens Chluba, Brajesh Gupt, Jakub Mielczarek, Javier Olmedo,  Patrick Peter, Jorge Pullin,  Sahil Saini, Parampreet Singh, Siddharth Soni, Lakshmanan Sriramkumar, and Edward Wilson-Ewing  for discussions. We acknowledge the use of high performance computing resources  provided by Louisiana State University (http://www.hpc.lsu.edu), Baton Rouge, U.S.A. This work was supported by the NSF CAREER grants PHY-1552603, NSF grant PHY-1603630,  and funds of the Hearne Institute for Theoretical Physics. BB acknowledges financial support from the Franco-American Fulbright Commission, ERC consolidator grant 725456 and ENS Lyon.

\section*{Appendix A : Explicit form of the constraints up to third order}

In this appendix we write the explicit form of the scalar and vector constraints of general relativity, written in equations (\ref{scons}), around a FLRW background, up to third order in perturbations. For simplicity, we only show terms involving scalar perturbations, and  after gauge fixing $\gamma_1=\gamma_2=0$. These expressions have been used in section \ref{sec:classperts} to derive the second and third order Hamiltonians for scalar perturbations.

\bea   \mathbb S^{(0)}&=&  -\f{\kappa\,\pi_a^2}{12\,a} +\f{\pp^2}{2\,a^3}\, +a^3\, V(\phi)=0\, . \nonumber \\
\mathbb V_i^{(0)}&=&0\, . \nonumber \\
\mathbb S^{(1)}(\v x)\, &=&\,   \f{\pp\,}{a^3}\, \dpp(\vec{x})\,   -\, \f{\kappa\, \, \pi_a}{\sqrt{3}\,a^2}\, \pi_1(\v x)\, +\, a^3\, V_{\phi}\, \delta\phi(\v x) \, .\nonumber \\
\mathbb V^{(1)}_i(\v x)\, &=&\, \pp\, \partial_i\delta\phi(\v x)\,   -\, \f{2}{\sqrt{3}}\,\partial_i\pi_1(\v x)\, -\, 2\,\sqrt{\f{2}{3}} \partial_i\pi_2(\v x)\,.\nonumber \\
\mathbb S^{(2)}(\v x) &=&\, \f{1}{2\, a^3}\dpp^2(\v x)\, 
 -\, \f{\kappa}{a^3}\, \pi_1^2(\v x)\, -\, \f{\kappa}{ a^3}\, \pi_2^2(\v x)\, 
 +\, \f{1}{2}\, a^3\, \partial_i\delta\phi(\v x)\,\partial^i\delta\phi(\v x)\, 
+\nonumber\\
 && +\, \f{3\, \kappa\, \partial_i\partial_j\partial^{-2}\pi_2(\v x)\, \partial^i\partial^j\partial^{-2}\pi_2(\v x)}{a^3}\,
 +\, \f{a^3\, V_{\phi\phi}\, }{2} \, \delta\phi^2(\v x)\, . \nonumber \\
\mathbb V^{(2)}_i(\v x)\,&=&\, \dpp(\v x)\,  \partial_i \, \delta\phi (\v x)\, . \nonumber \\
\mathbb S^{(3)}(\v x)\, &=&\, \f{a^3}{6}\, V_{\phi\phi\phi}\, \delta\phi^3(\v x)\label{eq:HC3}\, .\eea
In these expressions, the subscripts $\phi$ in the potential $V(\phi)$ indicate derivative with respect to $\phi$.
The third order vector constraint $\mathbb V_i^{(3)}(\v x)$ appears in the Hamiltonian multiplied by $\delta N^i$, which itself is linear in perturbations.  Therefore, $\mathbb V_i^{(3)}(\v x)$  does not contribute to the third order Hamiltonian in the spatially flat gauge.

\section*{Appendix B: Explicit expressions for $f_1(\eta)$, $f_2(\eta)$, and $f_3(\eta)$}

Expressions of the functions $f_1(\eta)$, $f_2(\eta)$, and $f_3(\eta)$ appearing in expression (\ref{Bphi}), section \ref{sec:bisp}:
\bea
f_1(\eta)\, &=&\, a\,\biggl[\, 2\, \biggl(\, \f{243\,\pp^7}{2\,\kappa\,a^8\,\pi_a^5}\, 
-\, \f{81\,\pp^5}{2\,a^6\,\pi_a^3}\, +\, \f{27\,\kappa\,\pp^3}{8\,a^4\,\pi_a}\, 
+\, \f{81\,\pp^4\,V_{\phi}}{\kappa\,a\,\pi_a^4} - \f{27\,a\,\pp^2\,V_{\phi}}{2\,\pi_a^2}\,+\, \f{27\,a^6\,\pp\,V_{\phi}^2}{2\,\kappa\,\pi_a^3}\biggr)\,\nonumber\\
&&\, \times\biggl(\, 1\, -\ \f{(\v k_1 \cdot \v k_2)^2}{k_1^2\,k_2^2}\,\biggr)\, 
+\, \f{3\,a^2\,\pp}{\pi_a}\,\v k_1 \cdot \v k_2\,+\, \f{9\,a\,\pp^2\,V_{\phi}}{\pi_a^2}\, -\, 
\f{3\,a^2\,\pp\,V_{\phi\phi}}{\pi_a}\, +\, \f{a^3\,V_{\phi\phi\phi}}{3}\,\biggr]\,, \\
f_2(\eta)\, &=&\, a^3\,\biggl[\, \biggl(\, \f{81\,\pp^5}{\kappa\,a^7\,\pi_a^4}\, -\, \f{27\,\pp^3}{2\,a^5\,\pi_a^2}\, 
+\,\f{27\,\pp^2\,V_\phi}{\kappa\,\pi_a^3}\,\biggr)\biggl(\, 2\, -\, \f{(\v k_1 \cdot \v k_3)^2}{k_1^2\,k_3^2}\, 
-\, \f{(\v k_2 \cdot \v k_3)^2}{k_2^2\,k_3^2}\,\biggr)\, -\, \f{9\,\pp^3}{a^5\,\pi_a^2}\, \nonumber\\
&&\, +\, \biggl(\, \f{-3\, \kappa\,\pp}{2\,a^3}\, +\, \f{9\,\pp^3}{a^5\,\pi_a^2}\, +\, \f{3\,a^2\,V_{\phi}}{\pi_a}\,\biggr)\,
\biggl(\, \f{\v k_1\cdot \v k_2}{k_1^2}\, +\, \f{\v k_1\cdot \v k_2}{k_2^2}\,\biggr)\,\biggr]\, ,\\
f_3(\eta)\, &=&\, a^5\,\biggl[\, \f{27\,\pp^3}{\kappa\,a^6\,\pi_a^3}\,\biggl(\,1\, -\, \f{(\v k_2\cdot \v k_3)^2}{k_2^2\,k_3^2}\,\biggr)\, 
-\, \f{3\,\pp}{a^4\,\pi_a}\, +\, \f{3\,\pp}{a^4\,\pi_a}\,\biggl(\, \f{\v k_1\cdot\v k_3}{k_3^2}\, +\, \f{\v k_1 \cdot \v k_2}{k_2^2}\biggr)\, \biggr]\, .
\eea

\bibliography{Refs}

\end{document}